\RequirePackage{etoolbox}
\csdef{input@path}{%
 {img/}
}%

\documentclass[ba]{imsart}
\pubyear{0000}
\volume{00}
\issue{0}
\doi{0000}
\firstpage{1}
\lastpage{1}

\usepackage{amsthm}
\usepackage{amsmath}
\usepackage{natbib}
\usepackage[colorlinks,citecolor=blue,urlcolor=blue,filecolor=blue,backref=page]{hyperref}
\usepackage{graphicx}

\startlocaldefs
\numberwithin{equation}{section}
\theoremstyle{plain}

\endlocaldefs

\begin{document}

\begin{frontmatter}
\title{Improving multilevel regression and poststratification with structured priors}
\runtitle{Improving MRP with structured priors}

\begin{aug}
\author{\fnms{Yuxiang} \snm{Gao}\thanksref{addr1}\ead[label=e1]{ygao@utstat.toronto.edu}},
\author{\fnms{Lauren} \snm{Kennedy}\thanksref{addr2}\ead[label=e2]{lak2183@columbia.edu}},
\author{\fnms{Daniel} \snm{Simpson}\thanksref{addr3}\ead[label=e3]{simpson@utstat.toronto.edu}}
\and
\author{\fnms{Andrew} \snm{Gelman}\thanksref{addr4}
\ead[label=e4]{gelman@stat.columbia.edu}}

\runauthor{Y. Gao, L. Kennedy, D. Simpson and A. Gelman}

\address[addr1]{Department of Statistical Sciences, University of Toronto, Canada, \printead{e1}
}

\address[addr2]{Columbia Population Research Center and Department of Statistics, Columbia University, New York, NY, \printead{e2}
}

\address[addr3]{Department of Statistical Sciences, University of Toronto, Canada, \printead{e3}
}

\address[addr4]{Department of Statistics and Department of Political Science, Columbia University, New York, NY, \printead{e4}
}

\end{aug}

\begin{abstract}
A central theme in the field of survey statistics is estimating population-level quantities through data coming from potentially non-representative samples of the population. Multilevel regression and poststratification (MRP), a model-based approach, is gaining traction against the traditional weighted approach for survey estimates. MRP estimates are susceptible to bias if there is an underlying structure that the methodology does not capture. This work aims to provide a new framework for specifying structured prior distributions that lead to bias reduction in MRP estimates. We use simulation studies to explore the benefit of these prior distributions and demonstrate their efficacy on non-representative US survey data. We show that structured prior distributions offer absolute bias reduction and variance reduction for posterior MRP estimates in a large variety of data regimes.
\end{abstract}



\end{frontmatter}

\section{Introduction}

Multilevel regression and poststratification (MRP) is an increasingly popular tool for adjusting a non-representative sample to a larger population. In particular, MRP appears to be effective in areas where conventional design-based survey approaches have traditionally struggled, notably small-area estimation \citep{rao2014small, pfeffermann2013new, zhang2014multilevel} and with convenience sampling \citep{wang2015forecasting}.

One difference between MRP and traditional poststratified design-based weights is that MRP uses partial pooling. Simple poststratification has difficulties with empty cells, in which case the usual practice is to poststratify only on marginals (thus ignoring interactions), or pool cells together. In contrast, the partial pooling of multilevel modeling automatically regularizes group estimates.

Although other options for regularization with poststratification have been explored \citep{rpp,barp2019}, applications of MRP typically assume independent group-level errors. For example, in a political poll modeling there are varying intercepts for states using a regression on region indicators, state-level predictors such as previous voting patterns in the state, plus independent errors at the state level.  In some applications though, there is potential benefit from including underlying structure not captured by regression predictors. We demonstrate that this structure can be captured through more complex prior specifications. For example, instead of independent errors for an ordered categorical predictor, we specify an autoregressive structure instead. Ordered predictors are just one example where we can introduce \textit{structured prior distributions}. 

We begin this manuscript by providing an overview of the existing literature on methods that adjust for nonrepresentative data, as well as a review on the common application areas of MRP.

\subsection{Existing literature on post-sampling adjustments for non-representativeness and MRP}

Post-sampling adjustments aim to correct for differences between a potentially biased sample and a target population. Poststratification is a commonly used weighting procedure for nonresponse in model-based survey estimates \citep{little1993post}. It can improve accuracy of estimates but is no silver bullet, since the quality of poststratified estimates depends on the quality of the known information about the population sizes of the strata, along with the assumption that the sample is representative of the population within each poststratification cell. An approximation to poststratification is raking, which is an iterative algorithm using marginal totals \citep{deming1940least, lohr2009sampling, skinner2017introduction}. When adjusting for many factors, raking can yield unstable estimates caused by high variability of the adjusted weights \citep{izrael2009extreme}. For a modern overview of current methods of inference and post-sampling adjustments for nonprobability samples, see \cite{elliott2017inference}. As the demands for small-area estimation increase, so too should the utility of MRP. We use structured priors in our proposed improvement for MRP, with the aim of more sensible shrinkage of posterior estimates that should ultimately reduce estimation bias.

MRP has been used in a broad range of applied problems ranging from epidemiology \citep{zhang2014multilevel, downes2018multilevel} to social science \citep{twins, wang2015forecasting, trangucci2018voting}. MRP's beginnings saw applications in political science \citep{gelman1997poststratification,park2004bayesian} for the estimation of state-level opinions from national polls. The breadth of its applications has since matured substantially, even to the extent of being used by data journalists \citep{economist}. One of MRP's appeals to applied researchers is the ability to produce reliable estimates for small areas in the population and simultaneously adjust for non-representativeness.

On the methodology front, \cite{ghitza2013deep} and \cite{gelman2016using} extended MRP to include varying intercepts and slopes for interactions, along with inference for time series of polls.

\subsection{Outline for this paper}

This work explores alternative regularization techniques with structured prior distributions that lead to absolute bias reduction in MRP estimates. Our methodology of structured priors should not be confused with that of \cite{si2017bayesian}, who define structured priors as a way to perform variable selection for higher-order interaction terms. Our improvements on estimation precision come from replacing independent distributions of varying coefficients with Gaussian Markov random fields \citep{rue2005gaussian}.

This paper is structured as follows: Section 1 provides an overview of the existing literature for MRP and nonrepresentative data. Section 2 gives a concise overview of MRP and what's required for the methodology. Section 3 describes our structured priors framework in detail, along with motivation for their use in MRP. Section 4 presents simulation studies of structured priors across various regimes of non-representative survey data. Explanation of the simulation setup and interpretation of the simulation results are given. Bias and variance comparisons are made between structured priors and the classical independent random effects in MRP in section 4. Section 5 contains the application of structured priors in MRP to a real survey data set that's non-representative. Section 6 is the conclusion. All computation carried out in this manuscript was done in R \citep{rteamcitation}.

\section{Overview of MRP}

Multilevel regression and poststratification \cite{gelman1997poststratification} proceeds by fitting a hierarchical regression model to survey data, and then using the population size of each poststratification cell to construct weighted survey estimates. More formally, suppose that the population contains $K$ categorical variables and that the $k^{\text{th}}$ has $J_k$ categories. Hence the population can be represented by $J=\prod_{k=1}^K J_k$ cells. Usually the population contains continuous variables, and in that case these variables will be discretized to form categorical variables. For example, age in a demographic study can be discretized into a finite number of categories. For every cell, there is a known population size $N_j$. Increasing the number of groups for a continuous variable will increase the number of cells $J$ and correspondingly decrease the individual cell population sizes $N_j$. 

Choosing the optimal group size for continuous variables is a difficult model selection problem, involving tradeoffs between accuracy and computational load, and this is something that we do not address in this manuscript. This area of MRP research is an active field and the best practice has yet to emerge from the research. 

Suppose that the response variable of individual $i$ is $y_i \in \{0, 1\}$. MRP for binary survey responses is summarized by the two steps below:

\textbf{Multilevel regression step}. Let $n$ be the number of individual observations in a dataset. Fit the hierarchical logistic regression model below to get estimated population averages $\theta_j$ for every cell $j \in \{1, \dots, J \}$. The hierarchical logistic regression portion of MRP has a set of varying intercepts $\{ \alpha_{j*}^k \}_{{j*}=1}^{J_k}$ for each categorical covariate $k$, which have the effect of partially pooling each $\theta_j$ towards a globally-fitted regression model, $X_j\beta$, with sparse cells benefiting the most from this regularization. $X_j$ is the row in the design matrix that corresponds to $\theta_j$, where $j \in \{1,\dots,J \}$. We follow a notation consistent with \cite{gelman2006data}.
\begin{align*}
    \mbox{Pr} (y_i = 1) &= \text{logit}^{-1} \left(X_i\beta + \sum_{k=1}^K \alpha^k_{j[i]} \right),\text{ for } i=1,\dots, n\\
    \alpha_{j}^k \mid \sigma^k &\overset{\text{ind.}}{\sim} \mbox{N}(0, (\sigma^k)^2), \text{ for } k=1,\dots,K ,\text{ } j=1,\dots, J_k\\
    \sigma^k &\sim \mbox{N}_+ (0,1), \text{ for }\text{ for } k = 1,\dots, K\\
    \beta &\sim \mbox{N}(0,1),
\end{align*}
where we are giving default weakly informative priors to the non-varying regression coefficients $\beta$.

\textbf{Poststratification step}. Using the known population sizes $N_j$ of each cell $j$, poststratify to get posterior preference probabilities at the subpopulation level. The poststratification portion of MRP adjusts for nonresponse in the population by taking into account the sizes of every cell $l$ relative to the total population size $N= \sum_{j=1}^J N_j$. Another way to interpret poststratification is as a weighted average of cell-wise posterior preferences, where the weighting scheme is determined by the size of each cell in the population. Smaller cells get downweighted and larger cells get upweighted. The final result is a more accurate estimate in the presence of non-representative data.

Let $S$ be some subset of the population defined based on the poststratification matrix. Then the poststratified estimand for $S$ is:

\begin{align*}
    \theta_S &:= \frac{\sum_{j\in S}N_j {\theta}_j}{\sum_{j \in S} N_j}
\end{align*}

For example, $S$ could correspond to the oldest age category in the lowest income bracket. Then $\theta_S$ would correspond to the proportion of people in this sub-population that would respond yes to the survey question of interest. It's important to note that, one can model at a finer scale than the poststratification scale.

\section{Proposed Approach and Motivation}

We consider structured prior distributions for MRP taking the form of Gaussian Markov random fields (GMRF), modeling certain structure of the underlying categorical covariate in the hierarchical regression. We proceed as follows for a covariate in the population of interest:

\textbf{Case 1:} If we do not want to model any structure in a categorical covariate, we model its varying intercepts as independently normally distributed. This would be what's described in the previous section, Overview of MRP.

\textbf{Case 2:} If there is underlying structure we would like to model in a covariate, and spatial smoothing using this structure seems sensible for the outcome of interest, then we use an appropriate GMRF as a prior distribution for this batch of varying intercepts.

We will specify informative hyperpriors when possible and model via a full Bayesian approach. For a detailed overview of principled hyperprior specification in GMRF models, we refer the reader to \cite{simpson2017penalising}. As well, we do not restrict structured priors to have directed or undirected conditional distributions \citep{rue2005gaussian}. Some examples of directed conditional distributions include the autoregressive and random walk processes with discrete time indices, which are frequently used in time series analysis. The CAR and ICAR processes \citep{besag1975statistical} are common undirected conditional distributions and are often used in specifying priors in spatial models.

More complex prior structure allows for nonuniform information-borrowing in the presence of non-representative surveys from a population. For example, it makes sense to partially pool inferences for the oldest age group toward data from the second-oldest group. An autoregressive prior placed on the ordinal variable age achieves this effect, without making the strong global assumptions involved in simply including age as a linear or quadratic predictor in the regression. The proposal of using structured priors aims to reduce bias for MRP estimates in extremely non-representative data regimes. 

Structured priors improve upon the multilevel aspect of MRP while maintaining the regression structure. Because MRP is a model-based survey estimation approach, the multilevel regression component can be replaced with other forms of regression modelling, for example with sparse hierarchical regression \citep{goplerud2018sparse} or Bayesian additive regression trees \citep{barp2019}. It is important, though, that the regression step be regularized in some way to preserve the ability of the method to account for a potentially large number of adjustment factors and their interactions \citep{rpp}. When compared to machine learning-style regularization methods as seen in \cite{barp2019} and \cite{goplerud2018sparse}, structured priors offer a lot more interpretability in the modelling step of MRP. For example, it's not as clear how Bayesian additive regression trees (BART) allow for information-borrowing for structured covariates despite BART and related regularized tree-based methods \citep{chen2016xgboost} being modern methods in out-of-sample prediction. In contrast to this, an AR(1) prior with autoregression coefficient $\rho \in [0,1)$ on the ordinal variable age has the clear interpretation that posterior estimates are regularized towards their previous first-order neighbouring age, where the amount of regularization is determined by the coefficient $\rho$.

GMRFs have a deep connection with Gaussian Processes (GPs). As the discretization of the underlying space gets finer, under certain technical conditions, a GMRF will converge to a specific GP  \citep{lindgren2011explicit}. More details on the style of convergence can be found in \cite{lindgren2011explicit}. These theoretical foundations show that structured priors in MRP are a discrete approximation to GP priors for structured covariates. GPs are universal function approximators, hence structured priors in MRP can be thought of as a flexible modeling strategy that simultaneously takes into account various data-generating processes and offers interpretable regularization.

\newpage

\section{Simulation Studies}

\subsection{Directed structured priors example: Models for partial pooling of group-level errors}

For the first simulation example, we work with a simple model of three poststratification categories---51 states, age in years ranging from 21--80, and income in 4 categories---and no other predictors. Age is further categorized into 12 groups. In practice, the number of categories that age is discretized into for MRP models is a lot smaller than the total number of possible integer ages 
\citep{twins,trangucci2018voting}. 12 categories were chosen since it was large enough to allow for prior distribution structure to introduce intelligent information-borrowing in the age covariate yet it was a lot less than the maximal number of age categories, 60. We define $\alpha_{j[i]}^\text{Age Cat.}$, $\alpha_{j[i]}^\text{Income}$ and $\alpha_{j[i]}^\text{Region}$ to be the varying intercepts for age category, income category and region respectively for the  $i^{th}$ survey respondent. Phone surveys will often have these three covariates on respondents, often times with the nonrepresentativeness of the survey driven by varying nonresponse rates across ages in the population.

For all three prior specifications of MRP, we use the link function,
\begin{equation}
    \begin{aligned}
\mbox{Pr} (y_i = 1) &= \text{logit}^{-1} \left(\beta^0 + \alpha^\text{State}_{j[i]} + \alpha_{j[i]}^\text{Age Cat.} + \alpha_{j[i]}^\text{Income} \right),\text{ for }i = 1,\dots,n.\\
    \end{aligned}
  \end{equation}
  For all three prior specifications we assume independent mean-zero normal distributions for the $\alpha_{j[i]}^\text{Region}$'s, $\alpha_{j[i]}^\text{Age Cat.}$'s and $\alpha_{j[i]}^\text{Income}$'s along with a weakly informative half-normal distribution for the corresponding scale parameter:
\begin{equation}
    \begin{aligned}
\alpha^{\text{State}}_j \mid \beta^\text{State-VS}, \beta^\text{Relig.}, ( \alpha_{j^*}^\text{Region} )_{j^*=1}^5, \sigma^\text{State}  &\overset{\text{ind.}}{\sim} \mbox{N}(\alpha_{m[j]}^\text{Region} + \beta^\text{Relig.} X_{\text{Relig.},j}\\
& + \beta^\text{State-VS} X_{\text{State-VS}, j}, (\sigma^\text{State})^2 ),\\
&\text{ for }j=1,\dots,51 \\
\alpha^\text{Region}_m \mid \sigma^\text{Region} &\overset{\text{ind.}}{\sim} \mbox{N}(0, (\sigma^\text{Region})^2),  \\
&\text{ for }m=1,\dots,5 \\
 \alpha_j^\text{Income} \mid \sigma^\text{Income}  &\overset{\text{ind.}}{\sim} \mbox{N}(0, (\sigma^\text{Income})^2),\\
 &\text{ for }j = 1,\dots,4 \\
    \sigma^\text{Income}, \sigma^\text{State}, \sigma^\text{Region}  &\sim \mbox{N}_+(0,1)\\
\beta^\text{State-VS}, \beta^\text{Relig.}, \beta^0 &\sim  \mbox{N}(0,1)\\
    \label{eq:incomepriors}
    \end{aligned}
\end{equation}
where $X_{\text{State-VS},j} \in [0,1]$ is the covariate that corresponds to the 2004 Democratic vote share for state $j$ and $X_{\text{Relig.},j} \in [0,1]$ is the percentage of conservative religion in state $j$, which is defined as the sum of the percentage of Mormons and percentage of Evangelicals in state $j$. The term $\alpha_{m[j]}^\text{Region} + \beta^\text{Relig.} X_{\text{Relig.},j} + \beta^\text{State-VS} X_{\text{State-VS}, j}$ are state-level predictors that utilize auxillary data accounting for structured differences among the states. 

The \textit{baseline specification} is the classical prior distribution used in MRP with independent normal distributions for the varying intercepts for age categories:
\begin{equation}
  \begin{aligned}
\alpha_j^\text{Age Cat.} \mid \sigma^{\text{Age Cat.}} &\overset{\text{ind.}}{\sim} \mbox{N}(0, (\sigma^\text{Age Cat.})^2) \text{, for }j = 1,\dots,12 \\
\sigma^\text{Age Cat.} &\sim \mbox{N}_+(0,1)
\label{eq:baselinemodel}
\end{aligned}
\end{equation}

The \textit{autoregressive specification} models the ordinal structure of age category as a first-order autoregression \citep{rue2005gaussian}. The prior distribution imposed on $\rho$ is restricted to the range $(-1,1)$, enforcing stationary for the autoregressive process.

\begin{equation}
  \begin{aligned}
\alpha^{\text{Age Cat.}}_1 \mid \rho, \sigma^\text{Age Cat.} &\sim \mbox{N}(0, \frac{1}{1-\rho^2}(\sigma^\text{Age Cat.})^2) \\
\alpha^{\text{Age Cat.}}_j \mid \alpha_{j-1}^\text{Age Cat.},\dots ,\alpha_{1}^\text{Age Cat.}, \rho, \sigma^\text{Age Cat.} &\sim \mbox{N}(\rho \alpha^{\text{Age Cat.}}_{j-1}, (\sigma^\text{Age Cat.})^2), \\
&\text{ for }j = 2,\dots, 12 \\ 
\sigma^\text{Age Cat.} &\sim \mbox{N}_+(0,1) \\
(\rho + 1)/2 &\sim \text{Beta}(0.5, 0.5)
\label{eq:armodel}
\end{aligned}
\end{equation}

Finally, we consider the \textit{random walk specification}, which is a special case of first-order autoregression with $\rho$ fixed as 1, although with a different parameterization to avoid the division by $1-\rho^2$ above. In addition, we introduce the sum-to-zero constraint $\sum_{j=1}^J \alpha_j^\text{Age Cat.} = 0$ to ensure that the joint distribution for the first-order random walk process is identifiable.
\begin{equation}
  \begin{aligned}
\alpha_j^\text{Age Cat.} \mid \alpha_{j-1}^\text{Age Cat.},\dots,\alpha_{1}^\text{Age Cat.}, \sigma^{\text{Age Cat.}} &\sim \mbox{N}(\alpha_{j-1}^\text{Age Cat.}, (\sigma^\text{Age Cat.})^2), \\
&\text{ for }j = 2, \dots, 12 \\
\sigma^\text{Age Cat.} &\sim \mbox{N}_+(0,1) \\
&\sum_{j=1}^J \alpha_j^\text{Age Cat.} = 0
\label{eq:rwmodel}
\end{aligned}
\end{equation}

The three prior specifications differ in the amount of information shared between neighbors in the age category random effect. In the baseline specification, no information is shared between $\alpha_j^\text{Age Cat.}$ and $\alpha_{j-1}^\text{Age Cat.}$ for $ j \in \{2, \dots, 12 \}$. In the autoregressive specification, partial information is shared and in the random walk specification the full amount of information is shared. The sharing of information between $\alpha_j^\text{Age Cat.}$ and $\alpha_{j-1}^\text{Age Cat.}$ is analogous to shrinkage of one posterior towards another. In this case, the posterior of $\alpha_j^\text{Age Cat.}$ shrinks toward the posterior of $\alpha_{j-1}^\text{Age Cat.}$ under both the random walk and autoregressive specifications. The amount of shrinkage is governed by the autoregressive coefficent $\rho$. The reason behind specifying first-order autoregressive processes as the structured prior for age is that individuals with similar ages should have similar opinions for the survey question of interest. First-order autoregressive processes incorporates the prior assumption that the opinion of an individual for a certain age is similar to the opinion of individuals with exactly the same demographics except with a slightly younger age. In the simulation studies below, we empirically show that the property of shrinking towards the previous neighboring variable in the autoregressive and random walk specifications result in decreased posterior bias of MRP estimates for every cell in the population.

\subsubsection{Simulated data}

\textbf{The sample}

We consider three scenarios of true $\mbox{E}(y)$ as a function of age:  U-shaped, cap-shaped, or monotonically increasing. We investigate the effects of non-representative data amongst elderly individuals (ages 61--80) in the simulation samples, and show that the random walk specification provides the lowest absolute bias in subpopulation level estimates when compared to the other two specifications. The likelihood of sampling from a subpopulation group given that individuals respond is dependent on the size of the subpopulation group along with the response probability of an individual in that group. 

The probability vector of sampling is defined as:
\begin{align*}
\frac{\text{(Probability of response)}\odot\text{($N_1,\dots,N_J$)}}{\sum_j \text{(Probability of response)}_j \cdot N_j }
\end{align*}
where $\odot$ is the Hadamard product. This probability vector is in reference to the poststratification matrix defined for this simulation study. A special case for the probability vector of sampling is when the probability of response is equal for all cells in the population, resulting in a probability vector of sampling that's fully representative of the population. The probability vector of sampling is used to generate a sample of binary responses along with covariates. Through this probability vector, one can augment it to get highly non-representative samples for certain subpopulation groups. In the case of a completely random sample for subpopulation groups of interest, all subpopulation groups of interest have the same probability of sampling. As an example, all 12 age categories would have equal probability of being sampled from in the scenario of completely random sampling for age categories.

\noindent \textbf{Assumed sample and population}

In the following simulation study we will assume that the population is sufficiently large so that sampling with replacement is equivalent to retrieving a random sample from the population.

To empirically validate the improvements that structured priors have on posterior MRP estimates, we construct various data regimes for age categories 9--12. More specifically, let $S $ be the index set corresponding to age categories 9--12. Summing the probability of sampling over $S$ will return the expected proportion of the sample who are older adults. We perturb this probability through 9 scenarios, ranging from 0.05 (under-representing older adults) to 0.82 (over-representing older adults). This section contains plots for the U-shaped true preference curve, with the appendix containing plots for the increasing-shaped true preference curve and the cap-shaped true preference curve. 

These three true preference curves capture the rough structure of the unseen truths in real survey data. Let $x$ represent age of an individual, and let $f(x)$ represent the preference curve for age. The three different preference curves with respect to age are defined as:\\

\textbf{Cap-shaped preference:}
\begin{align*}
f(x)=\frac{\Gamma (4)}{\Gamma (2) \Gamma (2)} x (1-x), \text{ } x \in [21,80]
\end{align*}

\textbf{U-shaped preference:}
\begin{align*}
f(x)=1 - \frac{\Gamma (4)}{\Gamma (2) \Gamma (2)} x (1-x), \text{ } x \in [21,80]
\end{align*}

\textbf{Increasing-shaped preference:}
\begin{align*}
    0.7 - 3 \text{exp}(-\frac{x}{0.2}), \text{ } x \in [21,80]
\end{align*}

True preferences for every poststratification cell $j \in \{1,\dots,J\}$ in the population are then generated with the following formula:
\begin{align*}
\theta_j = \text{logit}^{-1}\left( \beta^0 + f(X_{\text{Age}[j]}) + X_{\text{Income}[j]} + \beta^\text{State}X_{\text{State}[j]} + \beta^\text{Relig.}X_{\text{Relig.}[j]} \right)
\end{align*}

where $X_{\text{Age}[j]}$, $X_{\text{Income}[j]}$, $X_{\text{State}[j]}$, $X_{\text{Relig.}[j]}$ correspond to the age, income effect, state effect and religion effect respectively of poststratification cell $j$. $X_{\text{Income}[j]}$, $X_{\text{State}[j]}$, $X_{\text{Relig.}[j]}$ along with $\beta^0, \beta^\text{State}$ and $\beta^\text{Relig.}$ are defined in the appendix.

\subsubsection{Directed structured priors results}

We fit all models using the probabilistic programming language Stan \citep{osti_1430202, team2016rstan} to perform full Bayesian inference, using the default settings of 2000 iterations on 4 chains run in parallel, with half the iterations in each chain used for warmup.

\noindent \textbf{Impact of prior choice on bias of posterior preferences}

The first way we evaluate the impact of prior specification is by considering the impact of bias when we manipulate the expected proportion of the sample that are older adults. In Figure \ref{fig:combined_allmediansfacet_u} below, we plot the results for a sample size of $100$ and $500$. 

When the expected proportion of the sample that are older adults is equal to 0.33, this corresponds to a completely random sample for age categories (probability of sampling every age category is the same) \textit{and} a fully representative sample for age categories (probability of sampling every age category is proportional to the population sizes for every age category). In certain scenarios, a completely random sample may be more desirable than a fully representative sample of the population for modeling purposes. Certainly, oversampling a sparse subpopulation group in the population will return lower variance model estimates for that specific subpopulation group.

\begin{figure}[t]
    \centering
    \includegraphics[width=\textwidth,keepaspectratio]{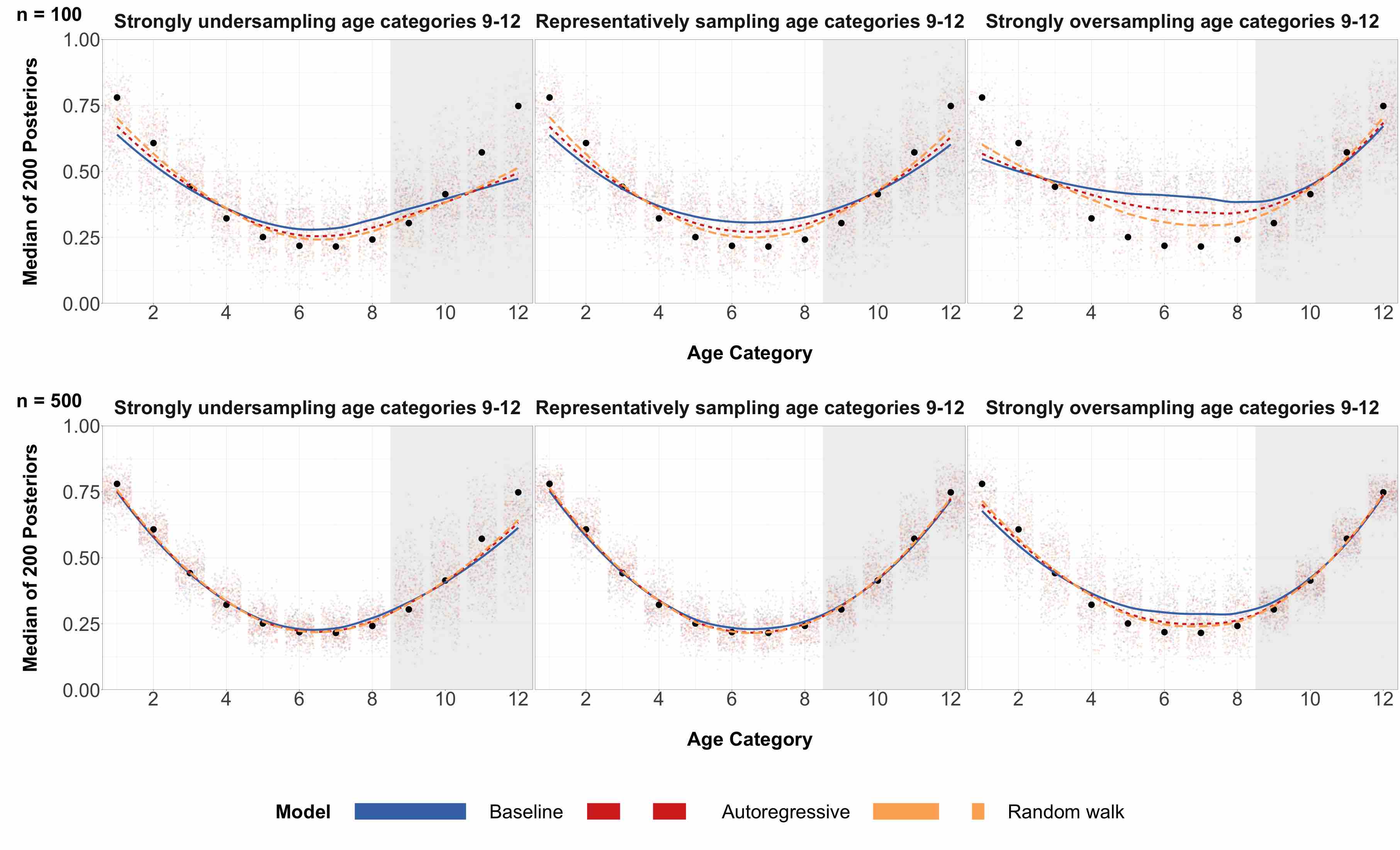}

\caption[short]{Posterior medians for 200 simulations for each age group under three different regimes of data, where true age preference is U-shaped. The top row corresponds to a sample size of 100 and the bottom row corresponds to a sample size of 500. Black circles are true preferences for each age group. The shaded grey region corresponds to the age categories of older individuals for which we over/undersample. The left column has a probability of sampling age categories 9-12 equal to 0.05. The middle column has a probability of sampling age categories 9-12 equal to 0.33, which is completely random sampling \textit{and} representative sampling for all age categories. The right column has a probability of sampling age categories 9-12 equal to 0.82. Local regression is used for the smoothed estimates amongst the three prior specifications. For the same plots involving different probabilities of sampling, refer to Table \ref{Tab:simscenarios} in the appendix.}
\label{fig:combined_allmediansfacet_u}
\end{figure}

We can see from Figure \ref{fig:combined_allmediansfacet_u} that the two structured prior specifications outperform the baseline prior specification by a few percentage points for almost all 12 age categories, and achieving the same performance for the remaining age categories.

When elderly individuals are undersampled relative to the rest of the population, the random walk prior specification outperforms the baseline prior specification in lower absolute bias by a few percentage points across all the age categories. 
    
When elderly individuals are oversampled relative to the rest of the population, the random walk prior specification outperforms the baseline prior specification in lower absolute bias by close to 10 percentage points for mid-aged individuals when sample size is 100. As expected, the three prior specifications produce essentially the same posterior estimates in the bottom row of Figure \ref{fig:combined_allmediansfacet_u}, due to the sample size being large in each of these age categories -- Increasing $n$ will increase the weight of the likelihood on the posterior in a statistical model. Regardless, absolute bias is reduced or stays the same for all age categories and all data regimes for the two structured priors specifications, as seen in Figure \ref{fig:combined_allmediansfacet_u}.

\begin{figure}[t]
    \centering
    \includegraphics[width=\textwidth,keepaspectratio]{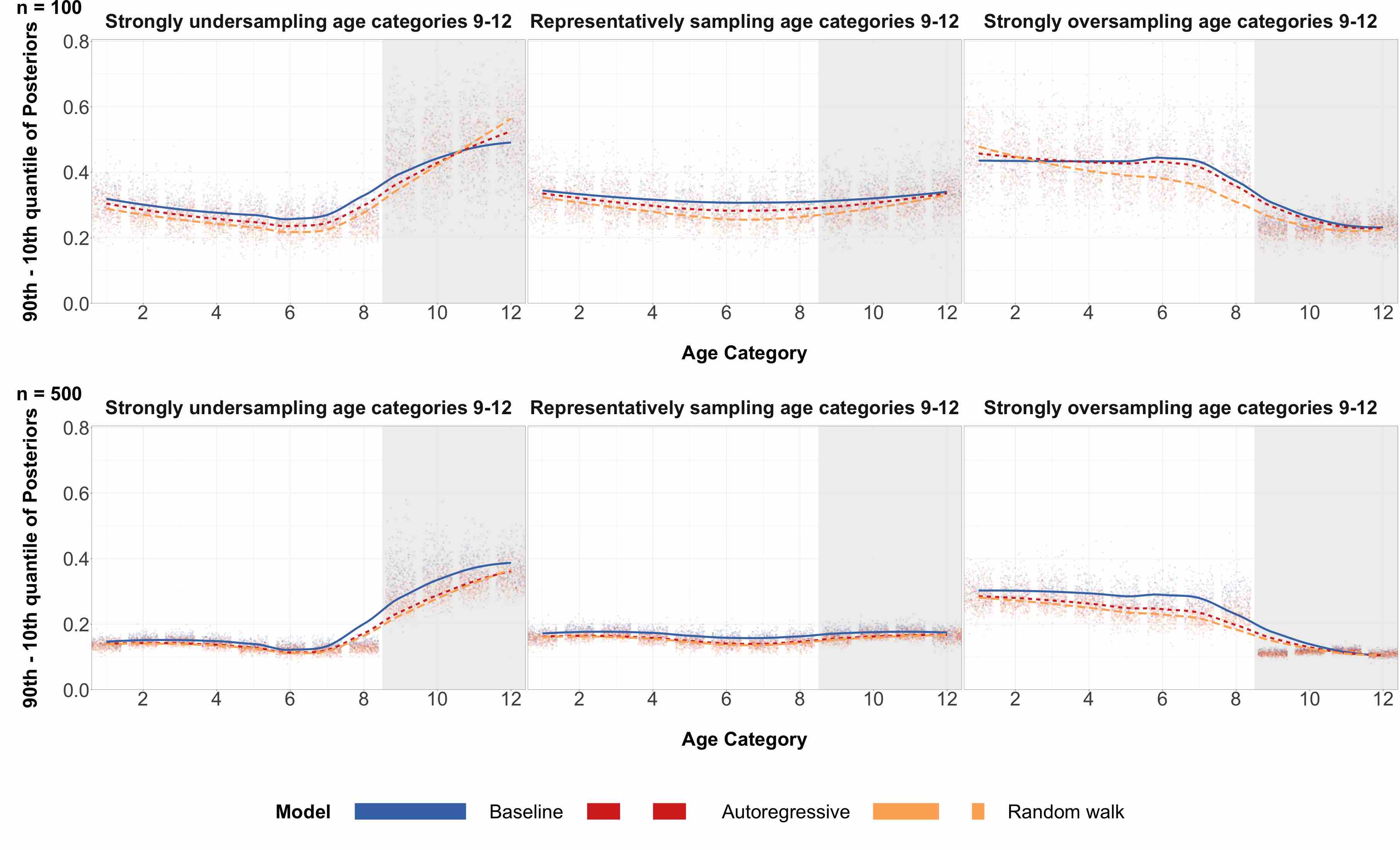}

\caption[short]{Differences in the $90^{th}$ and $10^{th}$ posterior quantiles for every age category when true preference is U-shaped for 200 simulations. The top row corresponds to a sample size of 100 and the bottom row corresponds to a sample size of 500. The shaded grey region corresponds to the age categories of older individuals for which we over/undersample. The left column has a probability of sampling age categories 9-12 equal to 0.05. The middle column has a probability of sampling age categories 9-12 equal to 0.33, which is completely random sampling \textit{and} representative sampling for all age categories. The right column has a probability of sampling age categories 9-12 equal to 0.82. Local regression is used for the smoothed estimates amongst the three prior specifications. For the same plots involving different probabilities of sampling, refer to Table \ref{Tab:simscenarioswidth} in the appendix.}
\label{fig:combined_allquantilefacet_u}
\end{figure}

As a secondary benefit of structured priors, averaging over all 200 runs, simulation studies had shown the difference of the $90^{th}$ and $10^{th}$ posterior quantiles for almost all age categories to be smaller when $n=100$. This is shown in Figure \ref{fig:combined_allquantilefacet_u}. This difference can be interpreted as a measure of posterior standard deviation. When $n=500$, reduction in posterior quantiles difference is even more apparent. Reduction in posterior standard deviations may not be ideal for estimators when the tradeoff is higher absolute bias, but for the case of structured priors, we see a reduction in both for every age category implying a decrease in $L_2$ risk for posterior estimates of every age category. 

\begin{figure}
    \centering
    \includegraphics[width=1\textwidth]{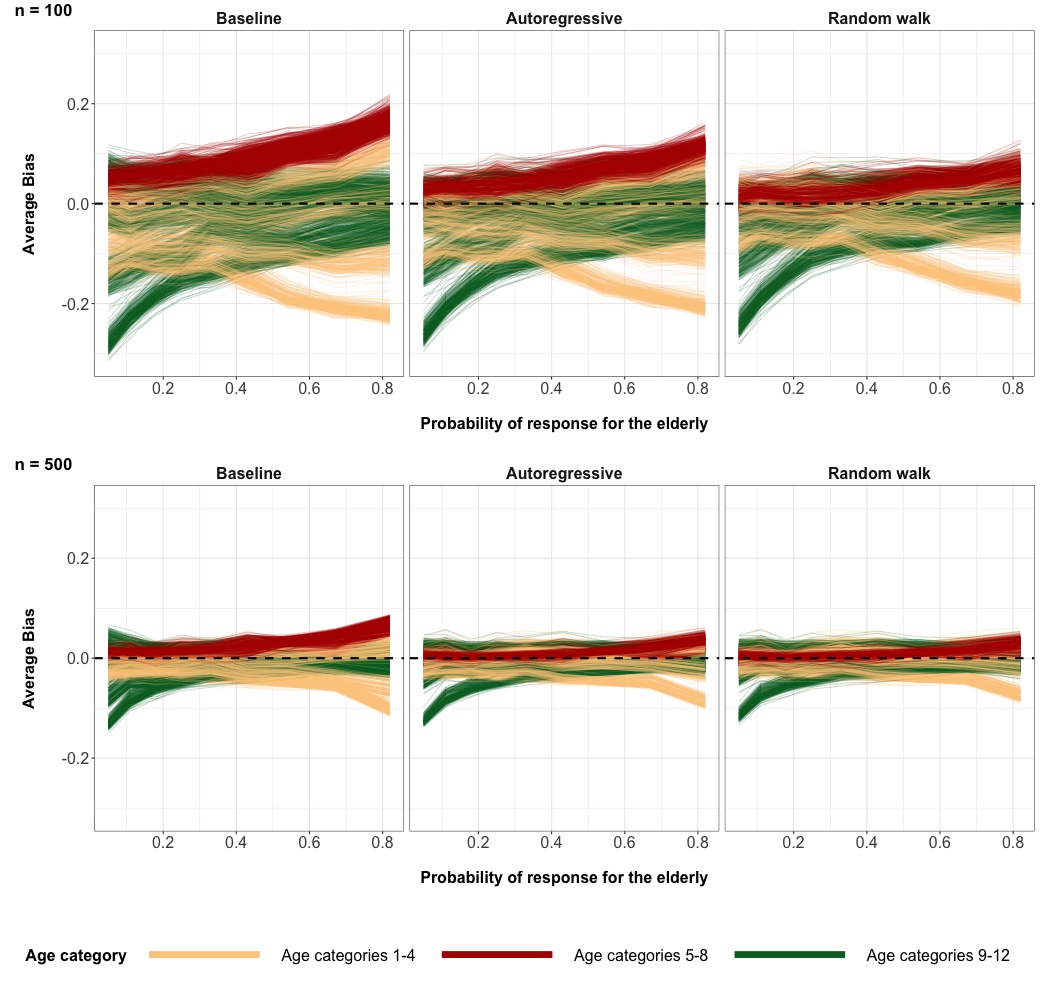}

\caption[short]{The average bias values coming from 200 simulations of posterior medians of the 2448 poststratification cells. The possible values of average bias are in the interval $(-1,1)$. Sample size $n = 100$ (top) and $n = 500$ (bottom). The true preference curve for age is U-shaped. The horizontal dashed line at $y=0$ represents zero bias.}
\label{fig:biasoverallu_100_500}
\end{figure}

Another visualization of bias reduction is based on Figure \ref{fig:biasoverallu_100_500}. It shows bias of posterior preferences for each cell in the population, the finest granularity, as the expected proportion of the sample that are older adults is perturbed. Absolute bias is significantly decreased when switching from the baseline specification to the random walk specification. The autoregressive specification also reduces absolute bias, but not as much when compared to the random walk specification. This is due to the prior $\rho$ defining before inference that the information being borrowed from the neighboring age category posteriors should be a value in $[-1,1]$.

The population preference estimates for the three prior specifications remain nearly the same across all probability of sampling indices when the true preference curve is U-shaped or cap-shaped. When the true preference curve is increasing-shaped, the population preference remains nearly the same for all probability of sampling indices except 0.05 and 0.82. In those cases, the first-order random walk prior produces more unbiased population estimates by a few percentage points. The advantage of structured priors appear to be more drastic when reducing to more granular sub-population levels. For additional bias plots on all three true preference curves, the reader can refer to the appendix.

In summary, based on the simulation studies on the U-shaped true preference along with the two other true preference curves, we see that structured priors decrease absolute bias for posterior MRP estimates more than the classical specification of priors in MRP, regardless of how representative the survey data are to the population of interest. This implies that posterior MRP estimates coming from structured priors are much more invariant to differential nonresponse and biased sampling when compared to the classical priors used in MRP. The main goal of this paper is to argue that structured priors offer an improvement to MRP, even in extremely non-representative data regimes. We indeed see that in the simulation studies as large decreases in absolute bias are seen when the probability of sampling age categories 9-12 are 0.05 and 0.82. A secondary benefit of structured priors is variance reduction on posterior estimates of the structured covariates.

The benefits of structured priors in this case are apparent even when sample size is 1000, but to a lesser degree than when sample size is 100 or 500. The absolute bias reduction and posterior standard deviation reduction due to structured priors when $n=1000$ are a few percentage points for the majority of under/over-represented age categories, based on all three true preferences. In contrast to this, the absolute bias reduction and posterior standard deviation reduction due to structured priors when $n=100$ are around 10 percent for the majority of under/over-represented age categories.  For results on this $n=1000$ simulation, refer to Tables \ref{Tab:simscenarios} and \ref{Tab:simscenarioswidth} with the configuration $n=1000$ for all three true preference shapes.

The three prior specifications start to converge in terms of having the same posterior bias and variance with an increase in sample size due to the likelihood dominating the prior distributions imposed. Regardless, with a sample size of 1000 and 12 age categories, it appears that structured priors are still able to reduce posterior variance and absolute bias in the nonrepresentative data regime -- where elderly individuals are oversampled and undersampled.

The efficacy of structured priors in this section were also evaluated based on the proportion of the time they outperformed baseline priors in bias and variance. The calculated metric, share of simulations where the posterior median of structured priors outperformed the posterior median of baseline priors, are shown in figures referenced by the Improvement Proportion column of Table \ref{Tab:simscenarios}. A similar metric, the share of simulations where posterior variance of structured priors outperformed posterior variance of baseline priors, is referenced by the Improvement Proportion column of Table \ref{Tab:simscenarioswidth}. In almost all cases, there was a strong improvement over the baseline priors more than half the time. This improvement was uniform across simulation scenarios, except for some simulations with sample size $n=100$ and severe undersampling of the highest age groups.

When there is complex and structured variation of the true preference across the age categories, it is important for the model to not surpress this variation. Standard MRP estimates assume exchangeability between age categories and therefore shrink cell estimates towards a global regression model in the absence of data. Structured priors for the age category covariate, while not perfectly recovering the true preference probabilities in the presence of limited data, are better able to capture much more of the variation in preference across age categories. This results in a better quantification of the variation across small areas. 

When the number of categories for the structured covariates of interest is sufficiently large, they start to show beneficial effects on posterior MRP estimates. To quantify "sufficiently large" is problem-dependent as every structured prior will be different depending on the covariates of the data set. Furthermore, there are multiple structured priors one can choose from for a covariate. This is something we will not address here. We previously ran the same set of experiments in this results section for 3 and 6 age categories and did not observe a significant difference in posterior estimates for all three prior specifications. 12 age categories and more for our simulation studies are when the beneficial effects of structured priors become obvious.

\subsection{Undirected structured priors example: Spatial MRP}

For the second simulation example, we will apply MRP for spatial data. More specifically, in the case of binomial regression to get subpopulation-level estimates for all 52 Public Use Microdata Areas (PUMA) in the state of Massachusetts. We simulate a nonrepresentative survey by purposefully oversampling certain spatial areas. 

In this section, we will fit the models using the \texttt{R-INLA} package \citep{rue2009approximate, rue2017bayesian, seppa2019estimating}, a fast approximate Bayesian inference package catered towards spatial modelling.

We will use penalized complexity (PC) prior specifications \citep{simpson2017penalising} for hyperpriors in the spatial MRP model. The response variable for an observation is count data $t_i$ and the total number of possible occurences is $T_i$. Let $n$ be the number of observations in the binomial response data set. Let $i = 1,\dots,n$ be the index of counts. We define the following model below:

\begin{equation}
    \begin{aligned}
    t_i &\sim \text{Binomial}(\mu_i, T_i)\\
    \mu_i &= \text{logit}^{-1} \left(\beta^0 + \alpha^\text{PUMA}_{j[i]} + \alpha_{j[i]}^\text{Education} + \alpha_{j[i]}^\text{Ethnicity} \right) \\
    \alpha_j^\text{Education} | \sigma^\text{Education} &\sim \mbox{N}(0,(\sigma^\text{Education})^2), \text{ for }j=1,\dots,6\\
    \alpha_j^\text{Ethnicity} | \sigma^\text{Ethnicity} &\sim \mbox{N}(0,(\sigma^\text{Ethnicity})^2), \text{ for }j=1,\dots,6\\
    \sigma^\text{Education}, \sigma^\text{Ethnicity} &\sim \text{PC-Prior}(1, 0.1)\\
    \beta^0 &\sim \mbox{N}(0,1)\\
    \label{eq:spatialmrpmodel}
    \end{aligned}
  \end{equation}
  
 The poststratified estimand based on Equation \ref{eq:spatialmrpmodel} for a subpopulation group $S \subseteq \{1, \dots, J \}$ is then $\sum_{j \in S} \frac{N_j}{N} \mu_j$.
  
 The \textit{BYM2 specification} models the prior spatial structure of PUMA effect, $\alpha_j^\text{PUMA}$, as a BYM2 model \citep{morris2019bayesian, riebler2016intuitive} specified in Equation $\ref{eq:bym2prior}$. Let $A_j$ be the set of first-degree neighbours for PUMA $j$. Let $d_j$ be the cardinality of $A_j$.
 
 The BYM2 spatial prior for PUMA is defined as:
 
 \begin{equation}
    \begin{aligned}
    \alpha_j^\text{PUMA} &= \frac{1}{\sqrt{\tau}} \left(\sqrt{1 - \rho} \theta^*_j + \sqrt{\frac{\rho}{s}} \phi^*_j \right) \\
    \phi^*_j &\sim \mbox{N}\left(\frac{\sum_{k \sim A_j} \phi_{k}^*}{d_j}, \frac{1}{d_j} \right) \\
    \theta^* &\sim \mbox{N}(0, I_{52}), \\
    \label{eq:bym2prior}
    \end{aligned}
  \end{equation}
  where we account for the multiple connected components using the method described in \cite{freni2018note}.
 
 From Equation \ref{eq:bym2prior}, we see that $\phi^*$ is an ICAR prior and $\theta^*$ is an isotropic multivariate normal prior. The scaling factor $s$ is computed so that the variance of $\sqrt{\frac{\rho}{\tau s}} \phi_j^*$ is approximately equal to 1 for all PUMA $j$. Additionally, we have the constraint $\sum_{j=1}^{52} \phi^*_j = 0$ since the joint distribution of $\phi^*$ is non-identifiable. We will specify $\tau \sim \text{PC-Prior}(1,0.1)$ \citep{simpson2017penalising}, which results in the precision $\tau$ to have density function $\frac{-\text{log}(0.1)}{2} \tau^{-3/2} e^{\frac{\text{log}(0.1)}{\tau^{1/2}}}$ and we'll specify the mixing parameter $\rho$ being the default BYM2 hyperprior specification in INLA.
 
 The \textit{IID specification} models the prior structure of PUMA effect, $\alpha^\text{PUMA}$, as an isotropic multivariate normal distribution with precision $\tau \sim \text{PC-Prior}(1,0.1)$.
    
The IID specification does not allow for any borrowing of information. However, the BYM2 prior specification allows for posterior estimates of true preference for every PUMA to borrow information with its first-degree neighbours, where the amount of information borrowed is controlled by the posterior mixing hyperparamter $\rho$. This borrowing of information with first-degree neighbours aims to capture the smooth spatial structure of true preference across the 52 PUMA in Massachusetts, if there indeed is one.

\subsubsection{Simulated data}

\textbf{The sample}

\begin{figure}
    \centering
    \includegraphics[width=1\textwidth]{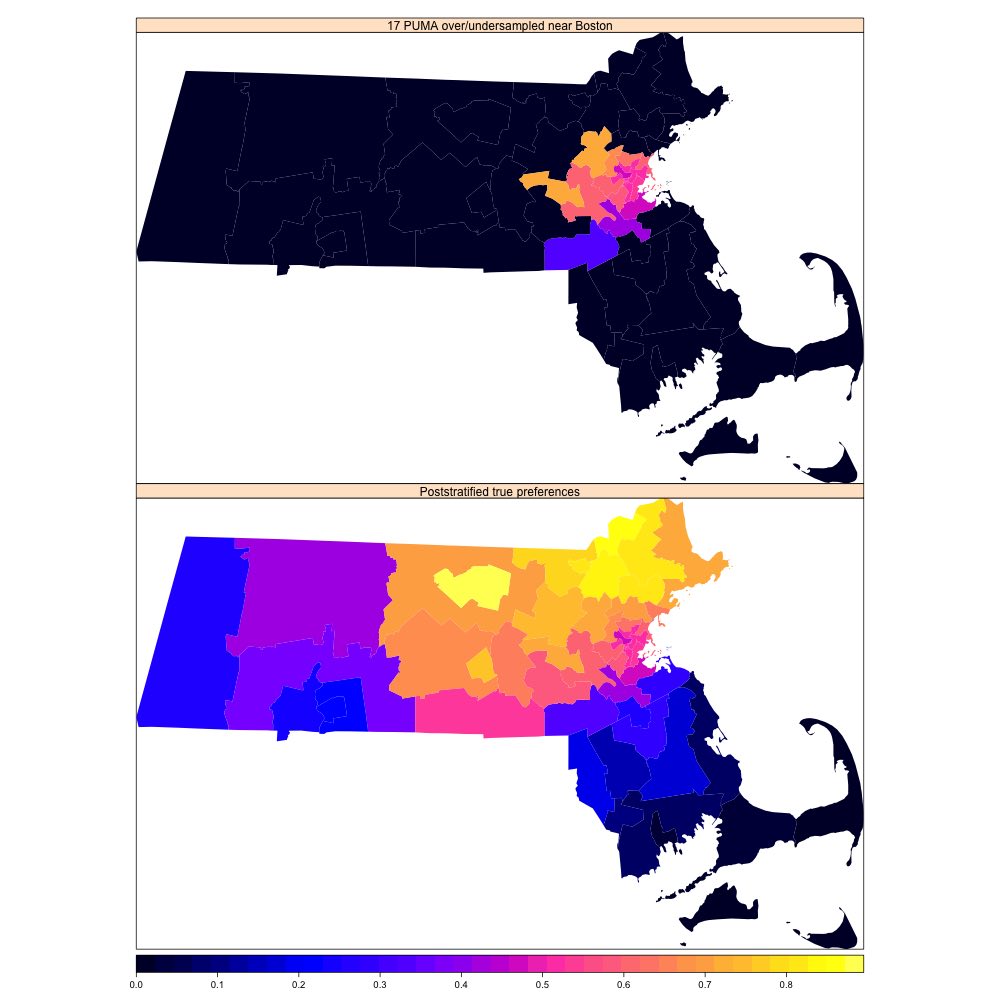}
    \caption[short]{True poststratified preference for 52 PUMA in Massachusetts is the bottom heatmap, which is the vector $X_\text{PUMA}$. The top heatmap corresponds to the 17 PUMA near Boston that are over/undersampled. The true poststratified preference for PUMA $j \in \{1,\dots,52\}$ is defined as $\frac{\sum_{k \in S_j}N_k \theta_k}{\sum_{k \in S_j}N_k}$, where $S_j$ corresponds to the index set for PUMA $j$ and $\theta_k$ is the true preference for poststratification cell $k$.}
    \label{fig:truepref_spatialmrp}
\end{figure}

We investigate the effects of over/undersampling 17 neighbouring PUMA areas near Boston. The true preference across 52 PUMA as well as the 17 cluster PUMA near Boston that are over/undersampled are visualized in Figure $\ref{fig:truepref_spatialmrp}$. The below simulations show that the spatial prior for PUMA effect, which is a BYM2 specification, outperforms the classical IID prior specification in MRP.

\textbf{Assumed sample and population}

We will again assume that the population is sufficiently large so that sampling with replacement is equivalent to retrieving a random sample from the population. 

Let $S \subseteq \{1,\dots,J \}$ be the index set corresponding to the group of 17 PUMA near Boston. We perturb the probability of sampling an individual in $S$ through 9 scenarios, ranging from $0.05$ (under-representing individuals in $S$) to $0.82$ (over-representing individuals in $S$).

True preferences for every poststratification cell $j \in \{1, \dots, J\}$ in the population are generated with the following formula:
\begin{align*}
\mu_j = \text{logit}^{-1}\left( \beta^0 + X_{\text{PUMA}[j]} + X_{\text{Ethnicity}[j]} + X_{\text{Education}[j]} \right)
\end{align*}
where $X_{\text{PUMA}[j]}$, $X_{\text{Ethnicity}[j]}$, $X_{\text{Education}[j]}$ corresponds to the PUMA, Race/Ethnicity and Education effect respectively for poststratification cell $j$. $X_{\text{PUMA}[j]}$, $X_{\text{Ethnicity}[j]}$, $X_{\text{Education}[j]}$ are defined in the appendix. $M$ individuals with binary responses are then sampled from this poststratification matrix and then grouped based on poststratification cell to generate the binomial responses $(t_i, T_i)_{i=1}^n$. It's important to note that $X_{\text{PUMA}[j]}$ was generated through one random sample of a Gaussian Markov random field to ensure that the true preference $\mu_j$ has some degree of smoothness across all PUMA.

\subsubsection{Undirected structured priors results}

\noindent \textbf{Impact of prior choice on bias of posterior preferences}

\begin{figure}
    \centering
    \includegraphics[width=1\textwidth]{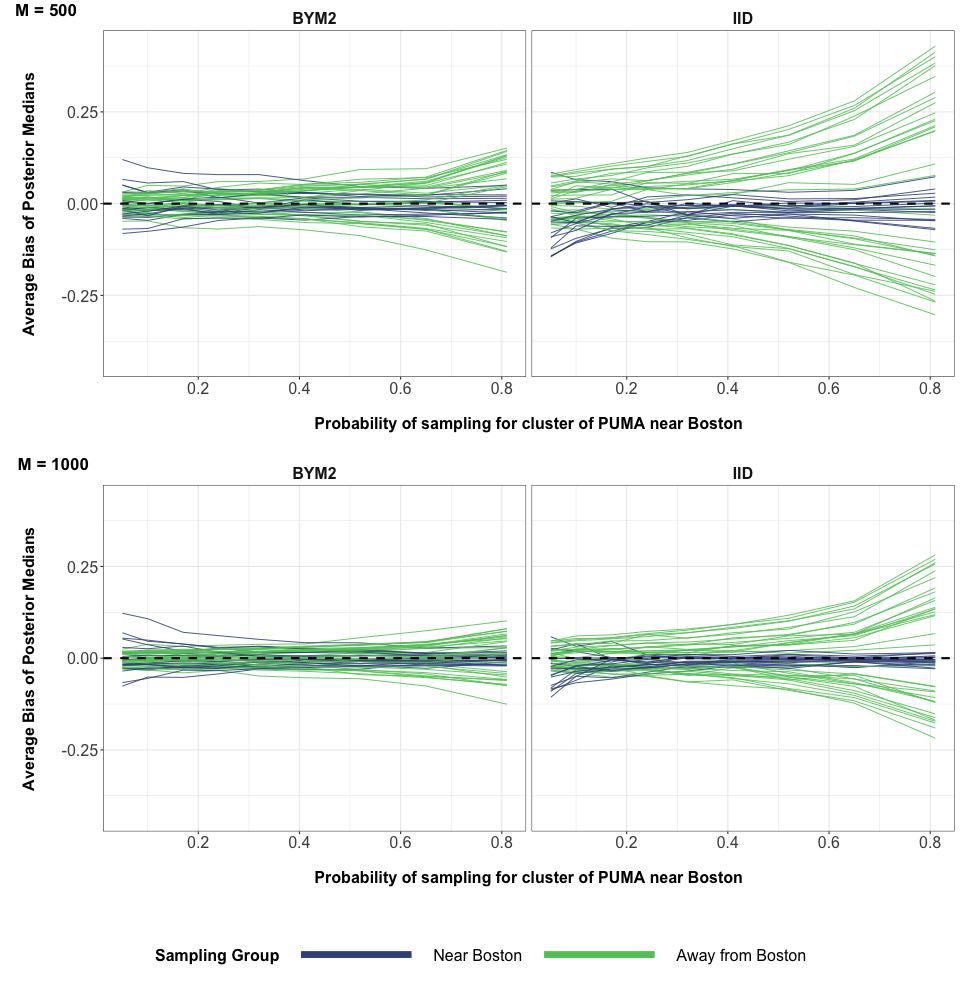}

\caption[short]{The average bias values coming from 200 simulations of poststratified estimates for the 52 PUMA areas. The possible values of average bias are in the interval $(-1,1)$. $M$ is the number of binary responses in every simulated data set. The top row corresponds to 500 binary responses used to define binomial responses for every simulation iteration. The bottom row corresponds to 1000 binary responses used to define binomial responses for every simulation iteration. The horizontal dashed line at y = 0 represents zero bias.}
\label{fig:biasfacetspatialpuma_200_500_1000}
\end{figure}

\begin{figure}
    \centering
    \includegraphics[width=1\textwidth ]{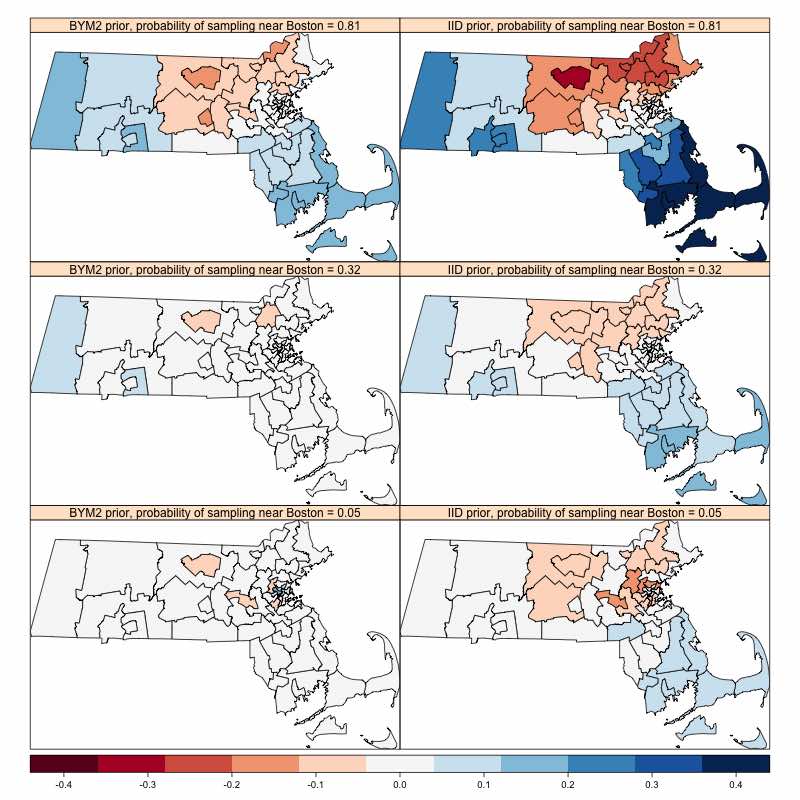}

\caption[short]{Average bias of posterior medians for every PUMA based off 200 simulations and binary response sample size is 500. The possible values of average bias are in the interval $(-1,1)$. The left column corresponds to the BYM2 spatial prior for PUMA effect. The right column corresponds to an IID prior for PUMA effect. The probabilities 0.81, 0.32 and 0.05 in the top, middle and bottom rows respectively correspond to the probability of sampling an individual in Group 1, the cluster of 17 PUMA around Boston.}
\label{fig:abs_avg_bias_postmedian_puma_forpaper200_500_}
\end{figure}

We will again evaluate the impact of prior specification through perturbations of data regime for individuals near Boston. We can see from Figure \ref{fig:biasfacetspatialpuma_200_500_1000} that the absolute bias for almost all 52 PUMA are significantly reduced when using a BYM2 specification instead of the IID specification, across all data regimes. The attenuation in absolute bias is apparent when the number of individuals sampled are 500 or 1000. This reduction in absolute bias is most apparent when individuals near Boston are over-represented (that is when the probability of response for $S$ is close to 0.82). In this case, the BYM2 spatial prior outperforms the IID spatial prior by over 10 percent in absolute bias reduction. A spatial heatmap visualization of Figure \ref{fig:biasfacetspatialpuma_200_500_1000} can be seen in Figures \ref{fig:abs_avg_bias_postmedian_puma_forpaper200_500_} and \ref{fig:abs_avg_bias_postmedian_puma_forpaper200_1000_} in the appendix. Based on these heatmaps, the PUMA that have the most absolute bias reduction when using the BYM2 spatial prior and when $S$ is over-represented are the regions near the south-eastern coast of Massachusetts.

\begin{figure}
    \centering
    \includegraphics[width=1\textwidth]{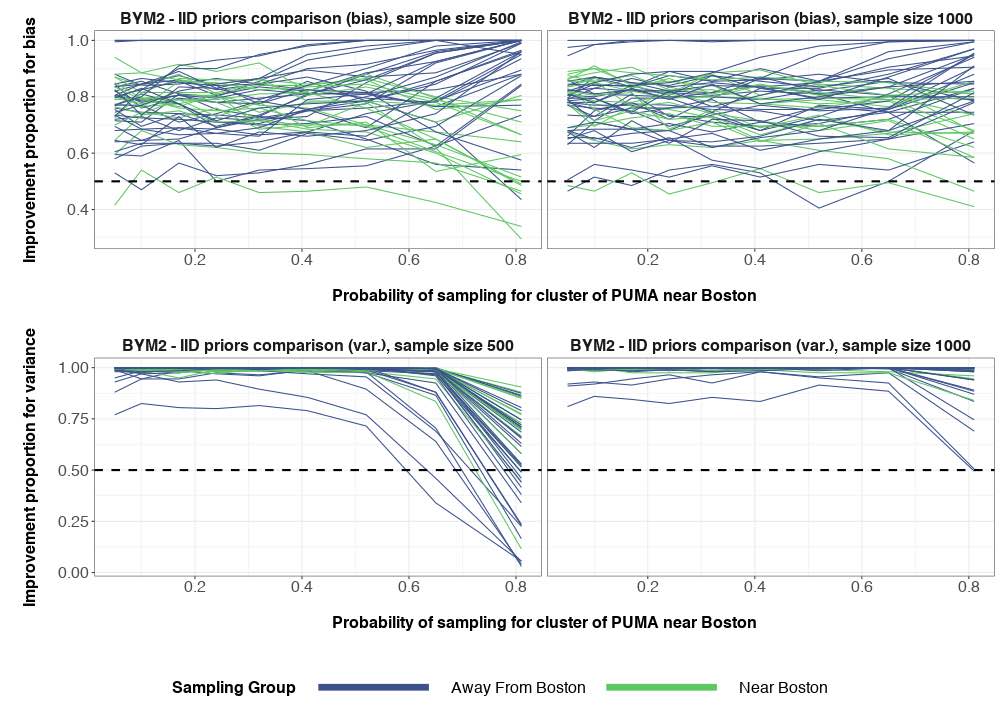}
    \caption{The top row corresponds to the proportion of the time that spatial BYM2 priors have lower absolute posterior median bias when compared to IID baseline priors, for each PUMA. The bottom row corresponds to the proportion of the time that spatial BYM2 priors have lower posterior variance when compared to IID baseline priors, for each PUMA. The left column corresponds to 500 binary responses in the sample. The right column corresponds to 1000 binary responses in the sample. The horizontal dashed line $y=0.5$ corresponds to equal proportion. The difference of the ${90}^{th}$ and ${10}^{th}$ posterior quantiles is used as a measure for posterior variance.}
    \label{fig:proportion_puma_sd_bias}
\end{figure}

The share of simulations when the structured BYM2 spatial priors outperformed the IID priors for PUMA are shown in Figure \ref{fig:proportion_puma_sd_bias}. Differences in absolute bias from posterior medians and differences in posterior width were used as measures of comparison. The improvement was uniform over the IID priors, except for when there was severe oversampling of areas near Boston. 

The average bias for every poststratification cell $\mu_j$ can be seen in Figure $\ref{fig:biasfacetspatial_200_500_1000}$ in the appendix. At this granularity, the BYM2 spatial prior is still outperforms the IID prior for PUMA for absolute bias reduction, regardless of sample size and what the probabillity of sampling $S$ is.

Based off these spatial MRP simulations, we had also observed that the posterior width of every poststratification cell remains nearly same when switching from the IID prior for PUMA to the BYM2 spatial prior for PUMA, as seen in the appendix Figure $\ref{fig:biasfacetspatial_sd_200_500_1000}$. This posterior width of every PUMA also remains nearly the same when switching from IID prior to BYM2 spatial prior, as seen in appendix Figure $\ref{fig:biasfacetspatialpuma_sd_200_500_1000}$. 

In summary, structured priors for covariates with an undirected spatial structure are shown to improve MRP estimates through reduction of absolute bias. In this particular spatial MRP with count data simulation, we saw noticeable improvements with lower absolute bias for every PUMA when using the BYM2 spatial prior. 

We had also run the same spatial MRP simulations but instead of having $X_\text{PUMA}$ coming from a Gaussian Markov Random Field, we generated $X_\text{PUMA}$ as a multivariate independent normal. The BYM2 spatial prior on PUMA produced nearly the same posterior estimates as the IID spatial prior on PUMA, indicating that the BYM2 spatial prior does not force spatial structure when it's not present.

\newpage

\section{Analysis on U.S. Survey Data}

Along with simulation studies that validate the benefit of structured priors, we further apply our approach to the National Annenberg Election Survey 2008 Phone Edition (NAES08-Phone) \citep{annenberg2008}. NAES08-Phone was a phone survey conducted over the course of the 2008 US Presidential Election and the sampling methodology was based on random telephone number generation. NAES08-Phone observed a response rate of 23 percent. The population comes from the 2006--2010 5-year American Community Survey (ACS, \cite{acs2008}). The response variable of interest is whether an individual favors gay marriage or not. In 2008, this question was discussed heavily in the political landscape, as some states had not legalized same-sex marriage yet. The covariates used in the Annenberg survey sample are sex, race/ethnicity, household income, state of residence, age, education. The same covariates in the 5-year ACS are used so that poststratification and more specifically MRP can be performed.

\subsection{National Annenberg Election Survey 2008}

Table \ref{Tab:annenbergsummary} contains the percentages of each factor for four of the covariates in the 2008 Annenberg phone survey and the 5-year ACS (excluding age and state of residence). A histogram summarizing the age covariate in the Annenberg phone survey is shown in the bottom plot of Figure \ref{fig:realdata_12_48_78}. Individuals with ages 88 or above in the Annenberg phone survey had their ages set to 88. The reason for doing this was to maintain the non-identifiability of such individuals since they were the least frequently observed individuals in the Annenberg phone survey. The size of the Annenberg phone survey is 24,387 respondents.

\begin{table}[ht]
\centering
\resizebox{\textwidth}{!}{
\begin{tabular}{rrrr}
Sex &             Education &            Race/Eth. & Income \\ 
  \hline
Male: 43.3 (48.1)      & Bachelor: 19.6 (16.9)   & White         : 80.2 (68.1) & Less than 10K: 4.8 (5.1) \\ 
Female: 56.7 (51.9)  & Bachelor+: 19.0 (9.3)  &  Black         : 8.6 (11.2) & [10K, 15K): 5.1 (4.1)  \\ 
   & Highschool: 28.5 (29.2)  & Hispanic          : 1.6 (5.2)  & (15K, 25K]: 8.8 (9.1) \\ 
   & No highschool: 7.1 (14.9)  & Asian         : 1.5 (4.8)  & (25K, 35K]: 10.4 (9.7) \\ 
   & Some college: 16.7 (22.5)  &  American Indian: 1.2 (1.0)  & (35K, 50K]: 15.3 (14.0) \\ 
   & 2 year college: 9.0 (7.3)  & Other         : 7.0 (10.3)  & (50K, 75K]: 19.1 (20.0) \\ 
  &  &  & (75K, 100K]: 14.4 (13.9) \\ 
  &  &  & (100K, 150K]: 12.3 (14.2) \\ 
  &  &  & More than 150K: 9.9 (9.9) \\
  \hline
\end{tabular}
}
\caption{\em Percentage of each factor in the Annenberg phone survey and the 2006-2010 5-year American Community Survey (in parentheses) for sex, education, race/ethnicity, household income. Percentages are rounded to one decimal.}
\label{Tab:annenbergsummary}
\end{table}

In the appendix, Figure \ref{fig:state_proportions} shows proportions of every state in the Annenberg survey and the differences in proportions between Annenberg survey and ACS. Ideally, there shouldn't be differences in the proportions for both surveys if both ACS and Annenberg survey are equally representative of the US population. Based on the bottom heatmap in Figure \ref{fig:state_proportions}, California and Texas are shown to be the most under-represented by the Annenberg survey whereas New Hampshire and Missouri are the most over-represented by the Annenberg survey.

\subsection{Poststratifying to the US population}

A smoothed density summarizing the age covariate in the ACS is shown in the bottom plot of Figure \ref{fig:realdata_12_48_78}. The continuous age covariate in both the 5-year ACS and the Annenberg survey is discretized into either 12, 48, or 72 age categories in our analysis. In theory, the number of poststratification cells for Table \ref{Tab:acsps} is $2 \times 6 \times 6 \times 9 \times 51 \times 78 = 2,577,744$. The cells left out by the expanded version of Table \ref{Tab:acsps} are assumed have a population size of $0$. 

The 2006--2010 5-year ACS is a weighted probability survey, with a weight assigned to every individual in the sample. Based on the weights of individuals in the 5-year ACS, we form a $929,082$-row poststratification matrix as seen in Table \ref{Tab:acsps}, which we will assume to be representative of the overall population for the 2008 Annenberg phone survey. We will use Table \ref{Tab:acsps} to poststratify the 2008 Annenberg survey estimates to the US population. The ACS aggregates monthly probabilistic samples to form 1, 3, and 5-year ACS data sets. It aims to capture the most current demographic information annually, and answering the survey is mandatory. For these reasons, we believe that it's the most accurate representation of the US population every year.



\begin{table}[ht]
\centering
\resizebox{\textwidth}{!}{
\begin{tabular}{rrrrrrrr}
 Age & Sex & Race/Eth. & Education & State & Income & N \\ 
  \hline
    18 & Male & American Indian & Highschool & Alabama & (15000, 25000] & 35 \\
    18 & Female & American Indian & Highschool & Arizona & Less than 10000 & 124 \\
    \vdots & \vdots & \vdots & \vdots & \vdots & \vdots & \vdots \\
    95 & Female & White & 2 year college & Rhode Island & (100000, 150000] & 15 \\
    \hline
\end{tabular}
}

\caption{\em Full poststratification matrix for the 5-year American Community Survey.}
\label{Tab:acsps}
\end{table}

\subsection{The models for the 2008 Annenberg phone survey}


For ages 18--40, the smooth ACS density in Figure \ref{fig:realdata_12_48_78} is higher than the Annenberg histogram, implying that the Annenberg survey underrepresents younger individuals. For the other demographic traits, relative to the 5-year ACS, Table \ref{Tab:annenbergsummary} show that the Annenberg survey overrepresents white individuals and women.

Let $y_i= 1$ if respondent $i$ favors same-sex marriage.  Then we model,
\begin{equation}
    \begin{aligned}
    \mbox{Pr} (y_i = 1) = \text{logit}^{-1} &\left(\beta^0 + \alpha^{\text{Age Cat.}}_{j[i]} + \beta^\text{Sex} X_{\text{Sex}, i} +\alpha^{\text{Ethnicity}}_{j[i]} \right. \\ 
    & \left. + \alpha^{\text{Education}}_{j[i]} + \alpha^{\text{State}}_{j[i]} + \alpha^{\text{Income}}_{j[i]} \right) \\
    \end{aligned}
\end{equation}
We define the \textit{baseline, autoregressive and random walk specifications} to have in common the priors distributions defined in Equation \ref{eq:commonpriorsrealdata} shown in the appendix.

Let $J^\text{Age Cat.}$ be the number of categories for the continuous covariate age. The \textit{baseline specification} has the prior distributions for age category:
\begin{equation}
  \begin{aligned}
\alpha^{\text{Age Cat.}}_j \mid \sigma^\text{\text{Age Cat.}} &\overset{\text{ind.}}{\sim} \mbox{N}(0, (\sigma^\text{Age Cat.})^2), \text{ for }j = 1, \dots, J_\text{Age Cat.} \\
\sigma^\text{Age Cat.} &\sim \mbox{N}_+(0,1)\\
\label{eq:blrealdataprior}
\end{aligned}
\end{equation}
The \textit{autoregressive specification} has the prior distributions:

\begin{equation}
  \begin{aligned}
\alpha^{\text{Age Cat.}}_1 \mid \rho, \sigma^\text{Age Cat.} &\sim \mbox{N}(0, \frac{1}{1-\rho^2}(\sigma^\text{Age Cat.})^2) \\
\alpha^{\text{Age Cat.}}_j \mid \alpha^{\text{Age Cat.}}_{j-1},\dots,\alpha^{\text{Age Cat.}}_{1}, \rho,\sigma^{\text{Age Cat.}} &\sim \mbox{N}(\rho \alpha^{\text{Age Cat.}}_{j-1}, (\sigma^\text{Age Cat.})^2), \\
&\text{ for }j = 2, \dots, J^\text{Age Cat.}  \\ 
\sigma^\text{Age Cat.} &\sim \mbox{N}_+(0,1)\\
(\rho + 1)/2  &\sim \text{Beta}(0.5, 0.5)
\label{eq:arrealdataprior}
\end{aligned}
\end{equation}
The \textit{random walk specification} has the prior distributions:
\begin{equation}
\begin{aligned}
\alpha^{\text{Age Cat.}}_j \mid \alpha^{\text{Age Cat.}}_{j-1},\dots,\alpha^{\text{Age Cat.}}_{1}, \sigma^\text{Age Cat.}  &\sim \mbox{N}(\alpha^{\text{Age Cat.}}_{j-1},( \sigma^\text{Age Cat.})^2), \\
&\text{ for }j = 2, \dots, J_\text{Age Cat.} \\
&\sum_{j=1}^{J_\text{Age Cat.}} \alpha^{\text{Age Cat.}}_j = 0.
\label{eq:rwrealdataprior}
\end{aligned}
\end{equation}
We treat age as an ordered categorical predictor. It is reasonable to believe that people of similar ages will have similar attitudes on same-sex marriage. Hence we propose autoregressive and random walk structures as the prior distributions for age category.



\subsection{Performing MRP with structured priors for the 2008 Annenberg phone survey}

\begin{figure}
    \centering
        \includegraphics[width=1\textwidth]{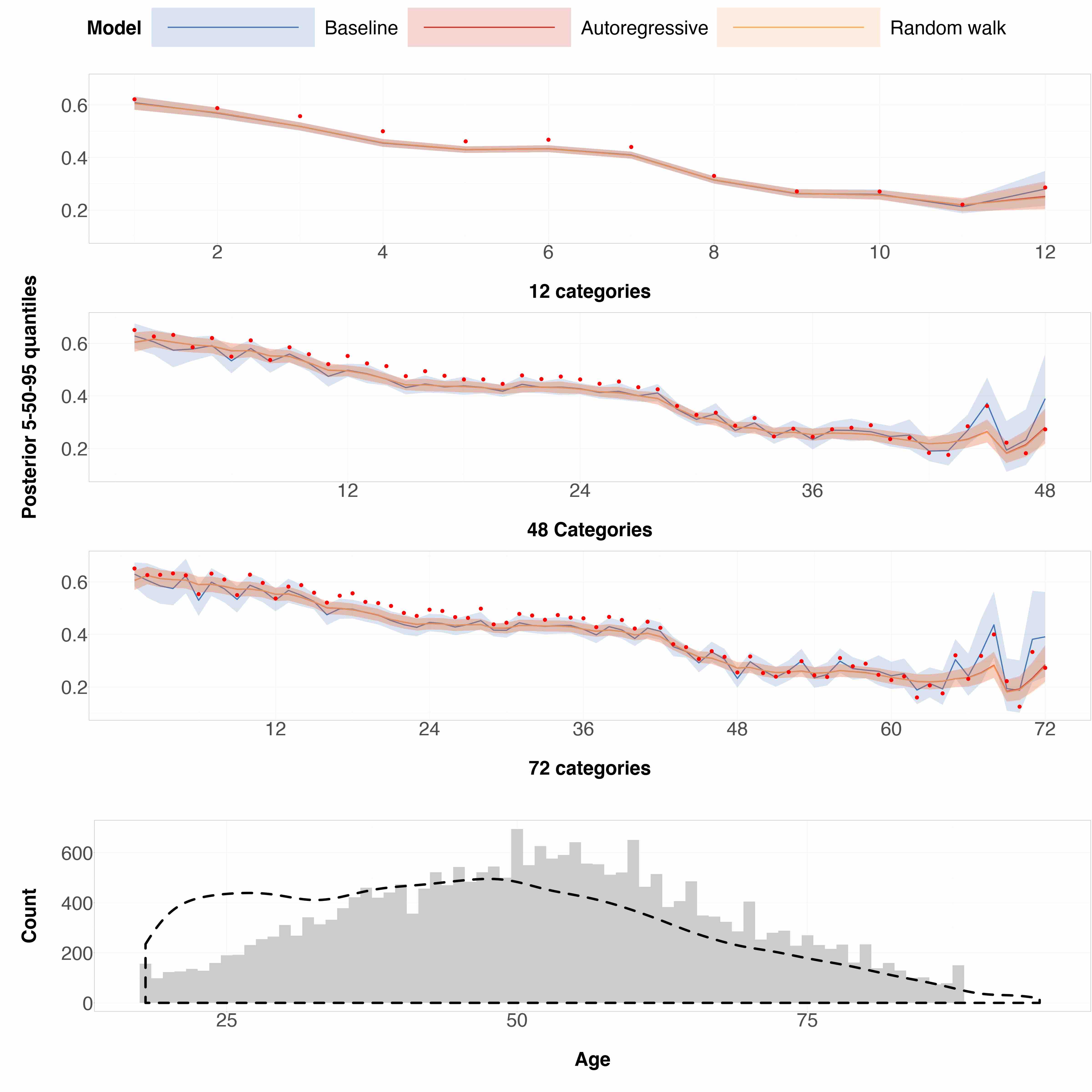}

\caption[short]{12 (top), 48 (middle) and 72 (bottom) age categories. Red points in the top three plots are the empirical mean. The upper and lower bands in the top three plots correspond to the 95-percent and 5-percent posterior quantiles for every age category, and the middle solid line contains the posterior median for every age category. The density plot of ages in the ACS are coming from a random sample based off the 5-year ACS, where sampling is conducted with replacement using person weights given by the ACS. This random sample size is the same size as the 2008 Annenberg phone survey, and is assumed to be representative of the overall population defined by the 5-year ACS. 2000 iterations for 4 chains were run, for each prior specification and for age discretized into 12, 48 and 72 categories. The burn-in was set to 50 percent.}
\label{fig:realdata_12_48_78}
\end{figure}

Hierarchical logistic regression with the two structured prior specifications and the baseline specification described previously are fit to the 2008 Annenberg phone survey. The poststratification matrix formed by the 5-year ACS is then used to poststratify posterior estimates for every age category. This is shown in Figure \ref{fig:realdata_12_48_78}.

When age is discretized into 12 categories, there are no noticeable differences among the three prior specifications for age categories 1--11. Only at age category 12 do we start seeing a difference between the baseline specification and the two structured prior specifications. As expected, this difference in posteriors is observed when the underlying age category is a sparse cell for the survey data set. When age is discretized into 48 and 72 categories, one starts to see differences between the structured prior specifications and the baseline specification in terms of posterior variance for every age category. Posterior variances for the baseline specification are wider based on the 5-95 percent quantiles, and they expand a significant amount for the oldest age categories. 

The baseline prior specification's posteriors become contracted towards their respective empirical means, which is not ideal since the empirical means swing more wildly for the older age categories. On the other hand, the autoregressive and random walk specifications are more smooth due to their property of having neighboring posterior random effects for age categories sharing information, and this is most noticeable when the number of age categories is 72. This smoothing effect is desirable for ordinal data as one may be interested in capturing a long-term trend when age increases.

What's also worth noting is that the baseline specification drastically changes the posterior variances when the number of categories for age changes from 12 to 48 to 72. Structured priors provide some stability in posterior variances despite how the input survey data is preprocessed through discretization of continuous variables.

The posterior population preferences for all three prior specifications remain nearly identical across the three age categories. This remains consistent with population preference results based on the simulation studies.



Based on the simulation studies, we had shown that structured priors reduce absolute bias and posterior variances of structured covariates. In our application of structured priors to the non-representative 2008 Annenberg phone survey, we see that structured priors reduce posterior variance on the structured covariate age as well.


 
\newpage

\section{Conclusion}

We proposed using priors that exploit underlying structure in the covariates of multilevel regression and poststratification. Defined as structured prior distributions, they aim to introduce more intelligent shrinkage of posterior estimates.

We found through simulation studies that structured priors, when compared to independent random effects, reduce posterior MRP bias regardless of nonresponse pattern if there is an underlying pattern. A secondary benefit of structured priors when compared to independent random effects is that they reduce posterior variances for MRP estimates at the subpopulation levels corresponding to structured covariates of interest. We found that structured priors weather even extreme nonresponse patterns when compared to traditional random effects used in MRP. This is as expected since structured priors enable intelligent information-borrowing and shrinkage in posterior MRP estimates. This allows for an improved quantification of variation across small areas. Our modeling strategy of using structured priors was also applied to the non-representative 2008 Annenberg phone survey. The structured priors we describe here have similar smoothing properties to nonparametric regression methods such as GP regression and kernel smoothing \citep{gelman2013bayesian, rasmussen2003gaussian}. 

Our investigations of using MRP for the Annenberg survey used ACS data to its full capacity through the usage of a 5-year ACS that covered the year 2008. Using a 1-year or a 3-year ACS which would have resulted in rougher information about the population. Indeed, the information used to build the poststratification can be a limiting factor for MRP. The accuracy of poststratification in MRP is dependent on whether the poststratification matrix used is a true representation of the target population or not. 

Based on both simulation studies and analysis on the Annenberg survey, we saw that more age categories resulted in lower posterior variance and bias for age category estimates. This comes at the tradeoff of coarser information about $N_j$, the size of the poststratification cell $l$. Another limitation one may have is deciding covariates to impose structured priors on. This choice is dependent on the modeller's knowledge of the problem and the data used. 

There is usually more than one set of structured priors to propose, and this model selection and comparison problem is not addressed in this paper. The method in the paper could also be extended to using structured priors on interaction terms \citep{ghitza2013deep}, which is another active area in MRP research. Furthermore, we do not analyze the scenario when a structured prior is used for a covariate with no apparent structure. 

In this manuscript we demonstrate improvements to MRP estimates through the use of structured priors when justified to do so. We believe that this is a contribution to the wider field considering other forms of regularization with MRP, but rather than employing black-box methods, using structured priors exploits methodologist and survey administrator knowledge. 

\newpage

\bibliographystyle{ba}
\bibliography{sample}

\begin{acknowledgement}
Daniel Simpson and Yuxiang Gao were funded by the Canadian Natural Sciences and Engineering Research Council and the Canadian Research Chairs program. Andrew Gelman was supported by the U.S. Office of Naval Research, the Institute for Education Sciences and the National Science Foundation. We thank Shira Mitchell for helpful feedback. We also thank three reviewers and the editor for insightful feedback.
\end{acknowledgement}

\newpage 

\begin{supplement}
\sname{Appendix A}\label{suppA} 
\stitle{Manuscript repository:}
\slink[url]{https://github.com/alexgao09/structuredpriorsmrp\_public}
\sdescription{This Github repository will fully reproduce the results and plots in this manuscript.}
\end{supplement}

\section*{Appendix A: Additional tables, definitions and figures for the simulation studies}

The various simulation conditions based on sample size $n$ and true preference curve based on age of an individual are given in Table \ref{Tab:simscenarios}. Table \ref{Tab:simscenarioswidth} below summarizes posterior quantile differences for the three true preference curves when the probability of sampling index is perturbed.

\begin{table}[ht]
\centering
\resizebox{\textwidth}{!}{
\begin{tabular}{rrrrr}
Sample size & Age preference curve & Poststrat. cell bias & Bias for each age category & Improvement proportion \\
    \hline
    100 & U-shaped & Figure \ref{fig:biasoverallu_100_500} & Figure  \ref{fig:agecatbiasu_100} & Figure \ref{fig:proportion_agecat_bias_u} \\
    500 & U-shaped & Figure \ref{fig:biasoverallu_100_500} & Figure \ref{fig:agecatbiasu_500}  & Figure \ref{fig:proportion_agecat_bias_u} \\
    1000 & U-shaped & Figure \ref{fig:biasfacet_100_12_1_1000_u}  & Figure \ref{fig:allmedians_facet_100_12_1000_u} & Figure \ref{fig:proportion_agecat_bias_u} \\
    100 & Cap-shaped & Figure \ref{fig:biasoverallcap_100_500} & Figure \ref{fig:agecatbiascap_100} & Figure \ref{fig:proportion_agecat_bias_cap} \\ 
    500 & Cap-shaped & Figure \ref{fig:biasoverallcap_100_500} & Figure \ref{fig:agecatbiascap_500} & Figure \ref{fig:proportion_agecat_bias_cap} \\
    1000 & Cap-shaped & Figure \ref{fig:biasfacet_100_12_1_1000_cap}  & Figure \ref{fig:allmedians_facet_100_12_1000_cap} & Figure \ref{fig:proportion_agecat_bias_cap} \\
    100 & Increasing-shaped & Figure \ref{fig:biasoverallincreasing_100_500} & Figure \ref{fig:agecatbiasincreasing_100} & Figure \ref{fig:proportion_agecat_bias_increasing} \\
 500 & Increasing-shaped & Figure \ref{fig:biasoverallincreasing_100_500} & Figure \ref{fig:agecatbiasincreasing_500} & Figure \ref{fig:proportion_agecat_bias_increasing} \\
 1000 & Increasing-shaped & Figure \ref{fig:biasfacet_100_12_1_1000_increasing}  & Figure \ref{fig:allmedians_facet_100_12_1000_increasing} & Figure \ref{fig:proportion_agecat_bias_increasing} \\
 \hline
\end{tabular}
}

\caption{\em Simulation scenarios for bias assessment}
\label{Tab:simscenarios}
\end{table}

\begin{table}[ht]
  \centering
  \resizebox{\textwidth}{!}{
  \begin{tabular}{rrrr}
Sample size & Age preference curve & Difference in posterior quantiles & Improvement proportion \\
    \hline
    100 & U-shaped & Figure \ref{fig:posteriorwidth_u_100} & Figure \ref{fig:proportion_agecat_sd_u}\\
     500 & U-shaped & Figure \ref{fig:posteriorwidth_u_500} & Figure \ref{fig:proportion_agecat_sd_u}\\
     1000 & U-shaped & Figure \ref{fig:allquantilediff_facet_100_12_1000_u} & Figure \ref{fig:proportion_agecat_sd_u}\\
     100 & Cap-shaped & Figure \ref{fig:posteriorwidth_cap_100} & Figure \ref{fig:proportion_agecat_sd_cap}\\ 
     500 & Cap-shaped & Figure \ref{fig:posteriorwidth_cap_500} & Figure \ref{fig:proportion_agecat_sd_cap}\\
     1000 & Cap-shaped & Figure \ref{fig:allquantilediff_facet_100_12_1000_cap} & Figure \ref{fig:proportion_agecat_sd_cap}\\
     100 & Increasing-shaped & Figure \ref{fig:posteriorwidth_increasing_100} & Figure \ref{fig:proportion_agecat_sd_increasing}\\
     500 & Increasing-shaped & Figure \ref{fig:posteriorwidth_increasing_500} & Figure \ref{fig:proportion_agecat_sd_increasing}\\
     1000 & Increasing-shaped & Figure \ref{fig:allquantilediff_facet_100_12_1000_increasing} & Figure \ref{fig:proportion_agecat_sd_increasing}\\
     \hline
  \end{tabular}
 }
  
\caption{\em Simulation scenarios for posterior standard deviation assessment}
\label{Tab:simscenarioswidth}
\end{table}

\textbf{Directed structured prior simulation studies}

In the directed structured prior simulation studies, $X_{\text{Income}} = (0.1, 0, -0.2, 0.2)$. $X_{\text{State}}$ and $X_{\text{Relig.}}$ are 51-length vectors that correspond to the 2004 Democratic vote share and 2004 percentage of Mormons + Evangelicals in every state respectively. These come from the data set used in \cite{kastellec2010estimating}. Finally, $\beta^0 = 0$ if the true preference curve is increasing and $-1.5$ otherwise. $\beta^\text{State} = 0.5$ and $\beta^\text{Relig.} = -0.5$.

\textbf{Undirected structured prior simulation studies}

In the undirected structured prior simulation studies, we have the following:

$X_{\text{Ethnicity}} = (-0.25, -0.15, -0.05, 0.05, 0.15, 0.25)$, 

$X_{\text{Education}}=(-0.675, -0.525, -0.375, -0.225, -0.075, 0.075)$,

$\beta^0 = 0$. Finally, $X_{\text{PUMA}}$ is a random sample coming an ICAR model $\{U_i\}_{i=1}^{52}$ with a sum-to-zero constraint.

\section*{Appendix B: Additional tables, definitions and figures for the data analysis}

The baseline, autoregressive and random walk specifications in the analysis of the 2008 Annenberg phone survey have in common the prior distributions defined in Equation \ref{eq:commonpriorsrealdata}.

\begin{equation}
\begin{aligned}
\alpha^{\text{Ethnicity}}_j \mid \sigma^\text{Ethnicity} &\overset{\text{ind.}}{\sim} \mbox{N}(0,(\sigma^\text{Ethnicity})^2),  \text{ for } j = 1,\dots, 6  \\
\alpha^{\text{Education}}_j \mid \sigma_\text{Education} &\overset{\text{ind.}}{\sim} \mbox{N}(0,(\sigma^\text{Education})^2), \text{ for } j = 1,\dots, 6 \\
\alpha^{\text{State}}_j \mid \beta^\text{State-VS}, \beta^\text{Relig.},( \alpha_{j^*}^\text{Region} )_{j^*=1}^5, \sigma^\text{State}  &\overset{\text{ind.}}{\sim} \mbox{N}(\alpha_{m[j]}^\text{Region} + \beta^\text{Relig.} X_{\text{Relig.},j} \\
&+ \beta^\text{State-VS} X_{\text{State-VS}, j}, (\sigma^\text{State})^2 ),\\
&\text{ for }j=1,\dots,51 \\
\alpha^\text{Region}_m \mid \sigma^\text{Region} &\overset{\text{ind.}}{\sim} \mbox{N}(0, (\sigma^\text{Region})^2),  \text{ for } m=1,\dots,5 \\
\alpha^{\text{Income}}_j \mid \sigma^\text{Income} &\overset{\text{ind.}}{\sim} \mbox{N}(0,(\sigma^\text{Income})^2), \text{ for } j = 1,\dots, 9 \\
\sigma^\text{Ethnicity}, \sigma^\text{Education}, \sigma^\text{State}, \sigma^\text{Region}, \sigma^\text{Income} &\sim \mbox{N}_+(0,1)\\
\beta^\text{Sex}, \beta^\text{State-VS}, \beta^\text{Relig.}, \beta^0 &\sim \mbox{N}(0,1) \\
\label{eq:commonpriorsrealdata}
\end{aligned}
\end{equation}

\begin{figure}
    \centering
    \includegraphics[width=1\textwidth]{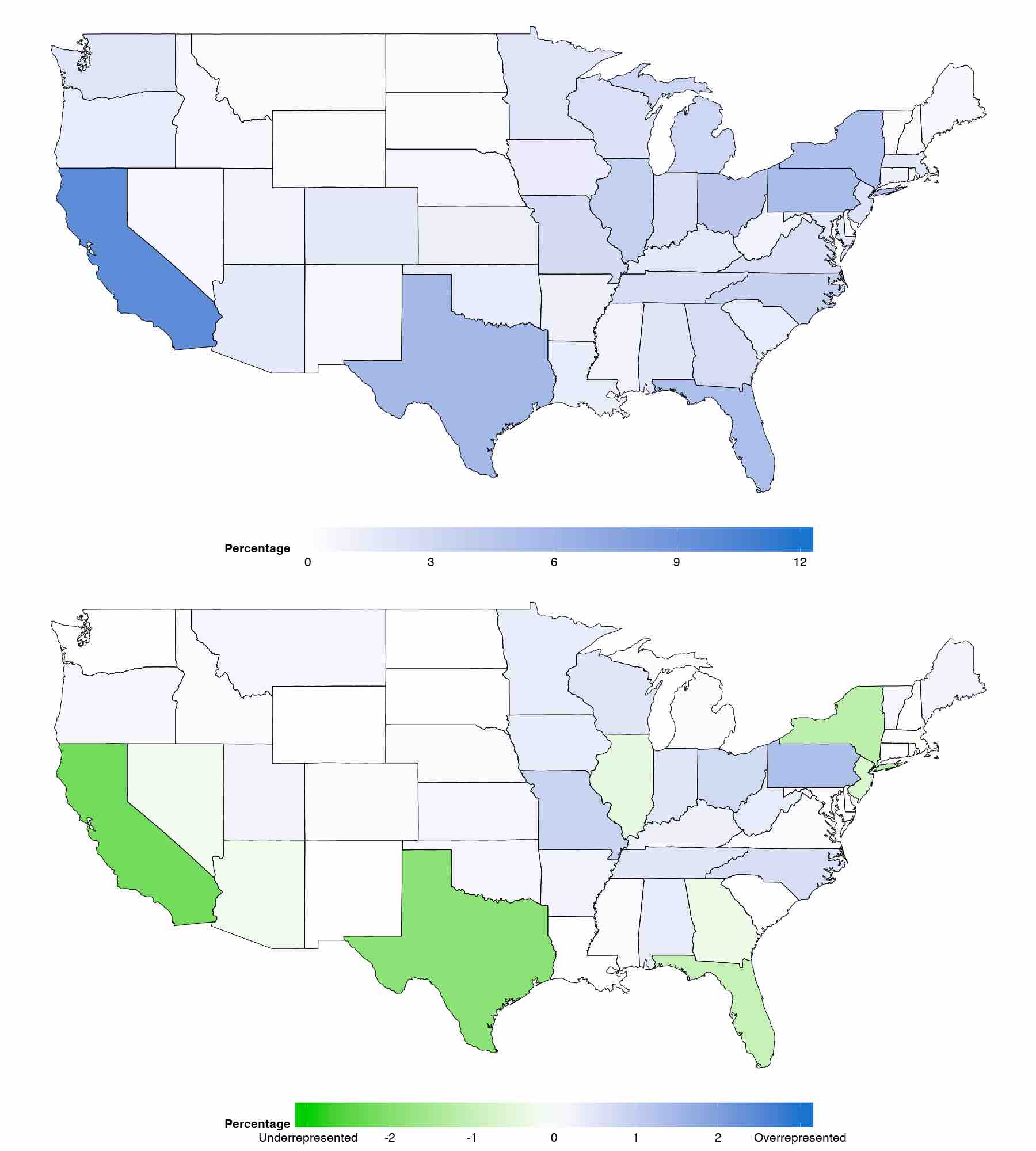}

\caption[short]{Percentage proportions of data in every state for the Annenberg survey (top) and Annenberg percentage proportions - ACS percentage proportions (bottom). The proportions in both surveys were rounded to two decimal places. A state with green hues in the bottom heatmap corresponds to the Annenberg survey underrepresenting that particular state. A state with blue hues in the bottom heatmap corresponds to the Annenberg survey overrepresenting that particular state.}
\label{fig:state_proportions}
\end{figure}

The proportion of observations for the Annenberg phone survey, broken down by state, is shown in the top heatmap of Figure \ref{fig:state_proportions}. The difference in proportions between Annenberg phone survey and the 2006-2010 5-year American Community Survey, broken down by state, is shown in the bottom heatmap of Figure \ref{fig:state_proportions}.

\begin{figure}
        
        \centering
        \includegraphics[width=1\textwidth]{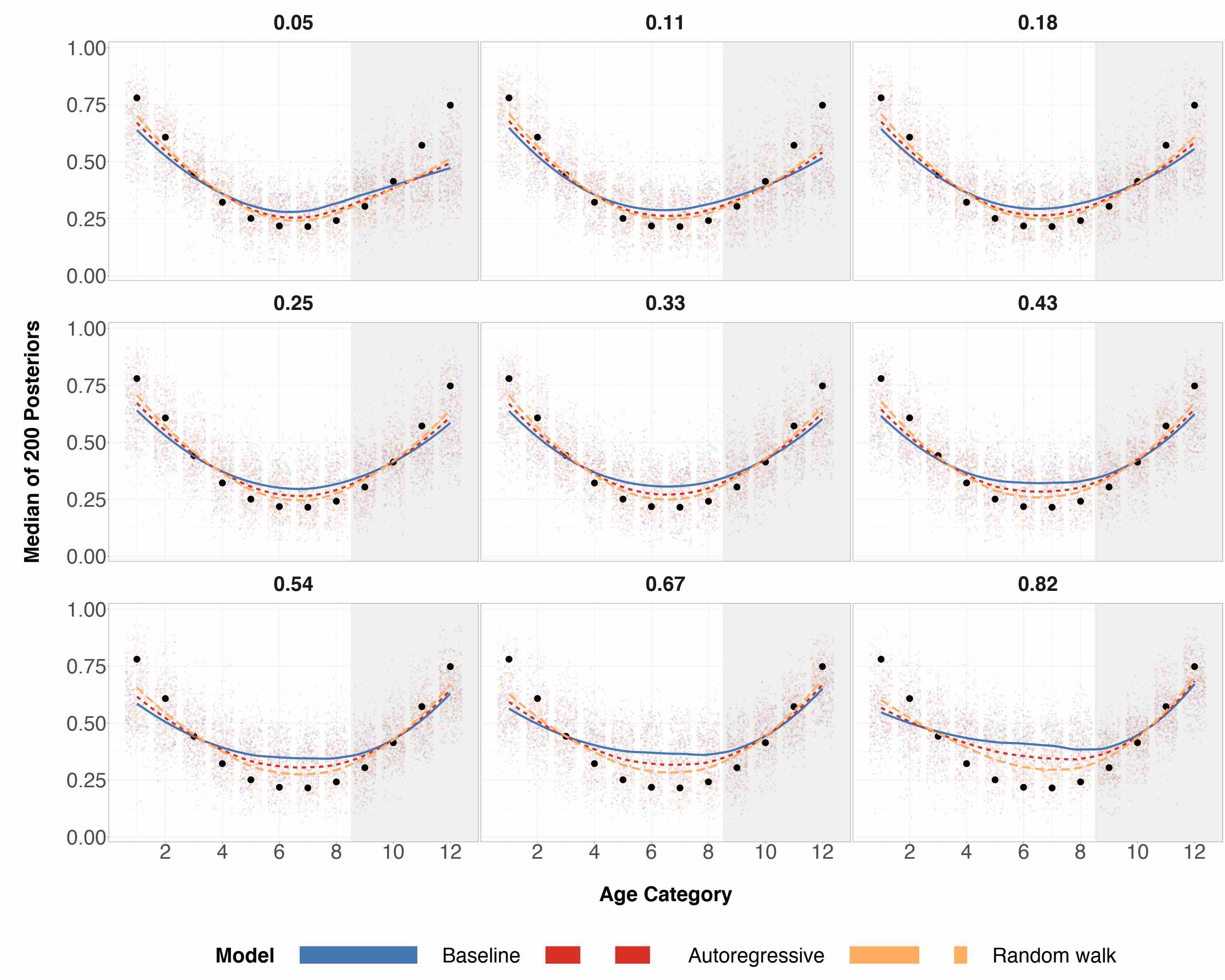}
    
    \caption[short]{Posterior medians for 200 simulations for each age group, where true age preference is U-shaped and sample size $n=100$. Black circles are true preference probabilities for each age group. The numerical index for the 9 plots correspond to the expected proportion of the sample that are older adults (also known as the probability of sampling the subpopulation group with age categories 9-12). The shaded gray region corresponds to the age categories of older individuals for which we over/under sample. The center of the grid represents completely random sampling \textit{and} representative sampling for age categories. Local regression is used for the smoothed estimates amongst the three prior specifications. 
    }
    
\label{fig:agecatbiasu_100}
\end{figure}

\begin{figure}
        
        \centering
        \includegraphics[width=1\textwidth]{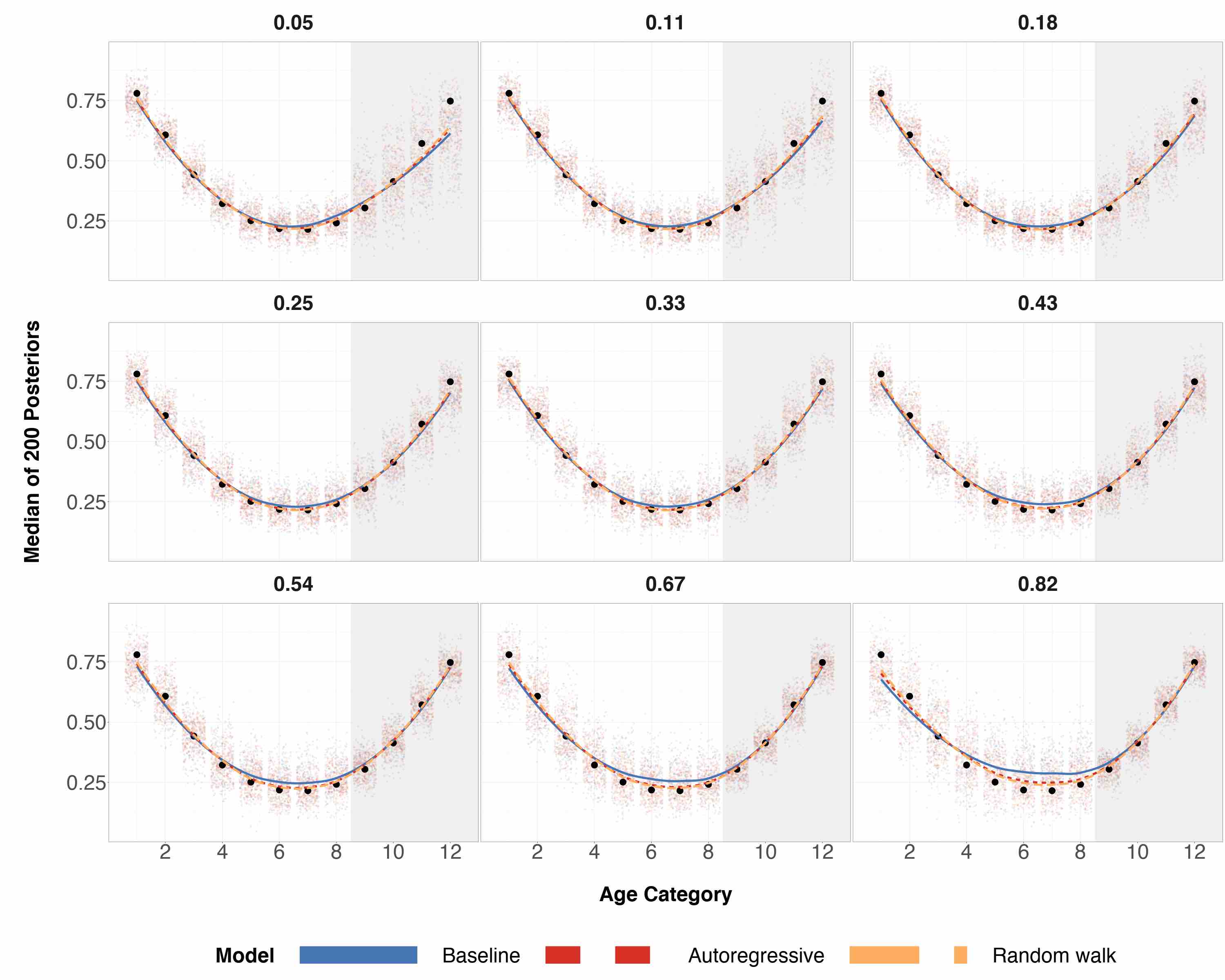}
    
    \caption[short]{Posterior medians for 200 simulations for each age group, where true age preference is U-shaped and sample size $n=500$. Black circles are true preference probabilities for each age group. The numerical index for the 9 plots correspond to the expected proportion of the sample that are older adults (also known as the probability of sampling the subpopulation group with age categories 9-12). The shaded gray region corresponds to the age categories of older individuals for which we over/under sample. The center of the grid represents completely random sampling \textit{and} representative sampling for age categories. Local regression is used for the smoothed estimates amongst the three prior specifications.
    }
\label{fig:agecatbiasu_500}
\end{figure}

\begin{figure}
    \centering
    \includegraphics[width=1\textwidth]{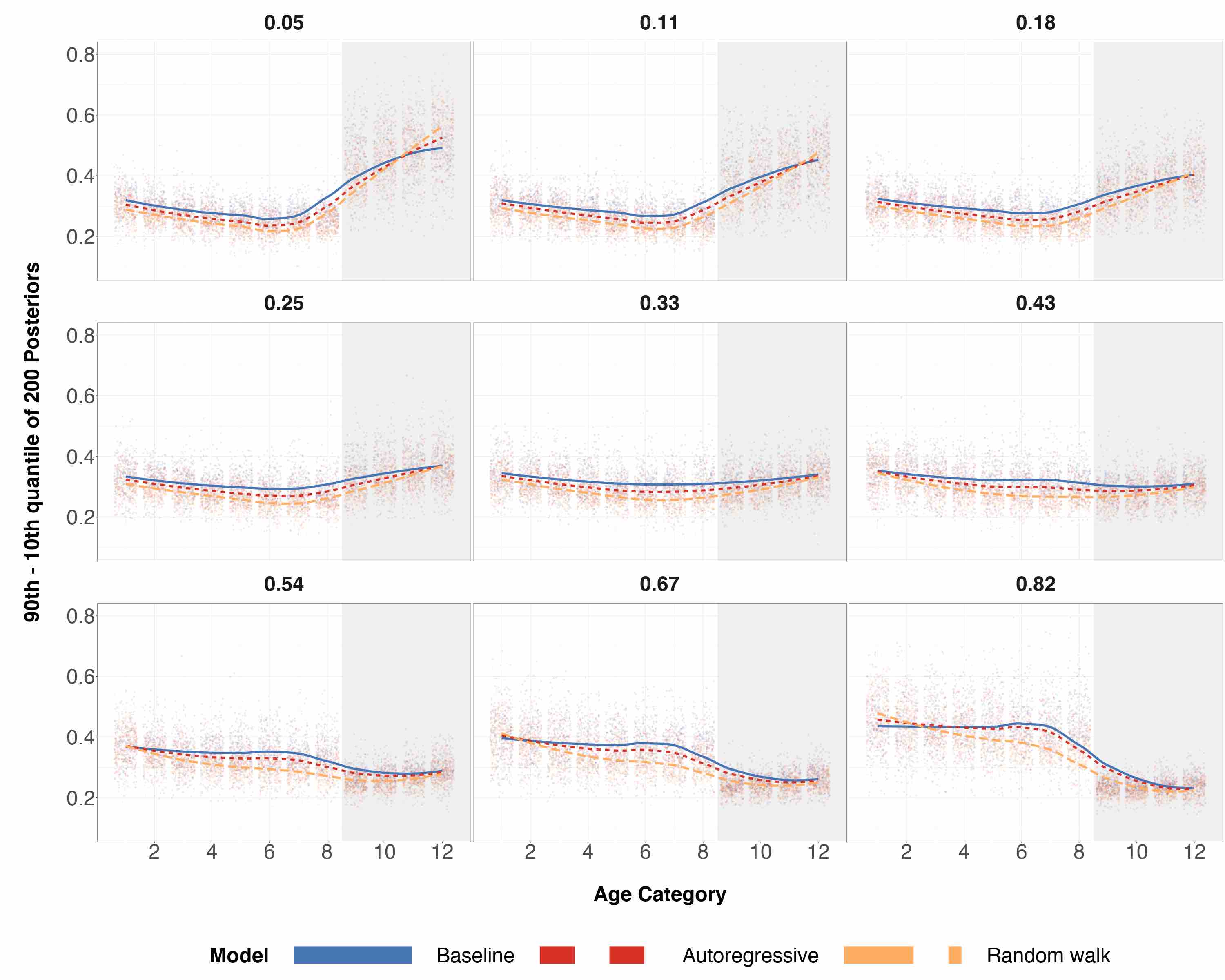}

\caption[short]{Differences in the $90^{th}$ and $10^{th}$ posterior quantiles for every age category when true age preference is U-shaped and $n=100$ for 200 simulations. The numerical index for the 9 plots correspond to the expected proportion of the sample that are older adults (also known as the probability of sampling the subpopulation group with age categories 9-12). The shaded gray region corresponds to the age categories of older  individuals for which we over/under sample. The center of the  grid represents completely random sampling \textit{and} representative sampling for age categories. Local regression is used for the smoothed estimates amongst the three prior specifications.}
\label{fig:posteriorwidth_u_100}
\end{figure}


\begin{figure}
    \centering
    \includegraphics[width=1\textwidth]{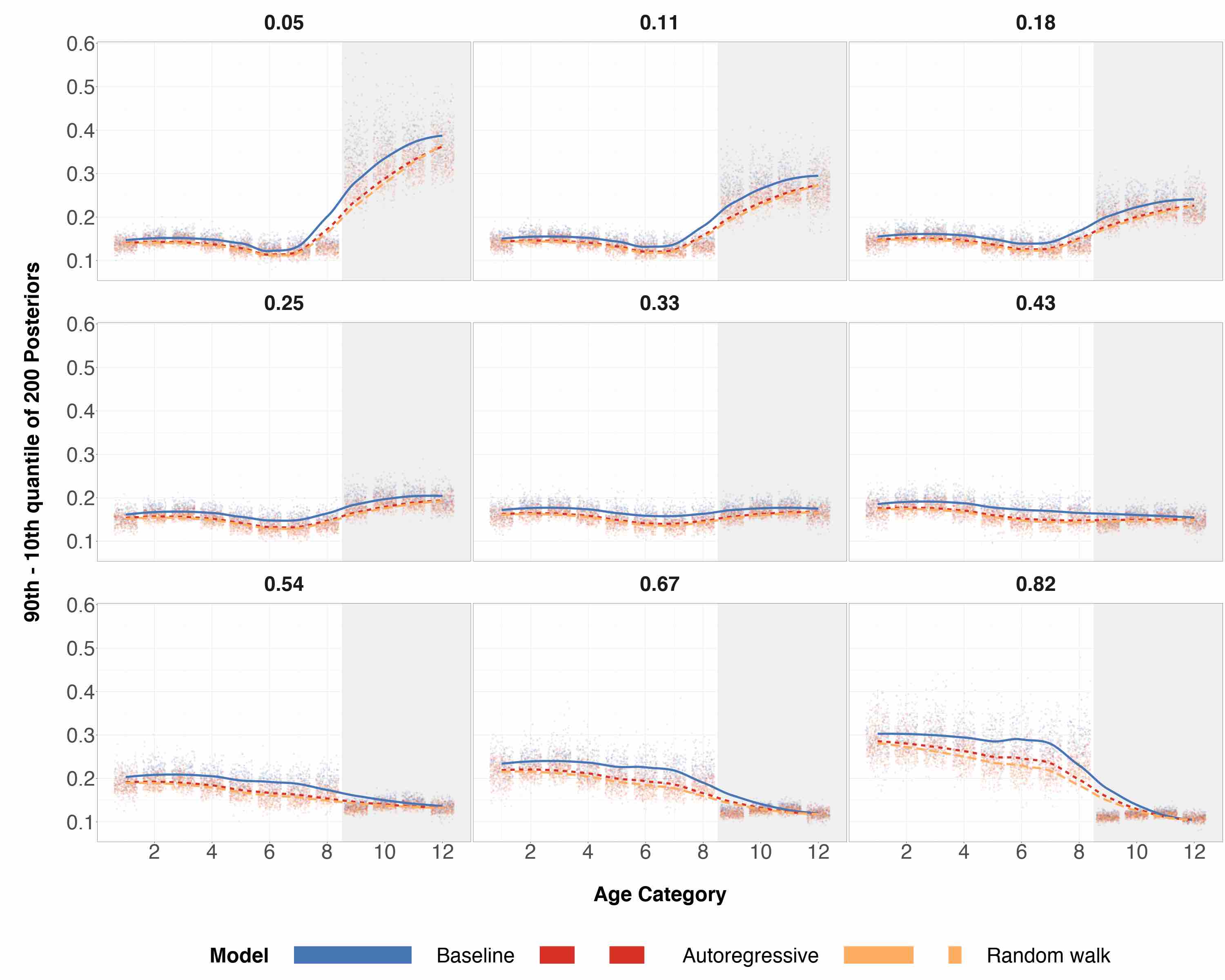}

\caption[short]{Differences in the $90^{th}$ and $10^{th}$ posterior quantiles for every age category when true age preference is U-shaped and $n=500$ for 200 simulations. The numerical index for the 9 plots correspond to the expected proportion of the sample that are older adults (also known as the probability of sampling the subpopulation group with age categories 9-12). The shaded gray region corresponds to the age categories of older  individuals for which we over/under sample. The center of the  grid represents completely random sampling \textit{and} representative sampling for age categories. Local regression is used for the smoothed estimates amongst the three prior specifications.}
\label{fig:posteriorwidth_u_500}
\end{figure}

\begin{figure}
    \centering
    
        \includegraphics[width=1\textwidth]{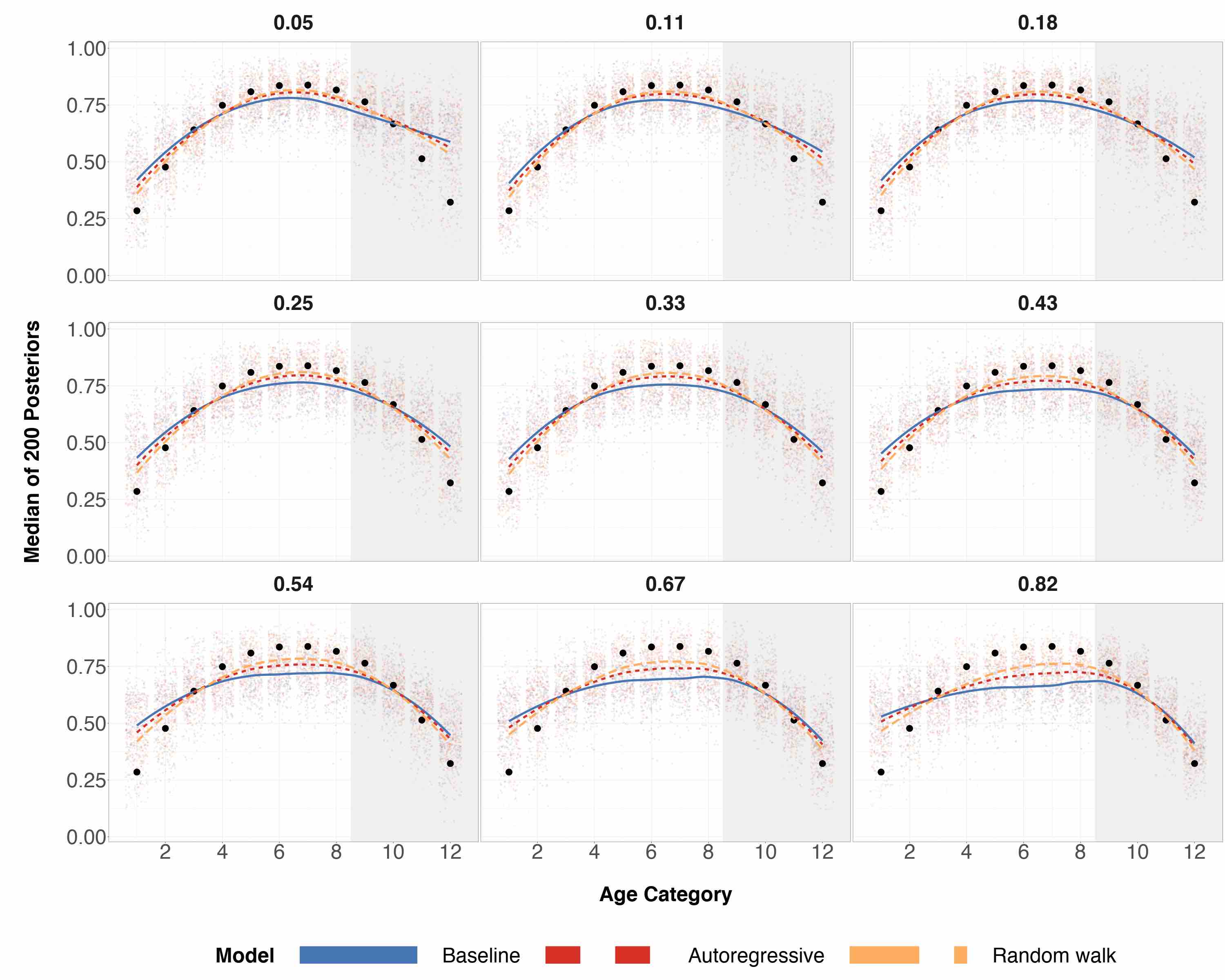}

\caption[short]{Posterior medians for 200 simulations for each age group, where true age preference is cap-shaped and sample size $n=100$. Black circles are true preference probabilities for each age group. The numerical index for the 9 plots correspond to the expected proportion of the sample that are older adults (also known as the probability of sampling the subpopulation group with age categories 9-12). The shaded gray region corresponds to the age categories of older individuals for which we over/under sample. The center of the grid represents completely random sampling \textit{and} representative sampling for age categories. Local regression is used for the smoothed estimates amongst the three prior specifications.}
\label{fig:agecatbiascap_100}
\end{figure}


\begin{figure}
    \centering

        \includegraphics[width=1\textwidth]{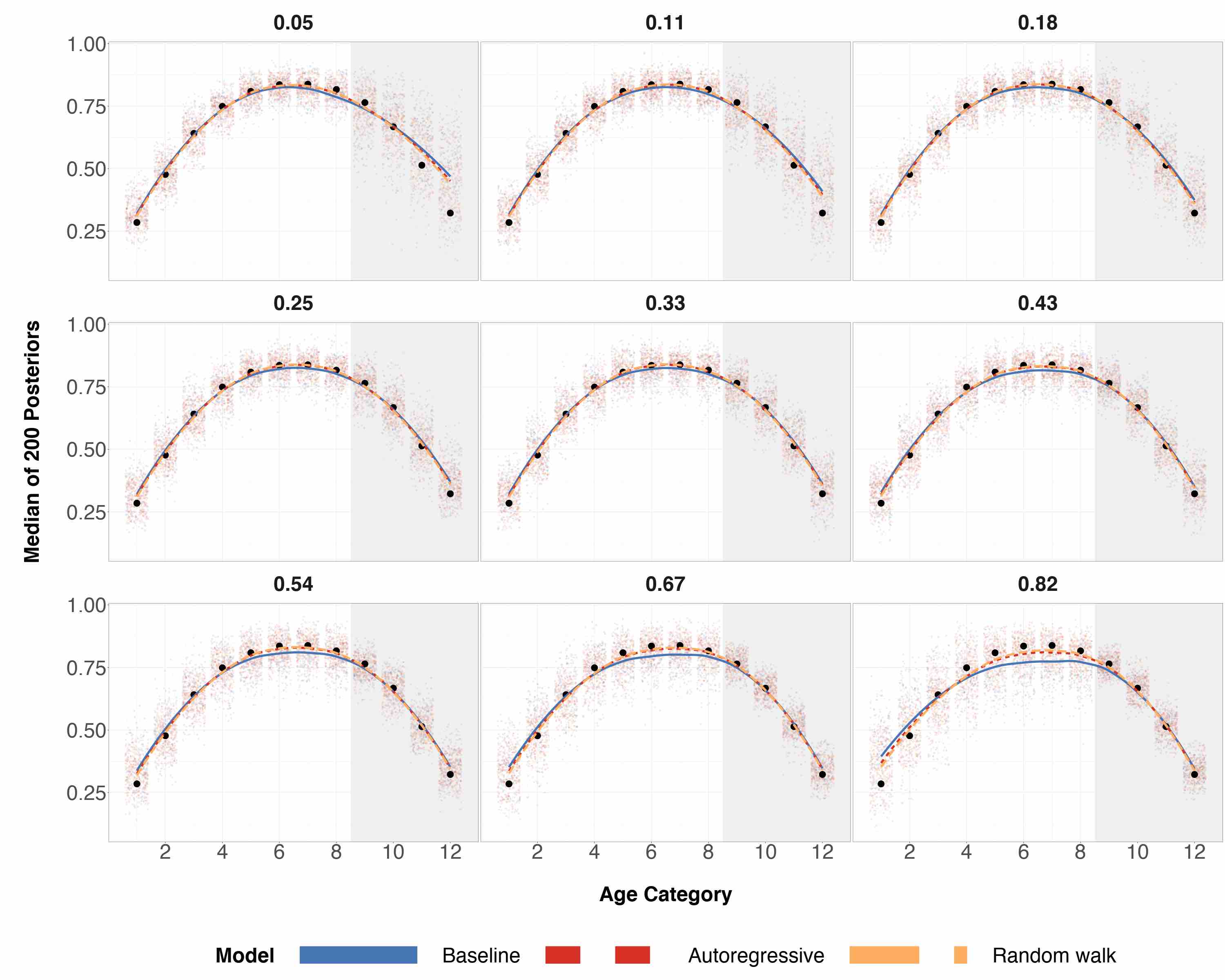}

\caption[short]{Posterior medians for 200 simulations for each age group, where true age preference is cap-shaped and sample size $n=500$. Black circles are true preference probabilities for each age group. The numerical index for the 9 plots correspond to the expected proportion of the sample that are older adults (also known as the probability of sampling the subpopulation group with age categories 9--12). The shaded gray region corresponds to the age categories of older individuals for which we over/under sample. The center of the grid represents completely random sampling \textit{and} representative sampling for age categories. Local regression is used for the smoothed estimates amongst the three prior specifications.}
\label{fig:agecatbiascap_500}
\end{figure}


\begin{figure}
    \centering
    \includegraphics[width=1\textwidth]{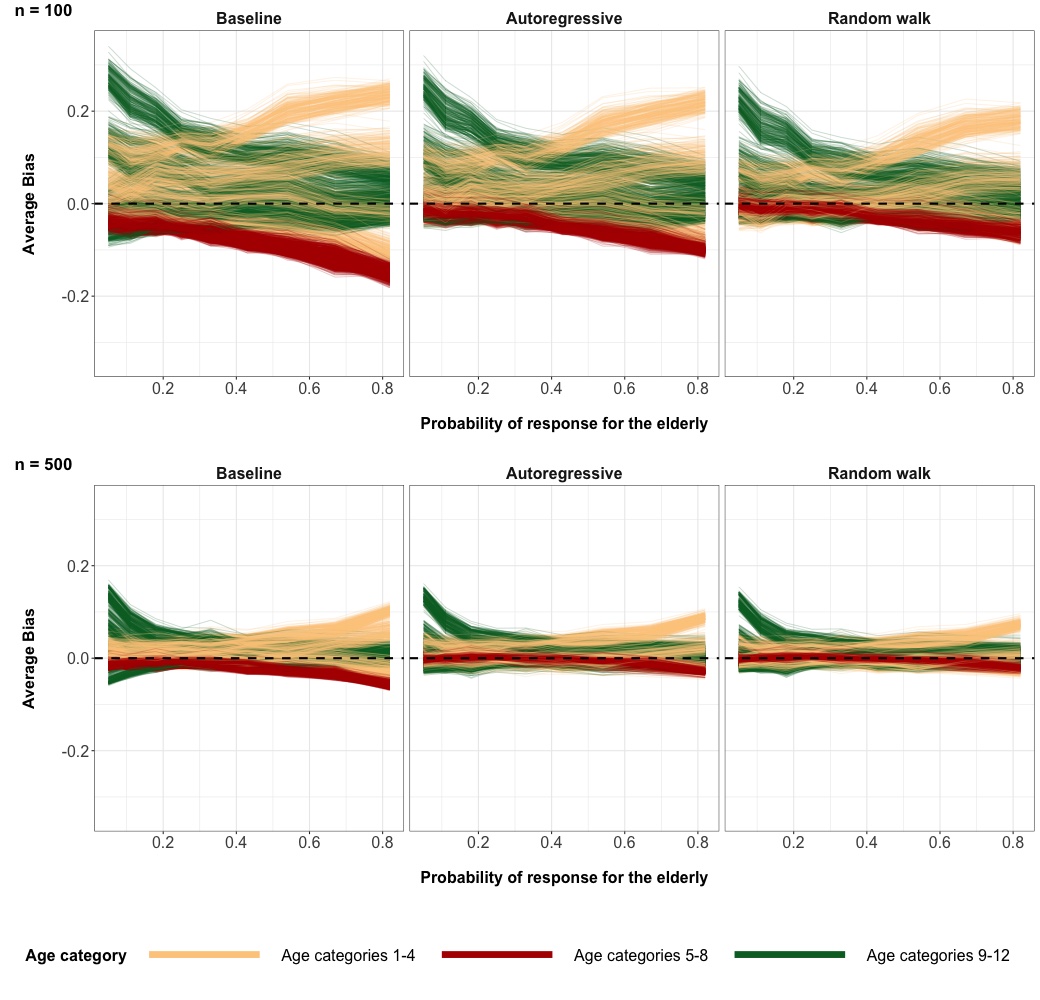}

\caption[short]{The average bias values coming from 200 simulations of posterior medians of the 2448 poststratification cells. The possible values of average bias are in the interval $(-1,1)$. Sample size $n = 100$ (top) and $n = 500$ (bottom). The true preference curve for age is cap-shaped. The horizontal dashed line at $y=0$ represents zero bias.}
\label{fig:biasoverallcap_100_500}
\end{figure}


\begin{figure}
    \centering
    \includegraphics[width=1\textwidth]{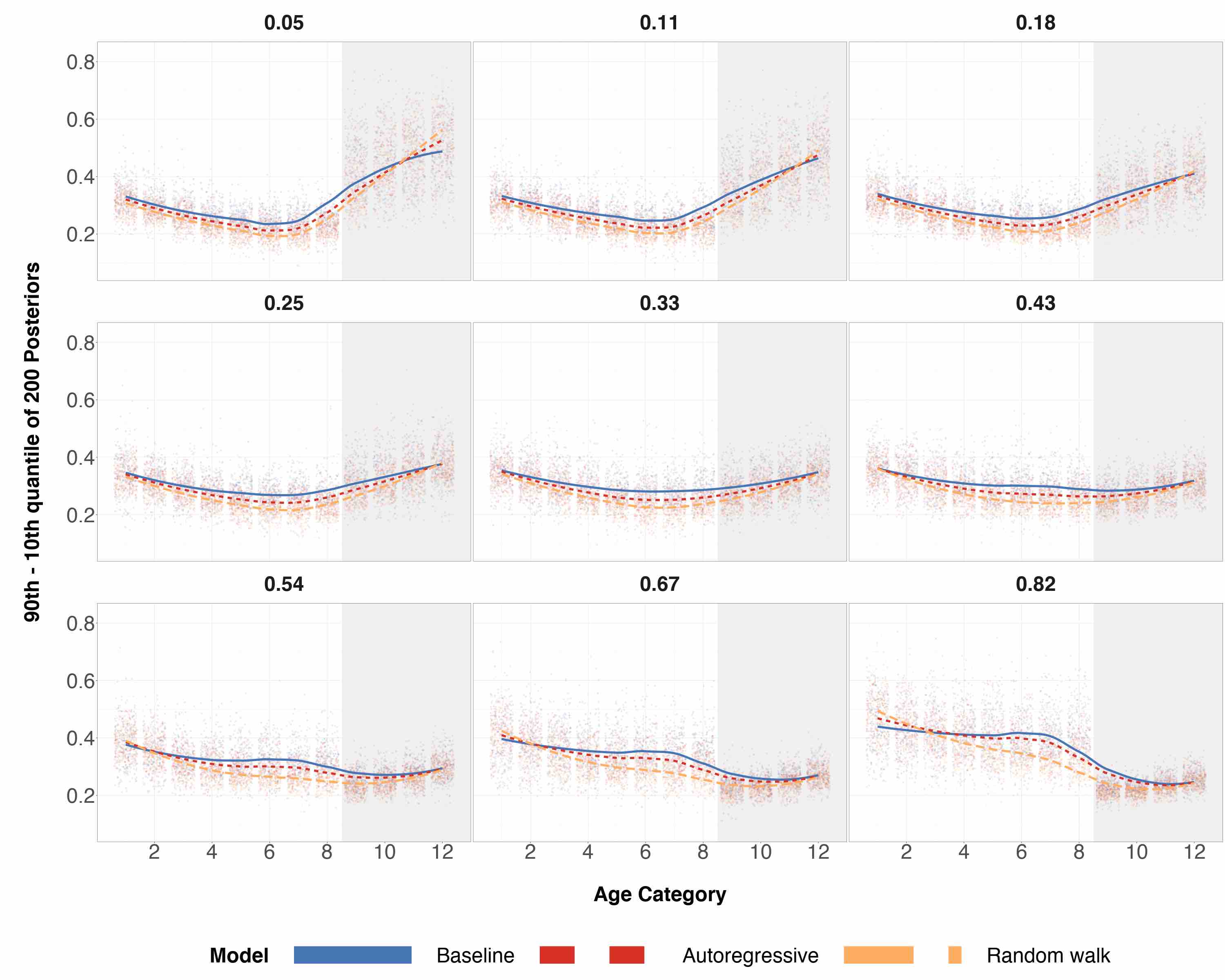}

\caption[short]{Differences in the $90^{th}$ and $10^{th}$ posterior quantiles for every age category when true age preference is cap-shaped and $n=100$ for 200 simulations. The numerical index for the 9 plots correspond to the expected proportion of the sample that are older adults (also known as the probability of sampling the subpopulation group with age categories 9--12). The shaded gray region corresponds to the age categories of older  individuals for which we over/under sample. The center of the  grid represents completely random sampling \textit{and} representative sampling for age categories. Local regression is used for the smoothed estimates amongst the three prior specifications.}
\label{fig:posteriorwidth_cap_100}
\end{figure}


\begin{figure}
    \centering
    \includegraphics[width=1\textwidth]{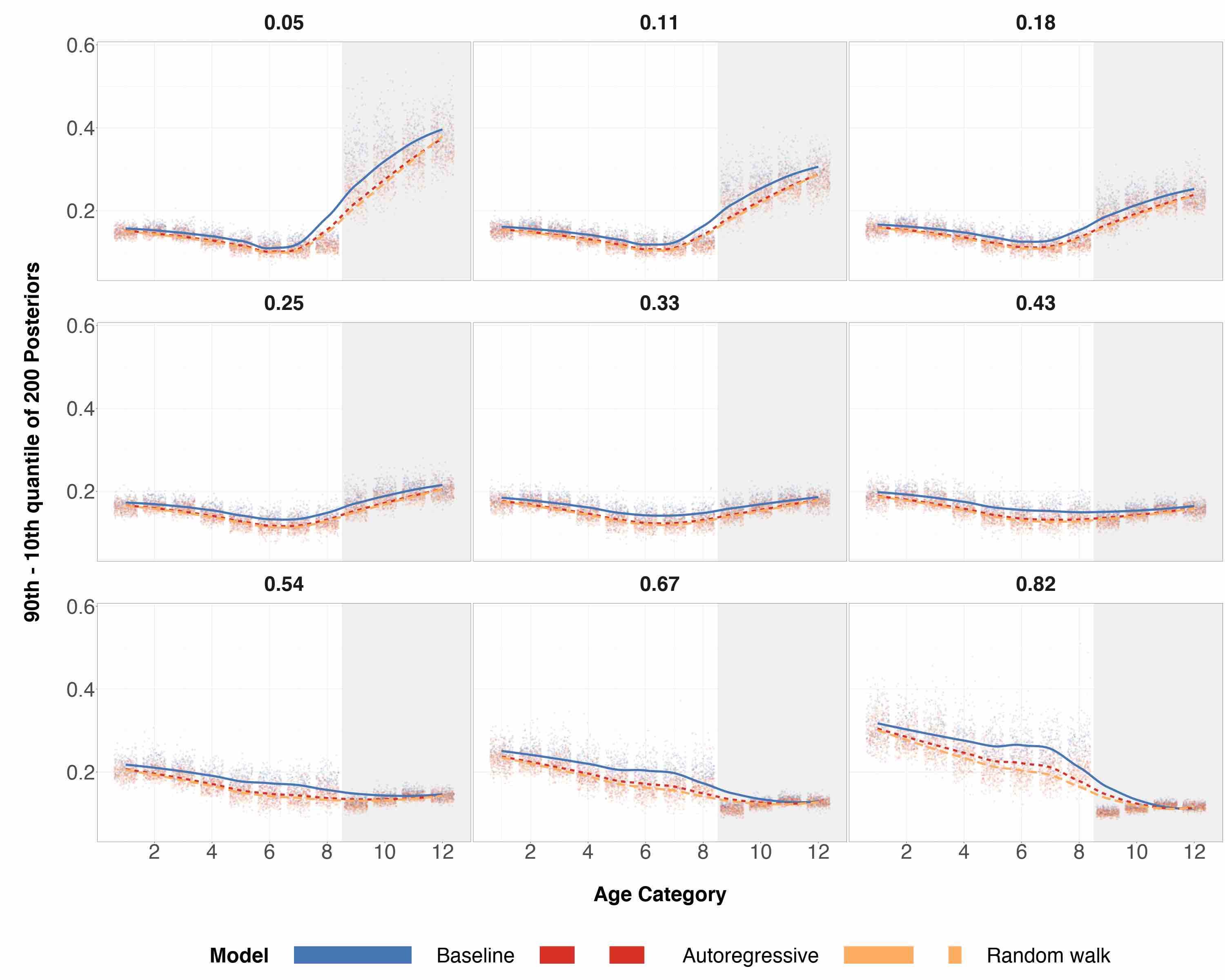}

\caption[short]{Differences in the $90^{th}$ and $10^{th}$ posterior quantiles for every age category, when true age preference is cap-shaped and $n=500$ for 200 simulations. The numerical index for the 9 plots correspond to the expected proportion of the sample that are older adults (also known as the probability of sampling the subpopulation group with age categories 9--12). The shaded gray region corresponds to the age categories of older  individuals for which we over/under sample. The center of the  grid represents completely random sampling \textit{and} representative sampling for age categories. Local regression is used for the smoothed estimates amongst the three prior specifications.}
\label{fig:posteriorwidth_cap_500}
\end{figure}


\begin{figure}
    \centering
    \includegraphics[width=1\textwidth]{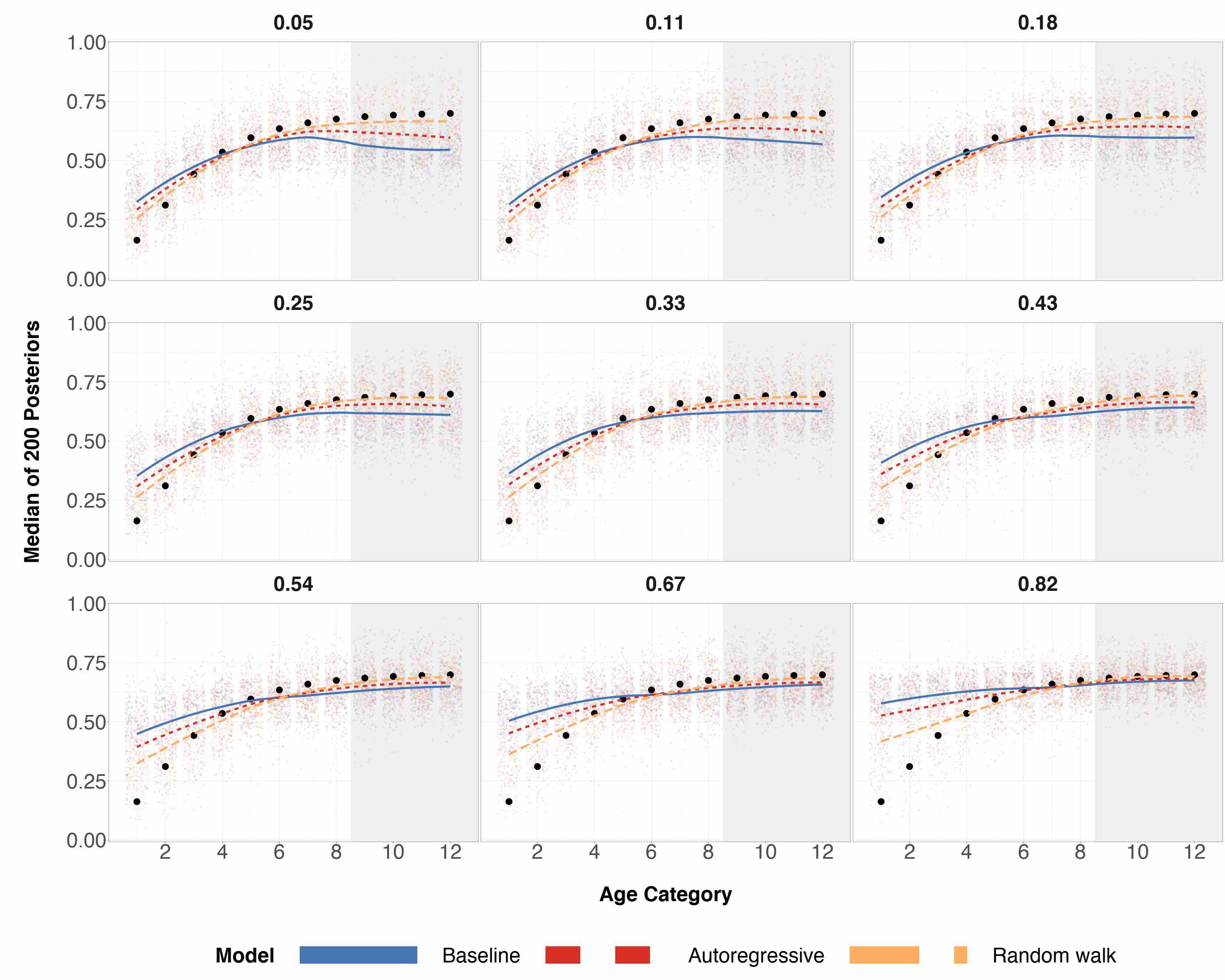}
    
\caption[short]{Posterior medians for 200 simulations for each age group, where true age preference is increasing-shaped and sample size $n=100$. Black circles are true preference probabilities for each age group. The numerical index for the 9 plots correspond to the expected proportion of the sample that are older adults (also known as the probability of sampling the subpopulation group with age categories 9--12). The shaded gray region corresponds to the age categories of older individuals for which we over/under sample. The center of the grid represents completely random sampling \textit{and} representative sampling for age categories. Local regression is used for the smoothed estimates amongst the three prior specifications.}
\label{fig:agecatbiasincreasing_100}
\end{figure}

\begin{figure}
    \centering
    \includegraphics[width=1\textwidth]{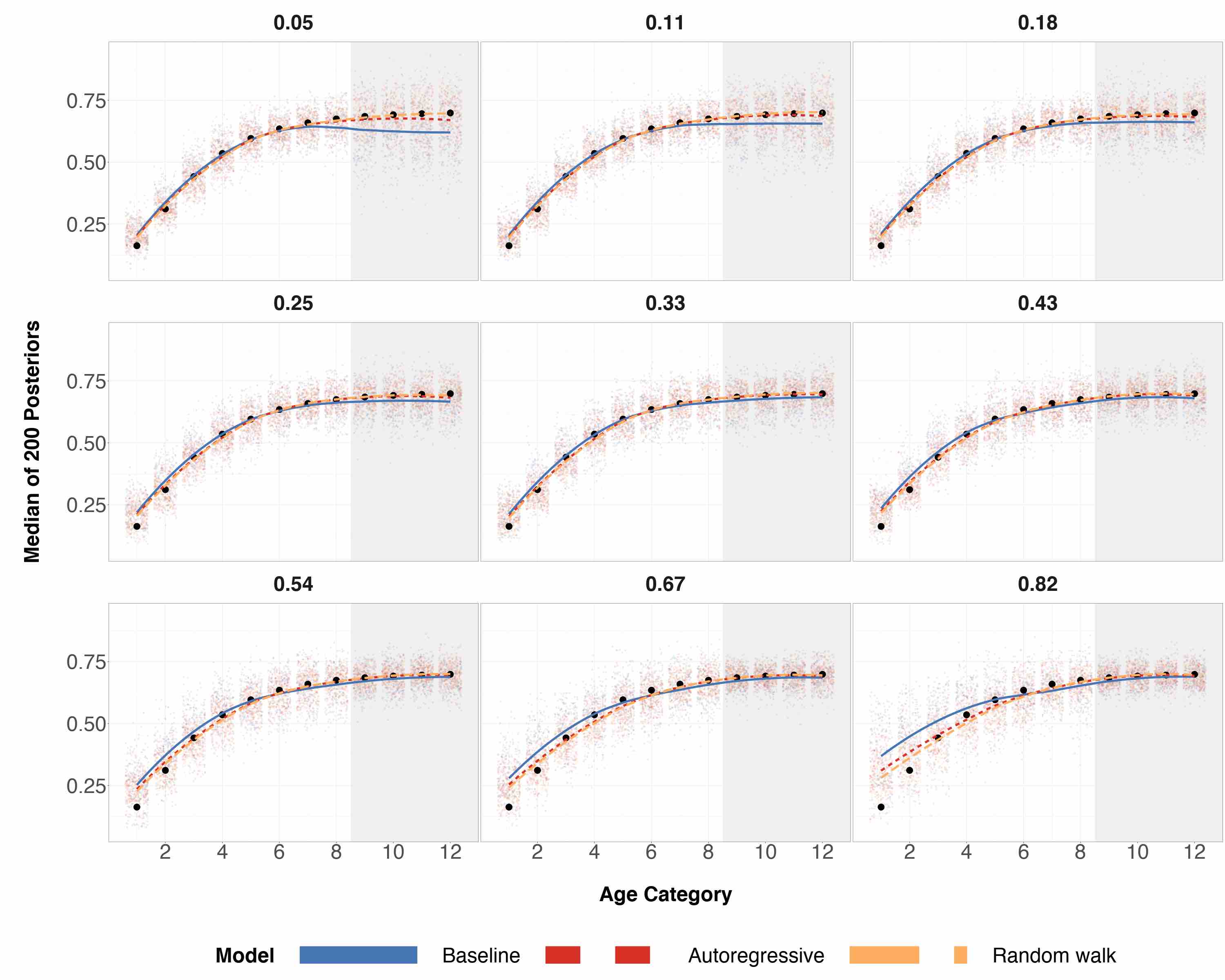}
    
\caption[short]{Posterior medians for 200 simulations for each age group, where true age preference is increasing-shaped and sample size $n=500$. Black circles are true preference probabilities for each age group. The numerical index for the 9 plots correspond to the expected proportion of the sample that are older adults (also known as the probability of sampling the subpopulation group with age categories 9--12). The shaded gray region corresponds to the age categories of older individuals for which we over/under sample. The center of the grid represents completely random sampling \textit{and} representative sampling for age categories. Local regression is used for the smoothed estimates amongst the three prior specifications.}
\label{fig:agecatbiasincreasing_500}
\end{figure}


\begin{figure}
    \centering
    \includegraphics[width=1\textwidth]{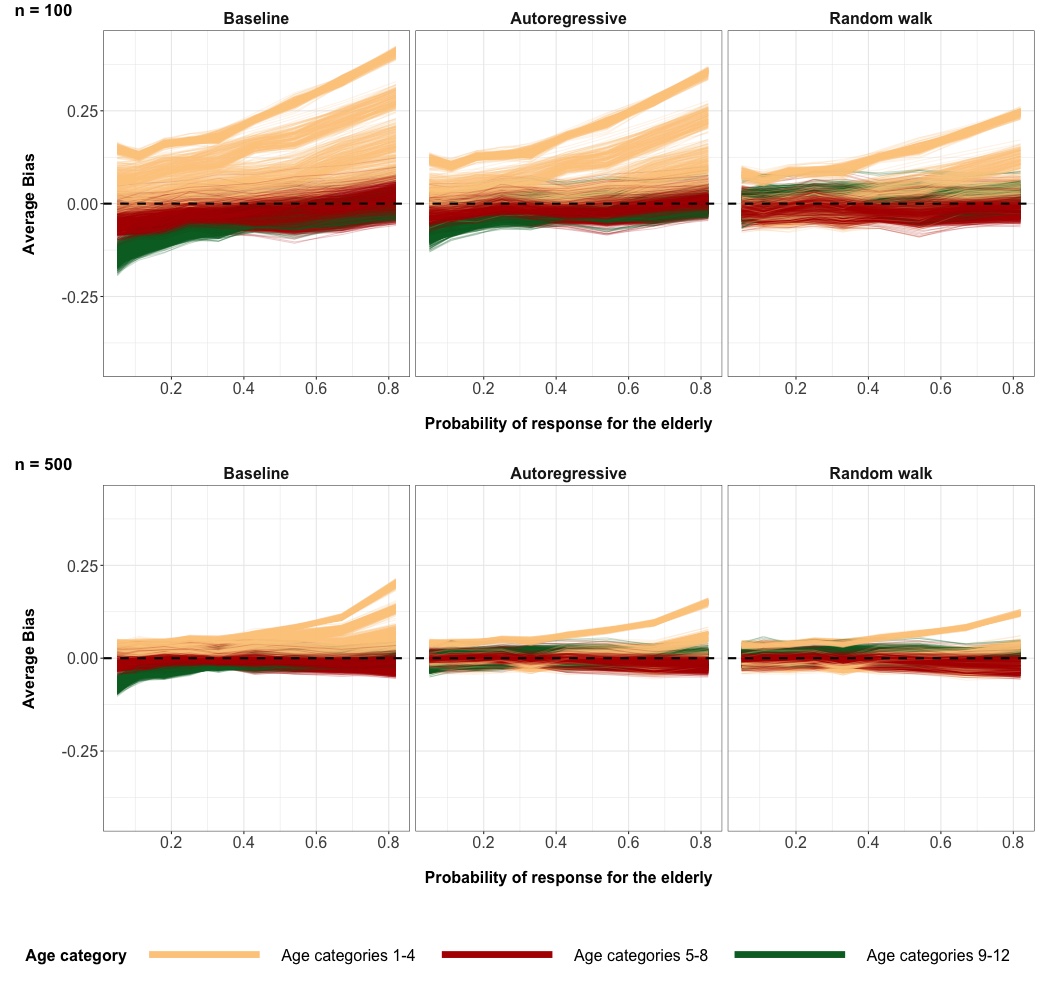}
\caption[short]{The average bias values coming from 200 simulations of posterior medians of the 2448 poststratification cells. The possible values of average bias are in the interval $(-1,1)$. Sample size $n = 100$ (top) and $n = 500$ (bottom). The true preference curve for age is increasing-shaped. The horizontal dashed line at $y=0$ represents zero bias.}
\label{fig:biasoverallincreasing_100_500}
\end{figure}


\begin{figure}
    \centering
    \includegraphics[width=1\textwidth]{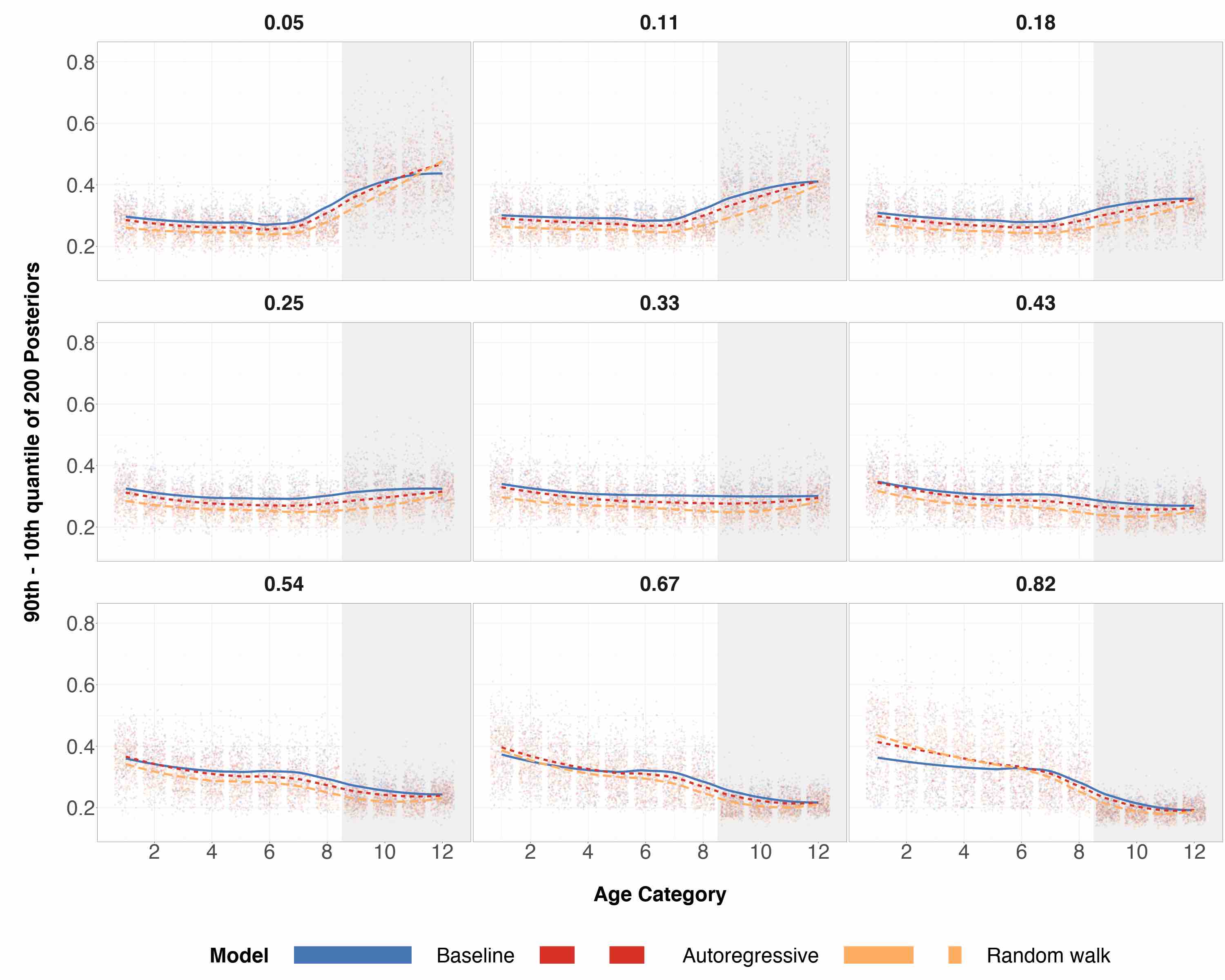}

\caption[short]{Differences in the $90^{th}$ and $10^{th}$ posterior quantiles for every age category when true age preference is increasing-shaped and $n=100$ for 200 simulations. The numerical index for the 9 plots correspond to the expected proportion of the sample that are older adults (also known as the probability of sampling the subpopulation group with age categories 9--12). The shaded gray region corresponds to the age categories of older  individuals for which we over/under sample. The center of the  grid represents completely random sampling \textit{and} representative sampling for age categories. Local regression is used for the smoothed estimates amongst the three prior specifications.}
\label{fig:posteriorwidth_increasing_100}
\end{figure}


\begin{figure}
    \centering
    \includegraphics[width=1\textwidth]{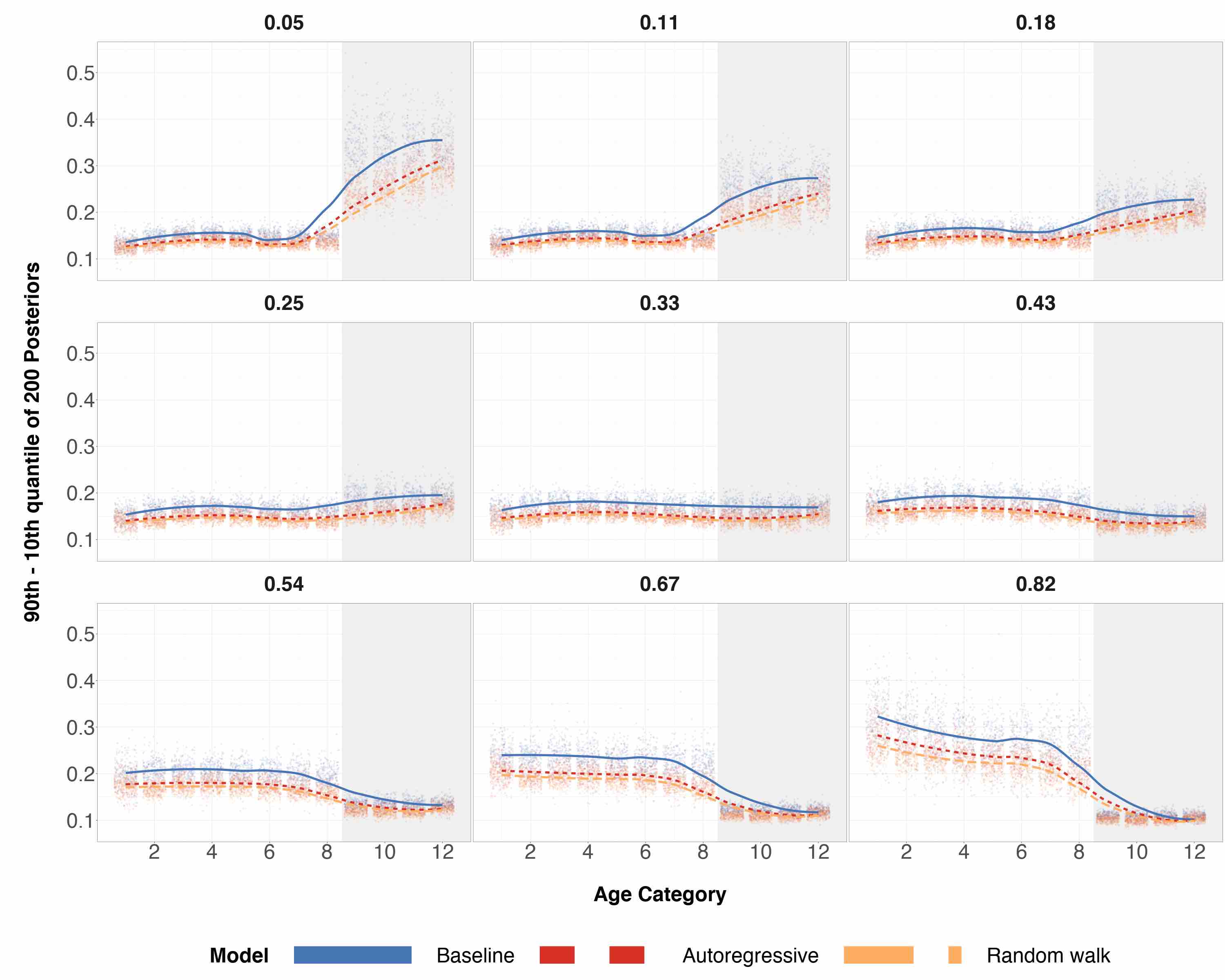}

\caption[short]{Differences in the $90^{th}$ and $10^{th}$ posterior quantiles for every age category when true age preference is increasing-shaped and $n=500$ for 200 simulations. The numerical index for the 9 plots correspond to the expected proportion of the sample that are older adults (also known as the probability of sampling the subpopulation group with age categories 9--12). The shaded gray region corresponds to the age categories of older  individuals for which we over/under sample. The center of the  grid represents completely random sampling \textit{and} representative sampling for age categories. Local regression is used for the smoothed estimates amongst the three prior specifications.}
\label{fig:posteriorwidth_increasing_500}
\end{figure}


\begin{figure}
    \centering
    \includegraphics[width=1\textwidth]{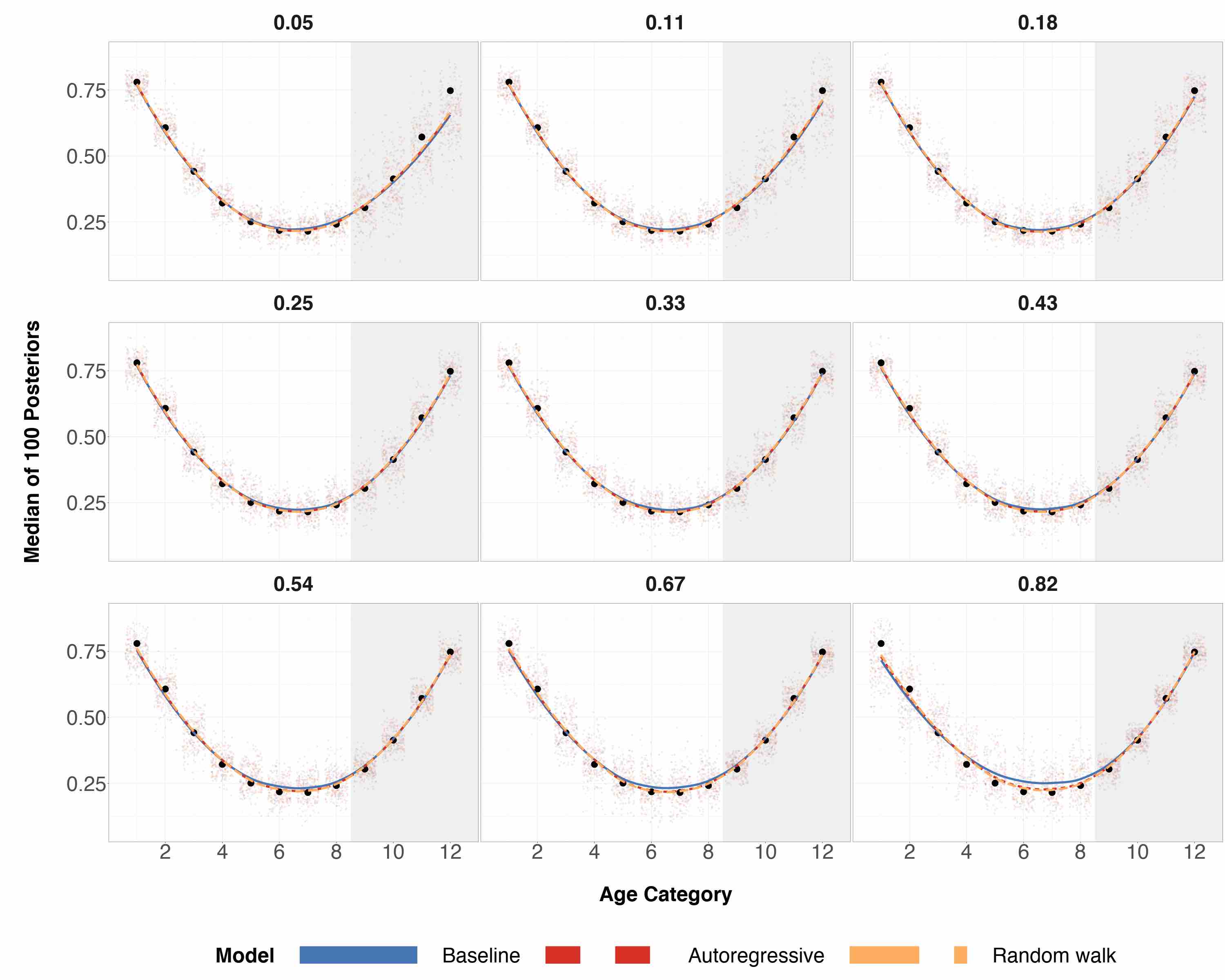}
    
\caption[short]{Posterior medians for 100 simulations with 12 age groups, where true age preference is U-shaped and sample size $n=1000$. Black circles are true preference probabilities for each age group. The numerical index for the 9 plots correspond to the expected proportion of the sample that are older adults (also known as the probability of sampling the subpopulation group with age categories 9--12). The shaded gray region corresponds to the age categories of older individuals for which we over/under sample. The center of the grid represents completely random sampling \textit{and} representative sampling for age categories. Local regression is used for the smoothed estimates amongst the three prior specifications.}
\label{fig:allmedians_facet_100_12_1000_u}
\end{figure}

\begin{figure}
    \centering
    \includegraphics[width=1\textwidth]{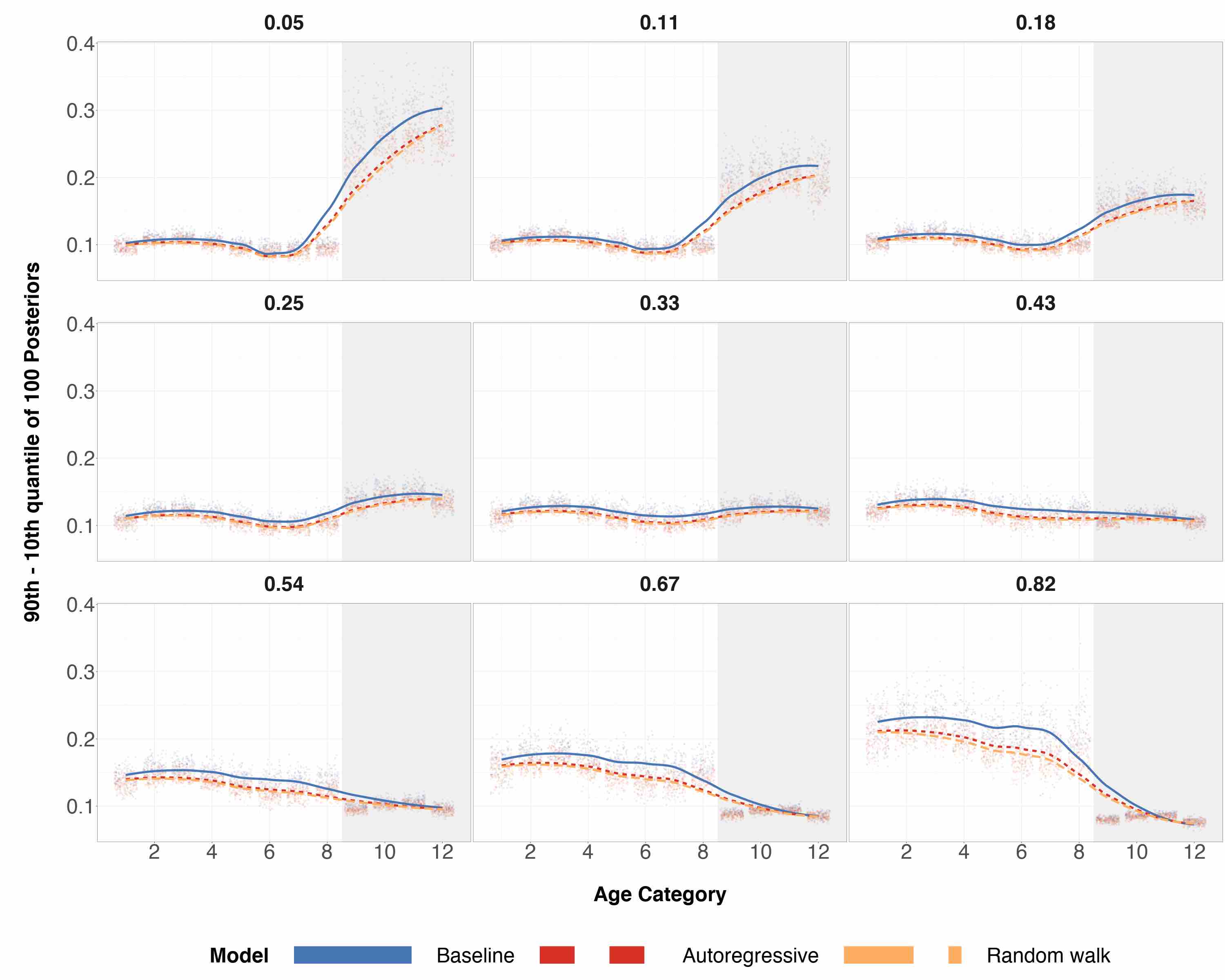}
    
\caption[short]{Differences in the $90^{th}$ and $10^{th}$ posterior quantiles for every age category when true age preference is U-shaped and $n=1000$ for 100 simulations. The numerical index for the 9 plots correspond to the expected proportion of the sample that are older adults (also known as the probability of sampling the subpopulation group with age categories 9--12). The shaded gray region corresponds to the age categories of older  individuals for which we over/under sample. The center of the  grid represents completely random sampling \textit{and} representative sampling for age categories. Local regression is used for the smoothed estimates amongst the three prior specifications.}
\label{fig:allquantilediff_facet_100_12_1000_u}
\end{figure}

\begin{figure}
    \centering
    \includegraphics[width=1\textwidth]{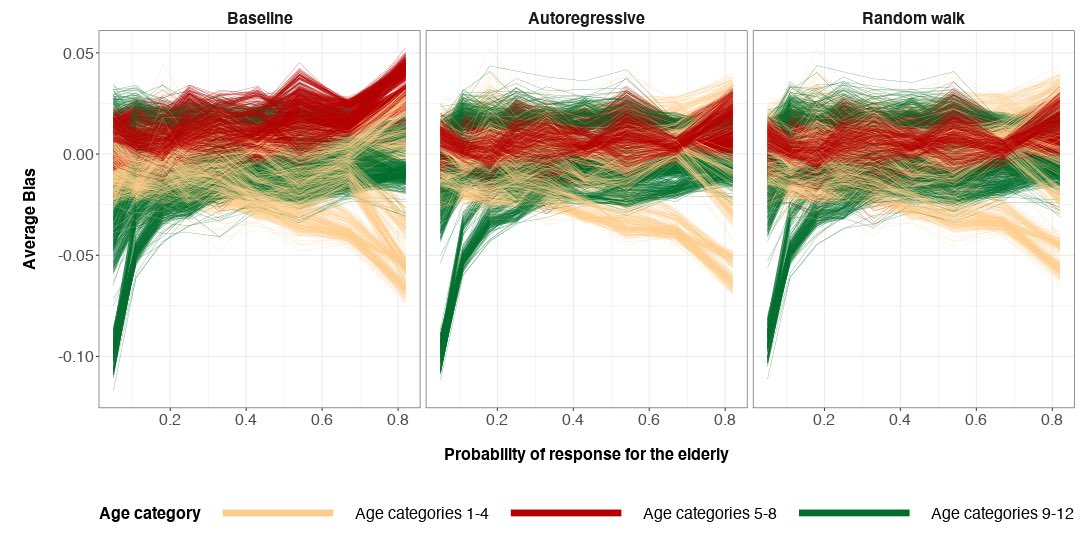}
    
    \caption[short]{The average bias values coming from 100 simulations of posterior medians of the 2448 poststratification cells. The possible values of average bias are in the interval $(-1,1)$. Sample size $n = 1000$. The true preference curve for age is U-shaped. $y=0$ represents zero bias.}
    \label{fig:biasfacet_100_12_1_1000_u}
\end{figure}


\begin{figure}
    \centering
    \includegraphics[width=1\textwidth]{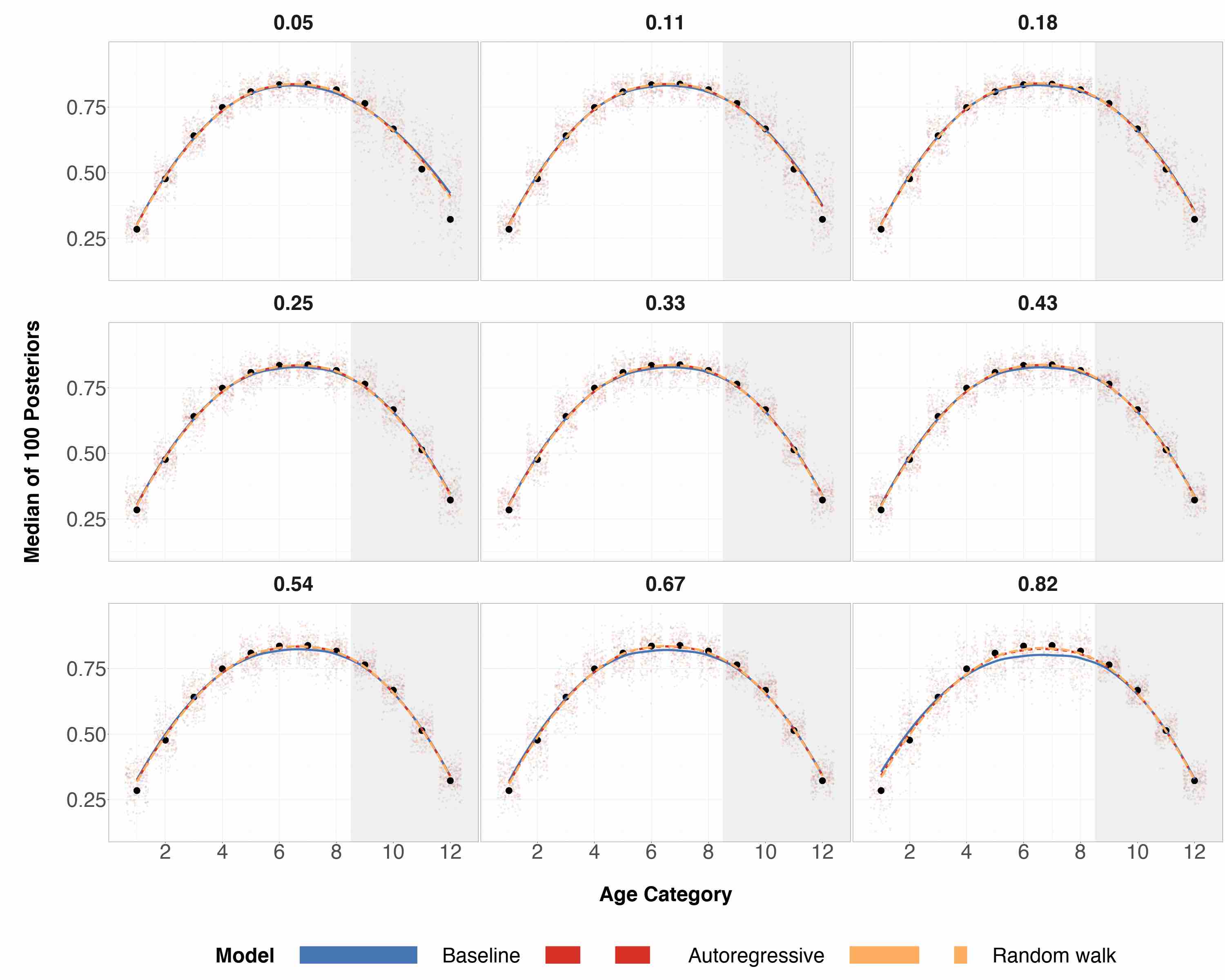}
    
\caption[short]{Posterior medians for 100 simulations with 12 age groups, where true age preference is cap-shaped and sample size $n=1000$. Black circles are true preference probabilities for each age group. The numerical index for the 9 plots correspond to the expected proportion of the sample that are older adults (also known as the probability of sampling the subpopulation group with age categories 9--12). The shaded gray region corresponds to the age categories of older individuals for which we over/under sample. The center of the grid represents completely random sampling \textit{and} representative sampling for age categories. Local regression is used for the smoothed estimates amongst the three prior specifications.}
\label{fig:allmedians_facet_100_12_1000_cap}
\end{figure}

\begin{figure}
    \centering
    \includegraphics[width=1\textwidth]{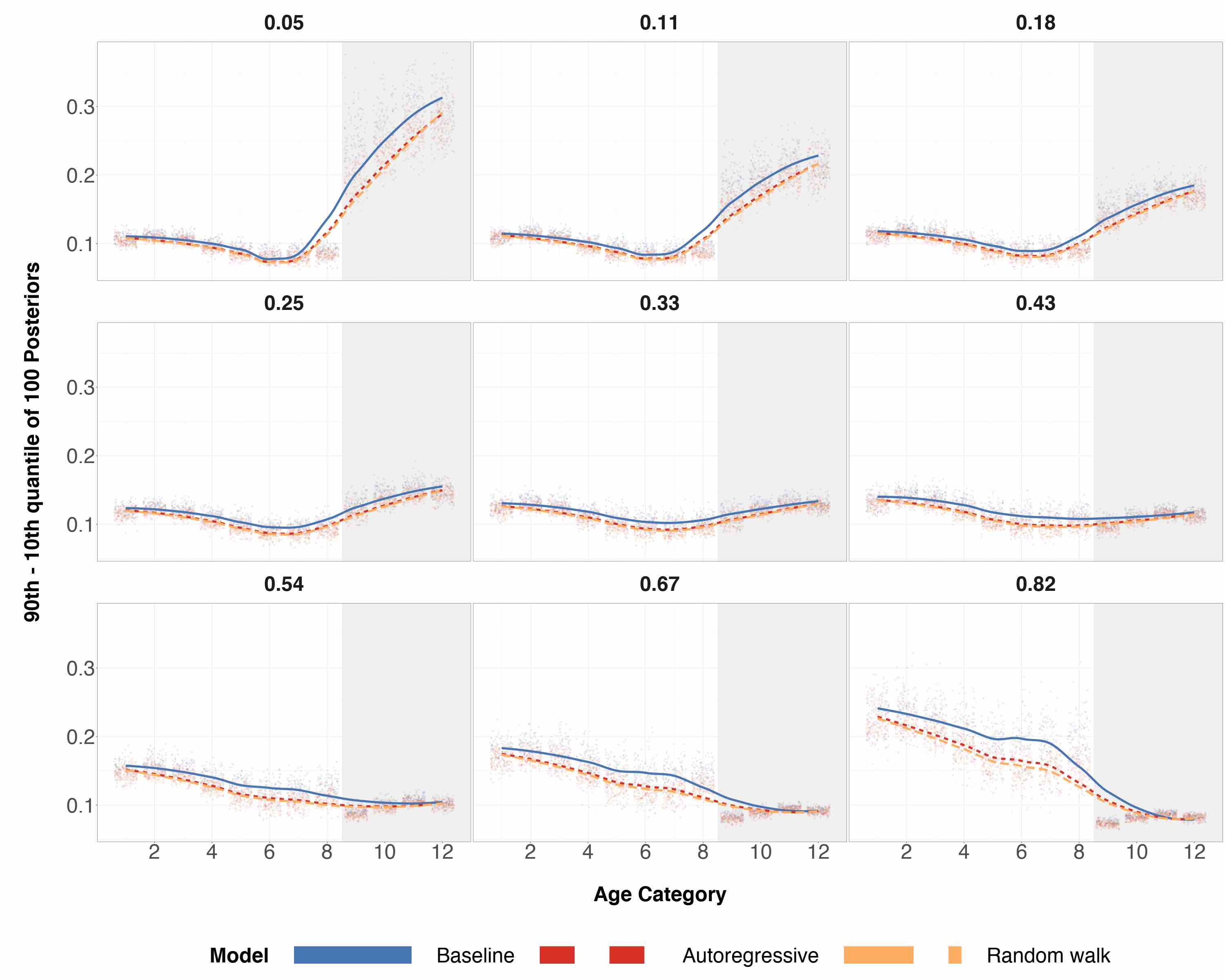}
    
\caption[short]{Differences in the $90^{th}$ and $10^{th}$ posterior quantiles for every age category when true age preference is cap-shaped and $n=1000$ for 100 simulations. The numerical index for the 9 plots correspond to the expected proportion of the sample that are older adults (also known as the probability of sampling the subpopulation group with age categories 9--12). The shaded gray region corresponds to the age categories of older  individuals for which we over/under sample. The center of the  grid represents completely random sampling \textit{and} representative sampling for age categories. Local regression is used for the smoothed estimates amongst the three prior specifications.}
\label{fig:allquantilediff_facet_100_12_1000_cap}
\end{figure}

\begin{figure}
    \centering
    \includegraphics[width=1\textwidth]{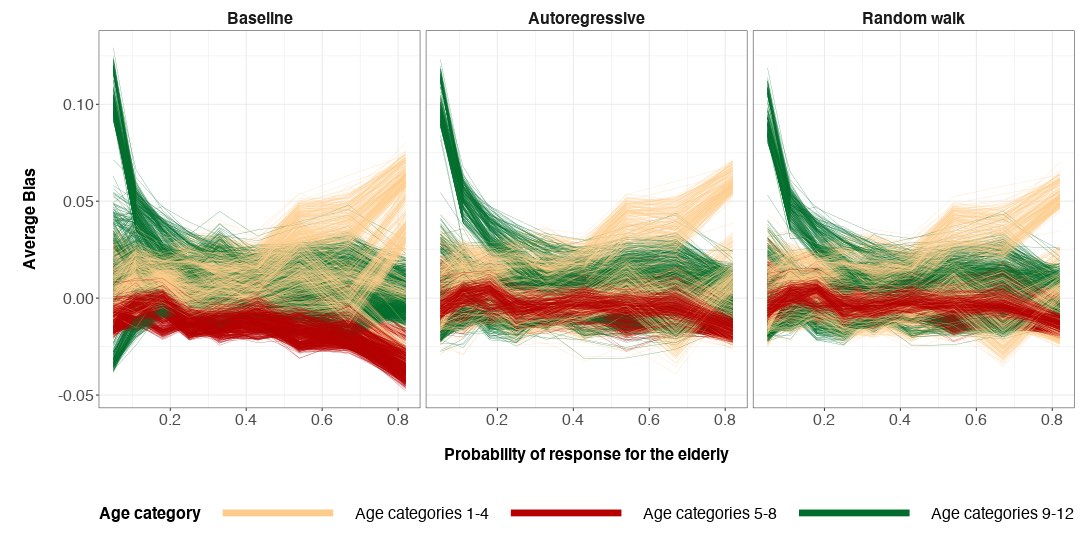}
    
    \caption[short]{The average bias values coming from 100 simulations of posterior medians of the 2448 poststratification cells. The possible values of average bias are in the interval $(-1,1)$. The possible values of average bias are in the interval $(-1,1)$. Sample size $n = 1000$. The true preference curve for age is cap-shaped. $y=0$ represents zero bias.}
    \label{fig:biasfacet_100_12_1_1000_cap}
\end{figure}


\begin{figure}
    \centering
    \includegraphics[width=1\textwidth]{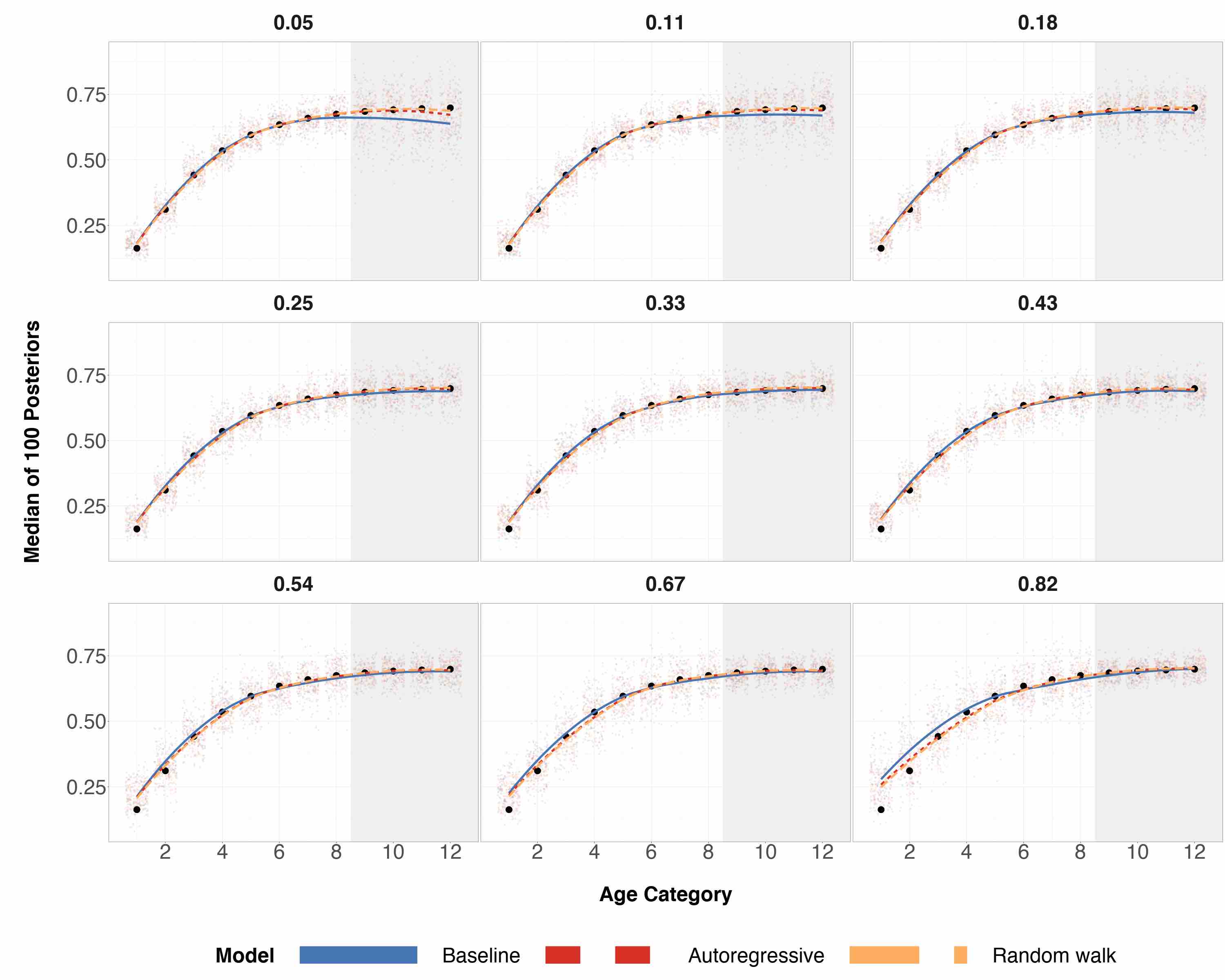}
    
\caption[short]{Posterior medians for 100 simulations with 12 age groups, where true age preference is increasing-shaped and sample size $n=1000$. Black circles are true preference probabilities for each age group. The numerical index for the 9 plots correspond to the expected proportion of the sample that are older adults (also known as the probability of sampling the subpopulation group with age categories 9--12). The shaded gray region corresponds to the age categories of older individuals for which we over/under sample. The center of the grid represents completely random sampling \textit{and} representative sampling for age categories. Local regression is used for the smoothed estimates amongst the three prior specifications.}
\label{fig:allmedians_facet_100_12_1000_increasing}
\end{figure}

\begin{figure}
    \centering
    \includegraphics[width=1\textwidth]{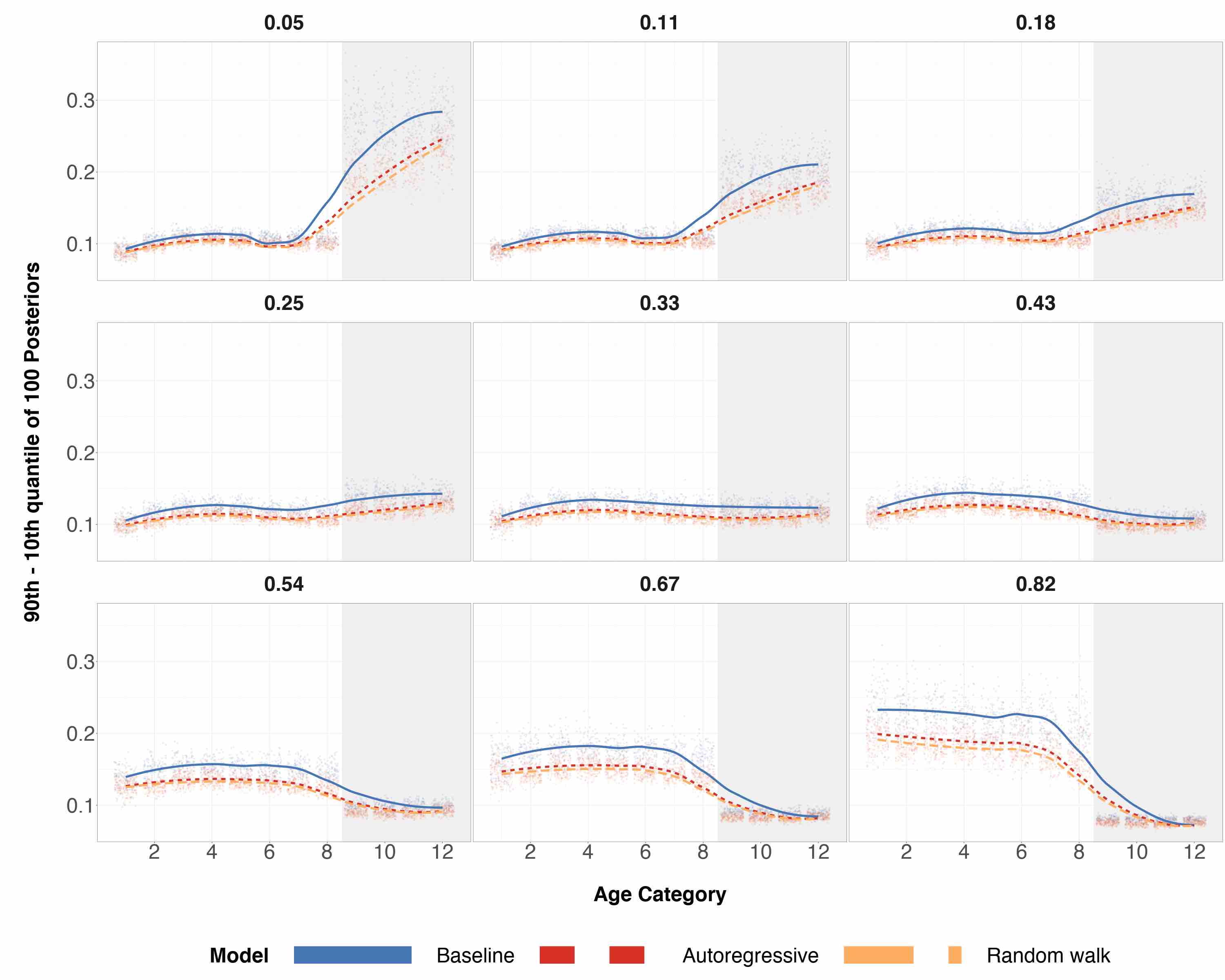}
    
\caption[short]{Differences in the $90^{th}$ and $10^{th}$ posterior quantiles for every age category when true age preference is increasing-shaped and $n=1000$ for 100 simulations. The numerical index for the 9 plots correspond to the expected proportion of the sample that are older adults (also known as the probability of sampling the subpopulation group with age categories 9--12). The shaded gray region corresponds to the age categories of older  individuals for which we over/under sample. The center of the  grid represents completely random sampling \textit{and} representative sampling for age categories. Local regression is used for the smoothed estimates amongst the three prior specifications.}
\label{fig:allquantilediff_facet_100_12_1000_increasing}
\end{figure}

\begin{figure}
    \centering
    \includegraphics[width=1\textwidth]{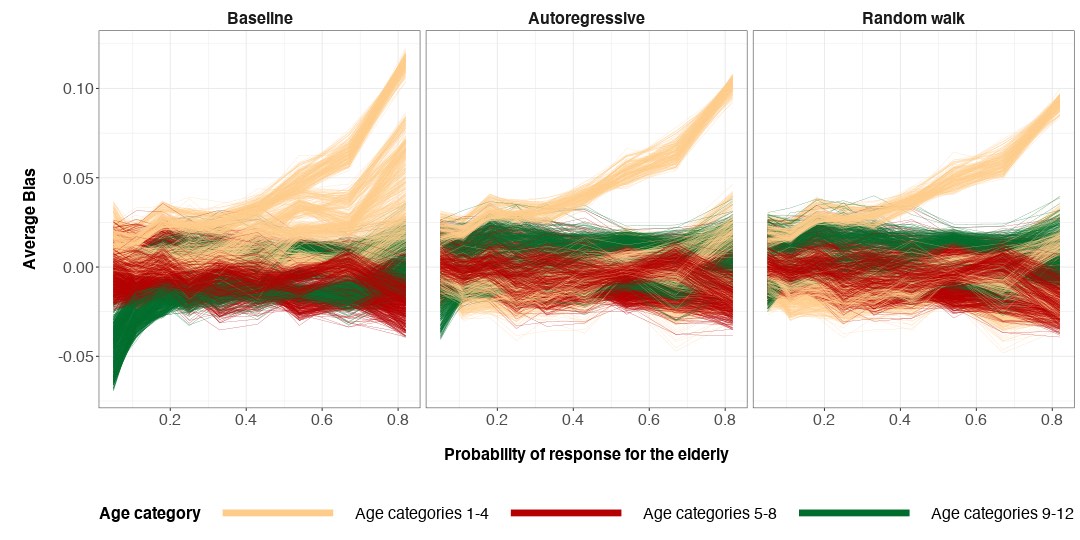}
    
    \caption[short]{The average bias values coming from 100 simulations of posterior medians of the 2448 poststratification cells. The possible values of average bias are in the interval $(-1,1)$. Sample size $n = 1000$. The true preference curve for age is increasing-shaped. $y=0$ represents zero bias.}
    \label{fig:biasfacet_100_12_1_1000_increasing}
\end{figure}


\begin{figure}
    \centering
    \includegraphics[width=1\textwidth]{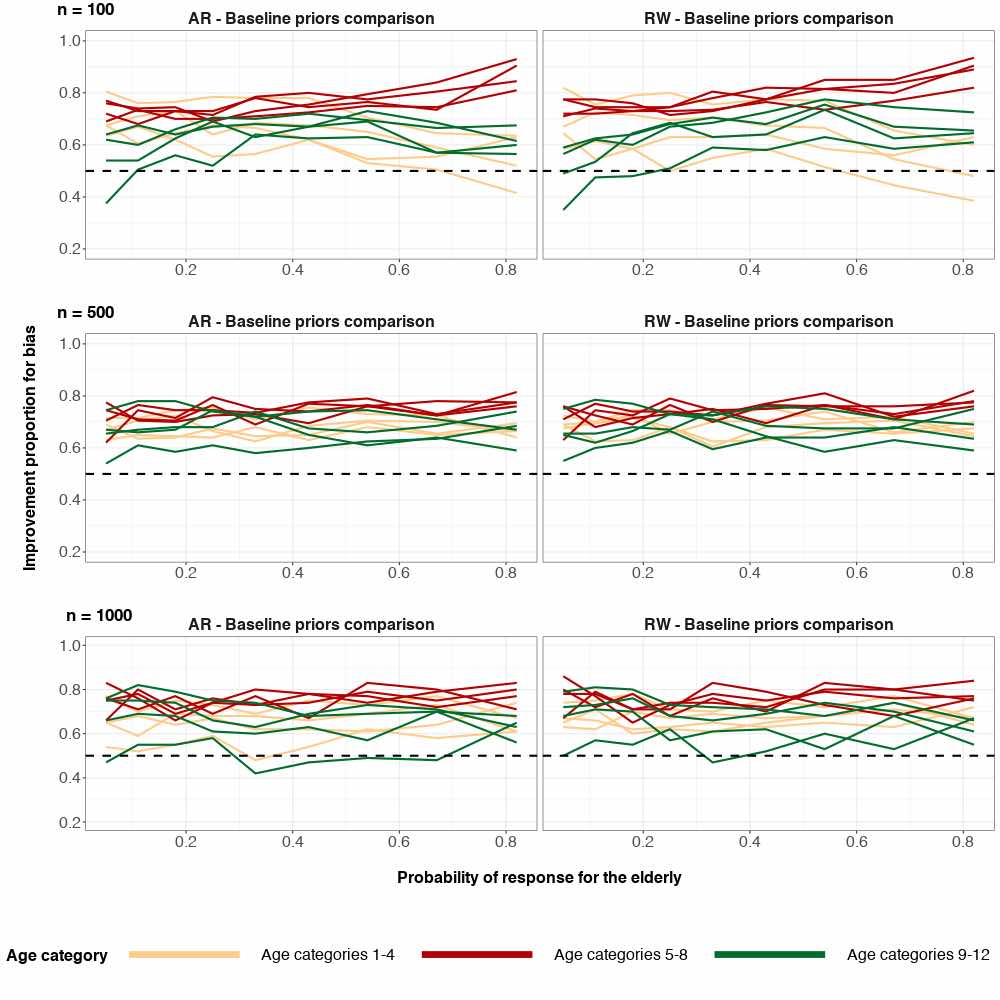}
    \caption{The proportion of the time that structured priors have lower absolute posterior median bias when compared to baseline priors, for each age category. True age preference is U-shaped. Sample size $n = 100$ (top), $n = 500$ (middle), $n = 1000$ (bottom). The top and middle rows are based on 200 simulation runs and the bottom row is based on 100 simulation runs. The right column corresponds to comparison of the random-walk prior and the baseline prior for age category. The left column corresponds to comparison of the autoregressive prior and the baseline prior for age category. The horizontal dashed line $y = 0.5$ represents equal proportion.}
    \label{fig:proportion_agecat_bias_u}
\end{figure}

\begin{figure}
    \centering
    \includegraphics[width=1\textwidth]{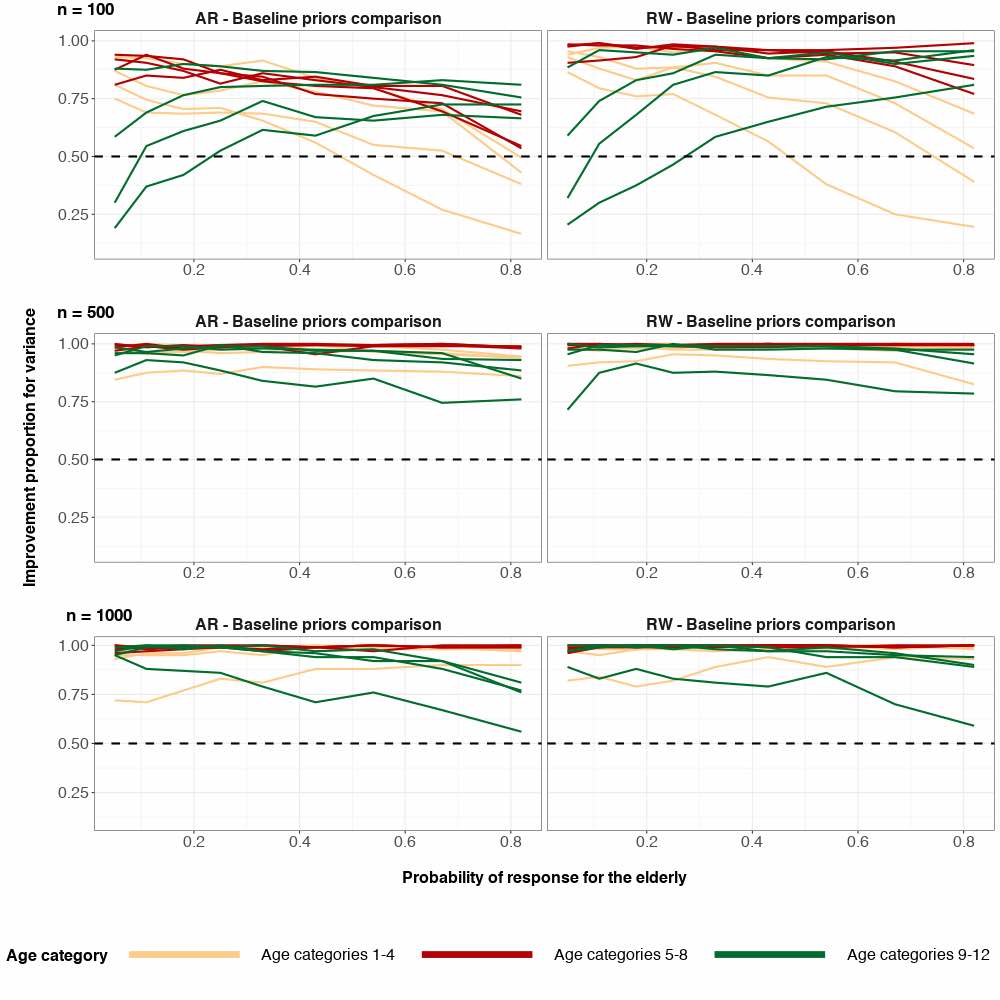}
    \caption{The proportion of the time that structured priors have lower posterior variance when compared to baseline priors, for each age category. True age preference is U-shaped. Sample size $n = 100$ (top), $n = 500$ (middle), $n = 1000$ (bottom). The top and middle rows are based on 200 simulation runs and the bottom row is based on 100 simulation runs. The right column corresponds to comparison of the random-walk prior and the baseline prior for age category. The left column corresponds to comparison of the autoregressive prior and the baseline prior for age category. The horizontal dashed line $y = 0.5$ represents equal proportion. The difference of the ${90}^{th}$ and ${10}^{th}$ posterior quantiles is used as a measure for posterior variance.}
    \label{fig:proportion_agecat_sd_u}
\end{figure}

\begin{figure}
    \centering
    \includegraphics[width=1\textwidth]{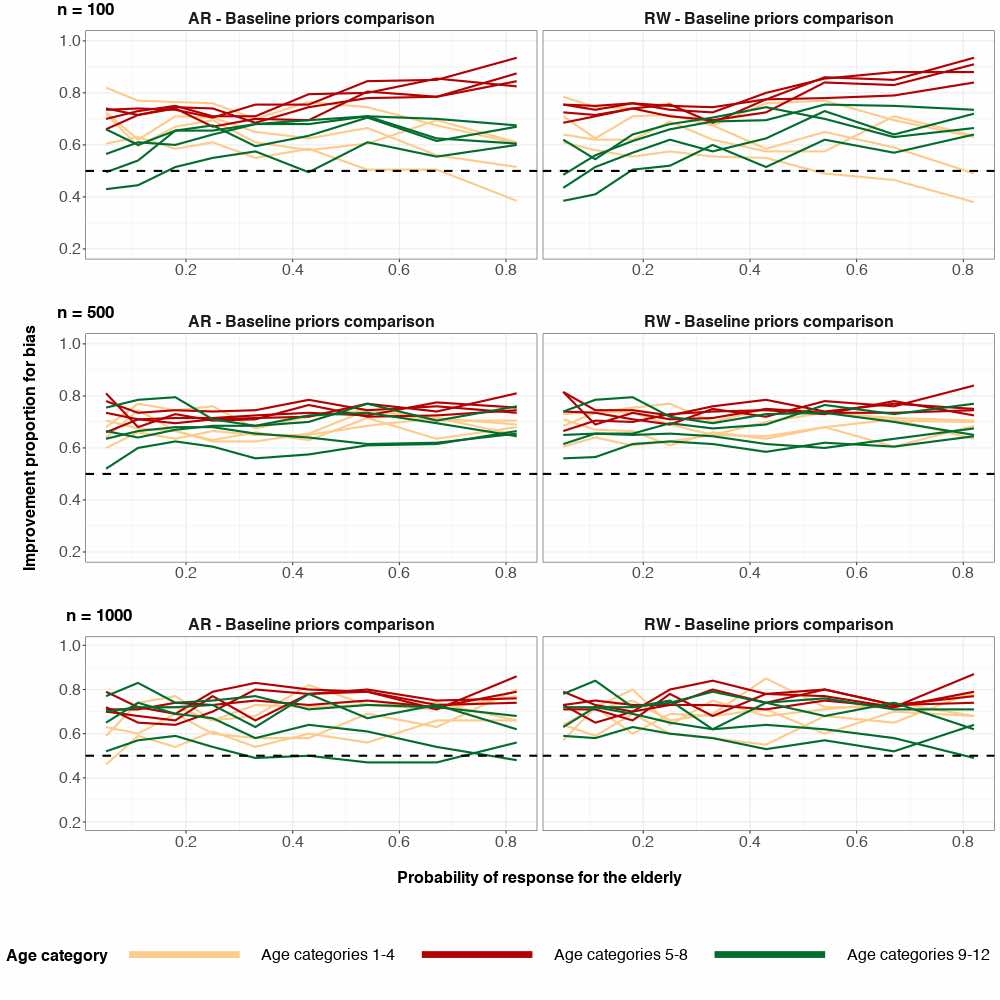}
    \caption{The proportion of the time that structured priors have lower absolute posterior median bias when compared to baseline priors, for each age category. True age preference is cap-shaped. Sample size $n = 100$ (top), $n = 500$ (middle), $n = 1000$ (bottom). The top and middle rows are based on 200 simulation runs and the bottom row is based on 100 simulation runs. The right column corresponds to comparison of the random-walk prior and the baseline prior for age category. The left column corresponds to comparison of the autoregressive prior and the baseline prior for age category. The horizontal dashed line $y = 0.5$ represents equal proportion.}
    \label{fig:proportion_agecat_bias_cap}
\end{figure}

\begin{figure}
    \centering
    \includegraphics[width=1\textwidth]{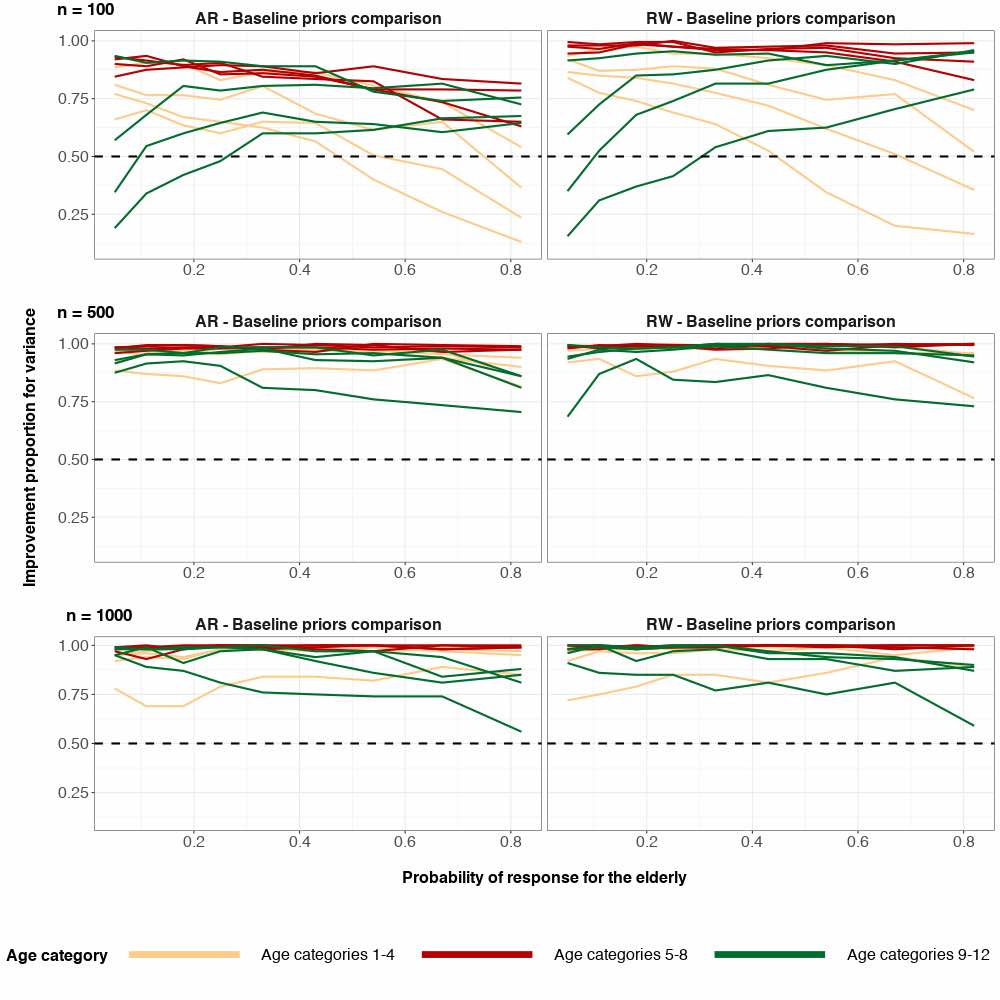}
    \caption{The proportion of the time that structured priors have lower posterior variance when compared to baseline priors, for each age category. True age preference is cap-shaped. Sample size $n = 100$ (top), $n = 500$ (middle), $n = 1000$ (bottom). The top and middle rows are based on 200 simulation runs and the bottom row is based on 100 simulation runs. The right column corresponds to comparison of the random-walk prior and the baseline prior for age category. The left column corresponds to comparison of the autoregressive prior and the baseline prior for age category. The horizontal dashed line $y = 0.5$ represents equal proportion. The difference of the ${90}^{th}$ and ${10}^{th}$ posterior quantiles is used as a measure for posterior variance.}
    \label{fig:proportion_agecat_sd_cap}
\end{figure}

\begin{figure}
    \centering
    \includegraphics[width=1\textwidth]{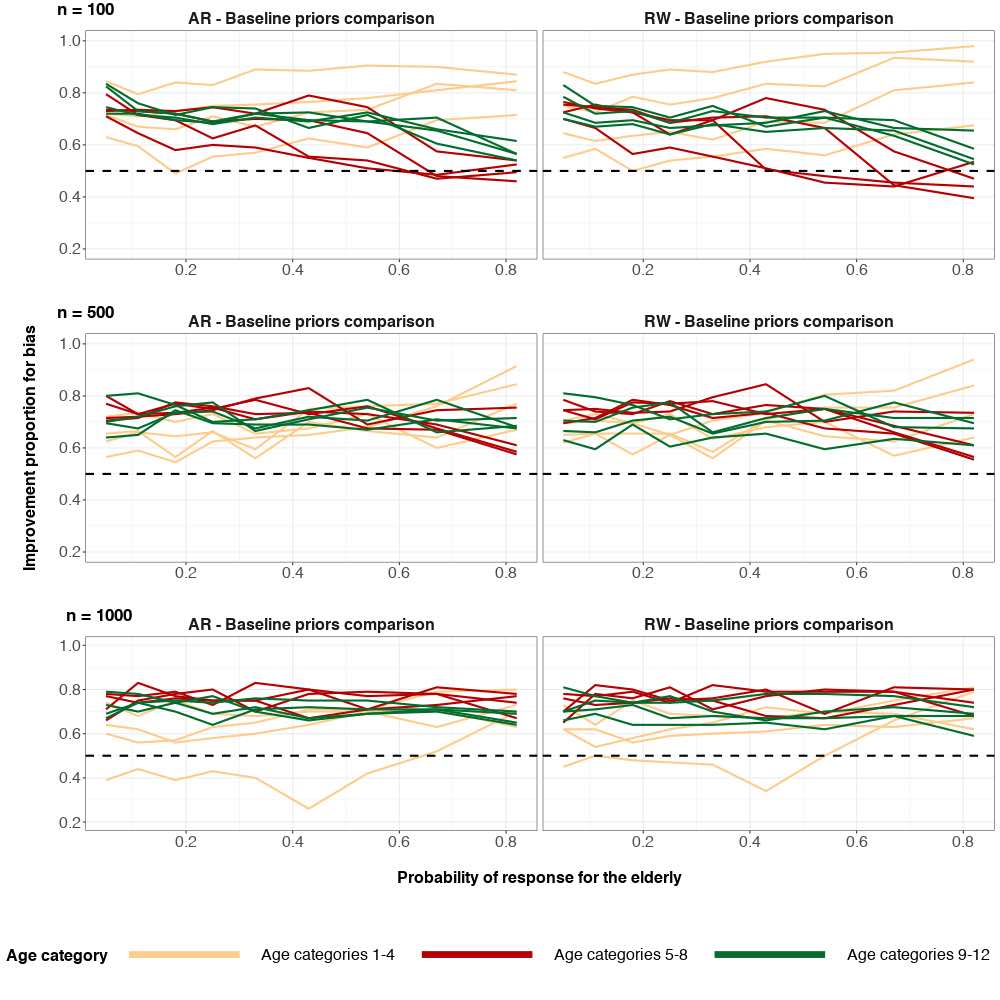}
    \caption{The proportion of the time that structured priors have lower absolute posterior median bias when compared to baseline priors, for each age category. True age preference is increasing-shaped. Sample size $n = 100$ (top), $n = 500$ (middle), $n = 1000$ (bottom). The top and middle rows are based on 200 simulation runs and the bottom row is based on 100 simulation runs. The right column corresponds to comparison of the random-walk prior and the baseline prior for age category. The left column corresponds to comparison of the autoregressive prior and the baseline prior for age category. The horizontal dashed line $y = 0.5$ represents equal proportion.}
    \label{fig:proportion_agecat_bias_increasing}
\end{figure}

\begin{figure}
    \centering
    \includegraphics[width=1\textwidth]{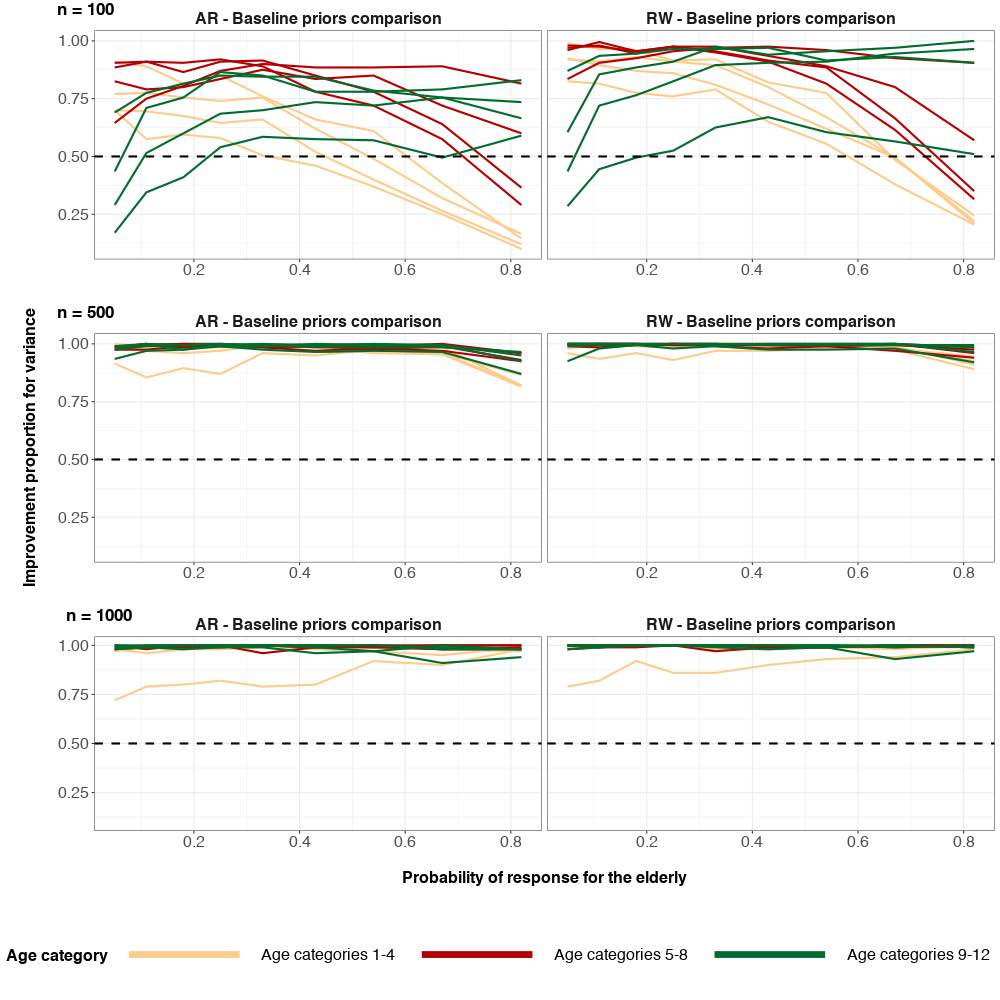}
    \caption{The proportion of the time that structured priors have lower posterior variance when compared to baseline priors, for each age category. True age preference is increasing-shaped. Sample size $n = 100$ (top), $n = 500$ (middle), $n = 1000$ (bottom). The top and middle rows are based on 200 simulation runs and the bottom row is based on 100 simulation runs. The right column corresponds to comparison of the random-walk prior and the baseline prior for age category. The left column corresponds to comparison of the autoregressive prior and the baseline prior for age category. The horizontal dashed line $y = 0.5$ represents equal proportion. The difference of the ${90}^{th}$ and ${10}^{th}$ posterior quantiles is used as a measure for posterior variance.}
    \label{fig:proportion_agecat_sd_increasing}
\end{figure}


\begin{figure}
    \centering
    \includegraphics[width=1\textwidth ]{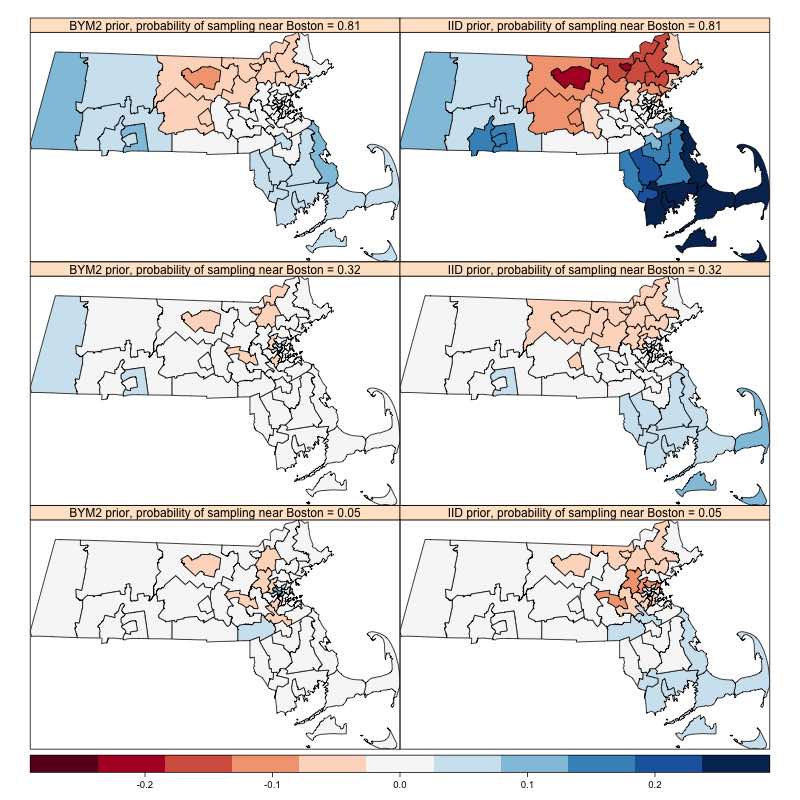}

\caption[short]{Average bias of posterior medians for every PUMA based off 200 simulations and binary response sample size is 1000. The possible values of average bias are in the interval $(-1,1)$. The left column corresponds to the BYM2 spatial prior for PUMA effect. The right column corresponds to an IID prior for PUMA effect. The probabilities 0.81, 0.32 and 0.05 in the top, middle and bottom rows respectively correspond to the probability of sampling an individual in the cluster of 17 PUMA near Boston.}
\label{fig:abs_avg_bias_postmedian_puma_forpaper200_1000_}
\end{figure}

\begin{figure}
    \centering
    \includegraphics[width=1\textwidth]{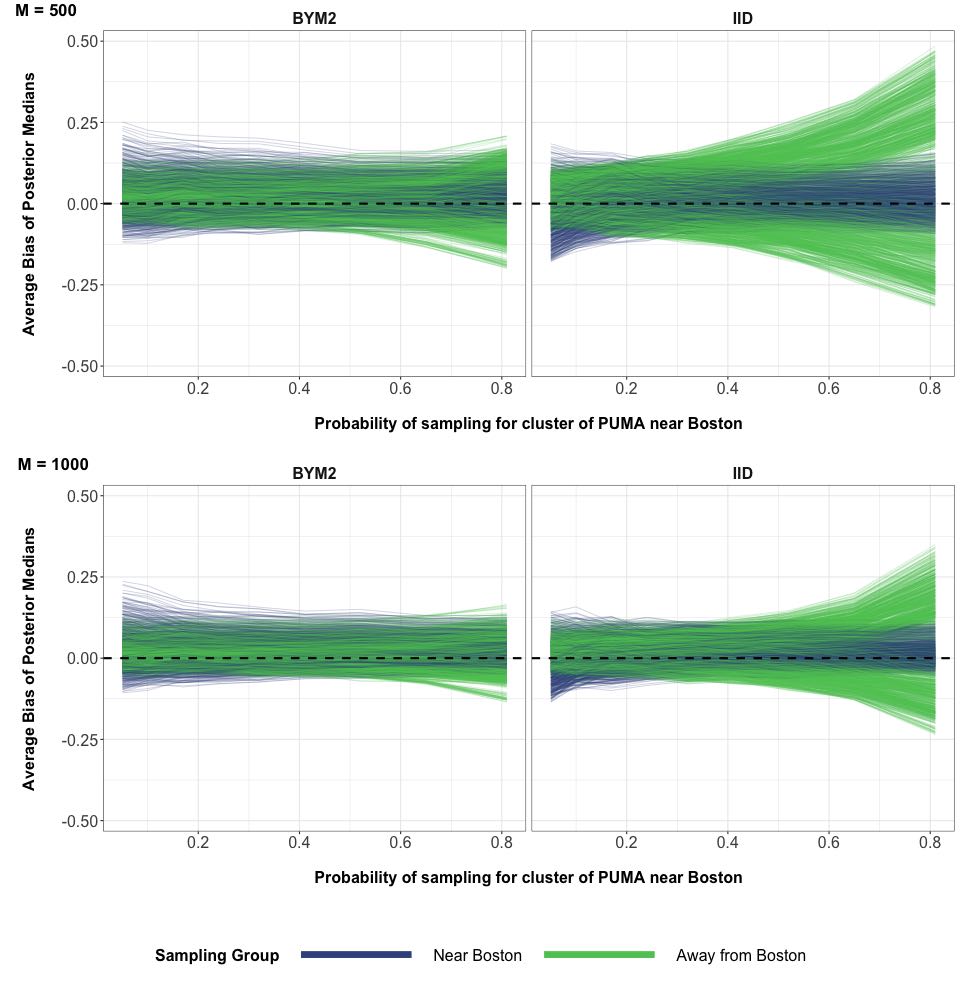}

\caption[short]{The average bias values coming from 200 simulations of posterior medians of the 1872 poststratification cells for the spatial MRP simulation. The possible values of average bias are in the interval $(-1,1)$. $M$ is the number of binary responses in every simulated data set. The top row corresponds to 500 binary responses used to define binomial responses for every simulation iteration. The bottom row corresponds to 1000 binary responses used to define binomial responses for every simulation iteration. The horizontal dashed line at y = 0 represents zero bias.}
\label{fig:biasfacetspatial_200_500_1000}
\end{figure}

\begin{figure}
    \centering
    \includegraphics[width=1\textwidth]{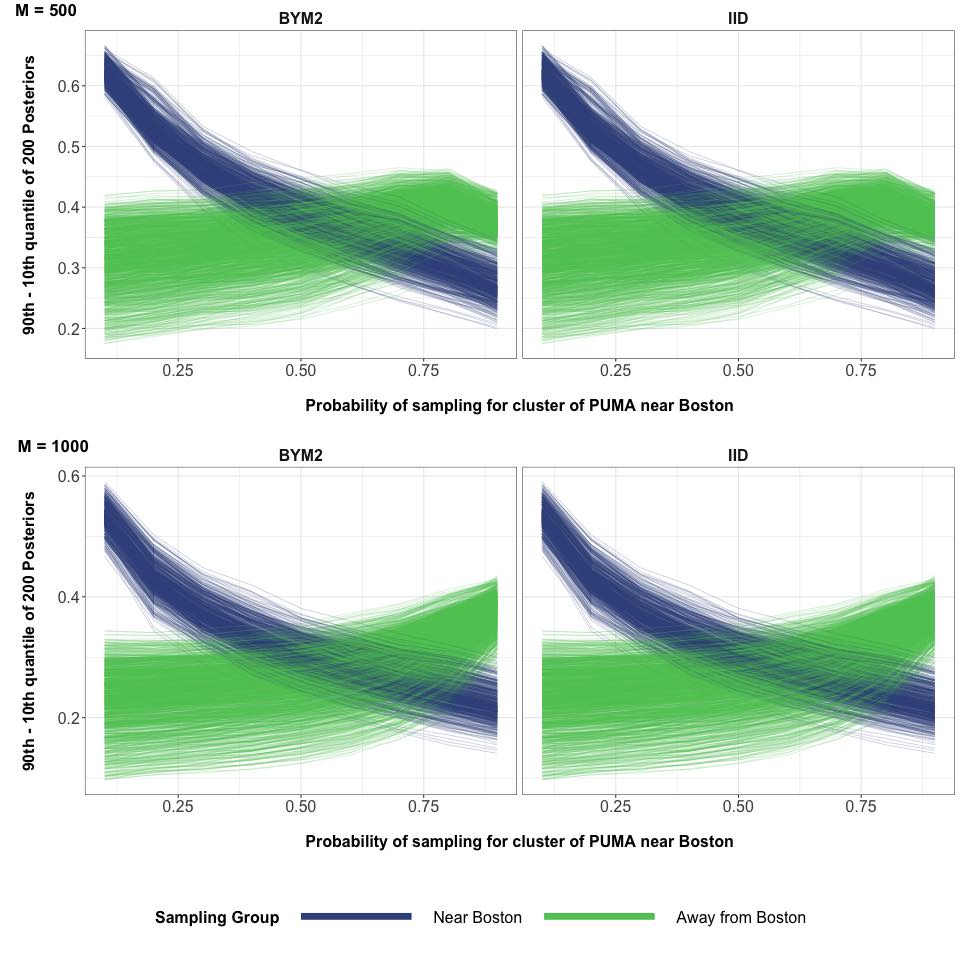}
    \caption{Average differences in the $90^{th}$ and $10^{th}$ posterior quantiles of the 1872 poststratification cells for the spatial MRP simulation. $M$ is the number of binary responses in every simulated data set. The top row corresponds to 500 binary responses used to define binomial responses for every simulation iteration. The bottom row corresponds to 1000 binary responses used to define binomial responses for every simulation iteration. }
    \label{fig:biasfacetspatial_sd_200_500_1000}
\end{figure}

\begin{figure}
    \centering
    \includegraphics[width=1\textwidth]{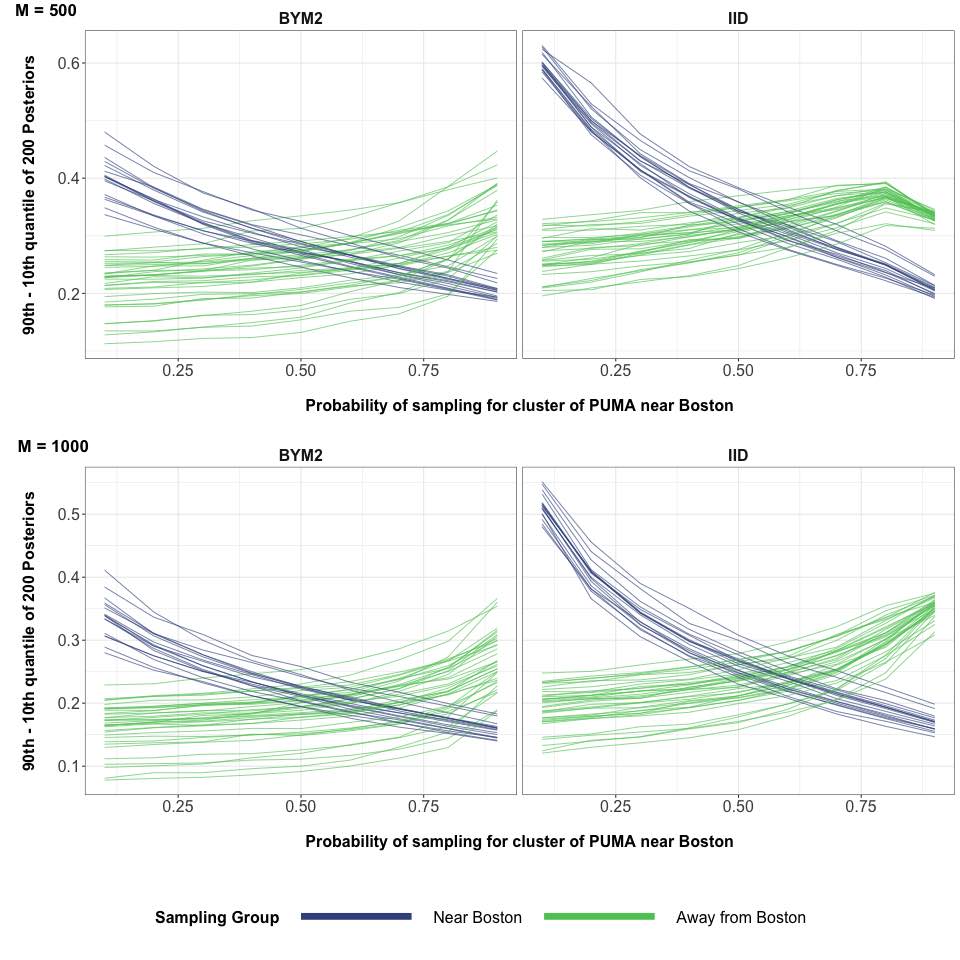}
    \caption{Average differences in the $90^{th}$ and $10^{th}$ posterior quantiles of the 52 PUMA for the spatial MRP simulation. $M$ is the number of binary responses in every simulated data set. The top row corresponds to 500 binary responses used to define binomial responses for every simulation iteration. The bottom row corresponds to 1000 binary responses used to define binomial responses for every simulation iteration.}
    \label{fig:biasfacetspatialpuma_sd_200_500_1000}
\end{figure}

\end{document}